\documentclass[12pt]{iopart}

\expandafter\let\csname equation*\endcsname\relax
\expandafter\let\csname endequation*\endcsname\relax
\usepackage{amsfonts}
\usepackage{amsmath}
\usepackage{amssymb}
\usepackage{graphicx}
\usepackage{color}
\usepackage{float}
\usepackage{mathtools}
\usepackage{algorithm}
\usepackage{algpseudocode}
\usepackage{wrapfig}
\usepackage{mathbbol}
\usepackage[hyperindex,breaklinks,colorlinks,linkcolor=blue, citecolor = blue, urlcolor = blue,linktocpage=false]{hyperref}
\usepackage{cite}
\usepackage{upgreek}
\usepackage[normalem]{ulem}

\newcommand{\Kd}{\mathcal{K}}
\newcommand{\I}{{\mathbb{i}}}
\newcommand{\hkl}{\texttt{h}\texttt{k}\texttt{l}}
\newcommand{\cc}{\text{c.c.}}
\newcommand{\dd}{\text{d}}

\newcommand{\no}{n_{\rm o}}

\newcommand{\tpar}{\texttt{t}}
\newcommand{\vpar}{\texttt{v}}
\newcommand{\DBpar}{\Delta\texttt{B}^0}
\newcommand{\Bxpar}{\texttt{B}^x}

\newcommand \Lap {\nabla^2}
\newcommand \Div {\nabla \cdot}

\newcommand \Grad { \nabla }

\newcommand \etai {\eta_{\hkl}}

\newcommand \Gi {\mathbf{G}_{\hkl}}

\newcommand \si {\sum_{\hkl}}

\newcommand \vepm {\varepsilon_{\rm m}}

\newcommand \rv {\mathbf{r}}
\newcommand \qv {\mathbf{q}}
\newcommand \av {\mathbf{a}}
\newcommand \kv {\mathbf{k}}
\newcommand \Gv {\mathbf{G}}
\newcommand \uv {\mathbf{u}}
\newcommand \bv {\mathbf{b}}
\newcommand \vv {\mathbf{v}}
\newcommand \BBv {\mathbf{B}}

\newcommand \pp[2] {\frac{\partial #1}{\partial #2}}

\newcommand \expF[1] {\e^{#1}} 

\begin{document}
\
{\textbf{TOPICAL REVIEW}}

\title[Amplitude Phase-Field Crystal: an overview]{Coarse-grained modeling of crystals by the amplitude expansion of the phase-field crystal model: an overview}

\author{Marco Salvalaglio}
\address{Institute of Scientific Computing, TU-Dresden, 01062 Dresden, Germany}
\address{Dresden Center for Computational Materials Science, TU-Dresden, 01062 Dresden, Germany}
\ead{marco.salvalaglio@tu-dresden.de}
\vspace{10pt}

\author{Ken R. Elder}
\address{Department of Physics, Oakland University, Rochester, Michigan 48309, USA.}
\ead{elder@oakland.edu}

\begin{abstract}
Comprehensive investigations of crystalline systems often require methods bridging atomistic and continuum scales. In this context, coarse-grained mesoscale approaches are of particular interest as they allow the examination of large systems and time scales while retaining some microscopic details. The so-called Phase-Field Crystal (PFC) model conveniently describes crystals at diffusive time scales through a continuous periodic field which varies on atomic scales and is related to the atomic number density. To go beyond the restrictive atomic length scales of the PFC model, a complex \textit{amplitude} formulation was first developed by Goldenfeld {\it et al.} [Phys. Rev. E 72, 020601 (2005)]. While focusing on length scales larger than the lattice parameter, this approach can describe crystalline defects, interfaces, and lattice deformations. It has been used to examine many phenomena including liquid/solid fronts, grain boundary energies, and strained films. This topical review focuses on this amplitude expansion of the PFC model and its developments. An overview of the derivation, connection to the continuum limit, representative applications, and extensions is presented. A few practical aspects, such as suitable numerical methods and examples, are illustrated as well. Finally, the capabilities and bounds of the model, current challenges, and future perspectives are addressed. 
\end{abstract}

%
\vspace{20pt}
%
\maketitle
%
%

\newpage
\tableofcontents
\markboth{Amplitude Phase-Field Crystal: an Overview}{}

\newpage
\section{Introduction}

The original phase-field crystal (PFC) model, introduced in 2002 \cite{Elder2002}, was developed 
as a simple way to incorporate 
elasticity and dislocations in continuum models in a manner similar to how 
interface and domain boundaries are introduced in traditional phase-field (PF)
models. In the latter case, the predictions of PF models can be shown to 
be consistent in the asymptotic limit of vanishing interface widths with well-known sharp interface 
(SI) models \cite{Langer80} that explicitly track the position of a given interface subject to various boundary conditions (such as, e.g., the Gibbs-Thomson condition (GTC)
for solidification or spinodal decomposition). PF models do not typically
provide quantitative predictions on small length scales, i.e., on the scale 
of interfacial widths or suitable correlation lengths. Usually, their parameters are chosen to match the ones entering SI
models \cite{Karma96,Karma98,Elder2001} 
(e.g., the capillary length and coefficient of kinetic undercooling 
that enter the GTC). Similarly, PFC models do not
quantitatively describe small length scale features, but in the appropriate
limit they reduce to standard results. It is straightforward to show that in the long-wavelength limit, the PFC free energy reduces to traditional 
continuum elasticity theory \cite{ElderPRE2010} and that the dynamics incorporate 
vacancy diffusion \cite{Elder2002,Elder2004}.  It has been shown, numerically in two dimensions, that GBs can form spontaneously and their energy is consistent
with the Read-Shockley equation \cite{Elder2002,Elder2004,mot11,Hirvonen2016}, 
that climb and glide of dislocations follow the Orowan equation \cite{mot3}, 
and in three dimensions that glide (climb) mediated sources of dislocation are consistent 
with Frank-Read (Bardeen-Herring) mechanisms \cite{Berry2012}.
More recently, it has been shown analytically that in PFC 
models the velocity of dislocations is determined by the Peach-Koehler force as expected
in pure \cite{SkaugenPRB2018} and binary systems \cite{SalvalaglioPRL2021}.
In addition, the predicted elastic
fields around a dislocation agree quantitatively 
with continuum elasticity theory, encoding additional features such as anisotropies and non-linearities
\cite{SalvalaglioNPJ2019,SkaugenPRL2018,SalvalaglioJMPS2020}.    
In many ways, the connection between PF and sharp interface approaches is analogous to the connection of PFC models with dislocation dynamics (DD) models \cite{DD0,DD1,DD2}, which
explicitly move dislocation lines due to Peach-Koehler forces that are 
generated by the elastic field of other dislocations, defects, or externally
applied forces. In particular, the coarse-grained PFC model referred to in the literature as \textit{amplitude expansion of the PFC}, \textit{complex amplitude phase-field crystal} or simply \textit{amplitude equations}, on which this review focuses, allows a description of defects without resolving atomistic length scales, closely resembling the basic features of DD models. The advantage of this approach over DD is that dislocations and their main phenomenology appear naturally, following from the considered free energy functional. Therefore, no external rules would be in principle needed to determine the interaction, annihilation, or creation of any type of defect. At the same time, the method is not restricted to a single-crystal sample with pre-defined glide planes. However, it is worth noting that quantitative description of specific phenomena and materials would require an extended parametrization compared to minimal PFC-like models typically reported in the literature. Such extensions may be achieved with later formulations \cite{Greenwood2010,xtal2} but to date, they have not been explored extensively in this regard.

The complex amplitude phase-field crystal (APFC) model was originally derived 
by Goldenfeld \etal \cite{Goldenfeld2005,Athreya2006}
from the PFC model, which describes the evolution of the atomic number density during crystallization and the related dynamic 
processes \cite{Elder2002,Elder2004,elder2007}.  While the PFC model can access diffusive time scales, the approach is limited by the need 
to incorporate density fluctuations on atomic length scales,  
thus requiring resolutions smaller than the lattice spacing. 
The main aspect of the APFC approach is to model the amplitude of 
the density fluctuations instead of the density itself.
The idea of describing liquid/solid 
transitions by amplitudes that are real has been exploited 
in the past \cite{Khachaturyan1983,Khachaturyan1996,cross1993pattern}. 
In Goldenfeld \etal's formulation \cite{Goldenfeld2005,Athreya2006}, 
density fluctuations are described by complex amplitudes, $\etai$, where 
$\hkl$ are Miller indices that describe specific crystallographic planes.
The magnitude of $\etai$ is finite in a crystal and zero in the liquid 
state. Thus, it can be used to characterize a liquid-solid transition. 
Gradients in the phase of $\etai$ occur when the crystal state is strained, 
which provides information about the elastic energy stored in the crystal. In addition, 
the phase can describe the rotation of the crystal, allowing for the 
study of polycrystalline states (although, as noted in 
Sec.~\ref{sec:limits_extensions}, there exist limitations). 
Finally, the combination of the magnitude and phase can 
describe dislocations in which large gradients in the phase do 
not lead to huge increases in the elastic energy as the magnitude of
$\etai$ goes to zero. While the APFC model is formally derived from the PFC model, it is in principle possible to phenomenologically write down an APFC model as long as it has the correct long-wavelength behavior as has been done for PF models of various phenomena.

One of the most important features of the APFC model is that it provides a natural bridge between atomic and mesoscopic continuum length scales. In a single crystal state, the amplitudes vary slowly in space (depending on the orientation) but can be used to reconstruct the underlying atomic density fluctuations completely. On long length scales, it is straightforward to derive standard continuum elasticity through the phase of the amplitudes. Significant variations of amplitudes occur at defects and solid-liquid interfaces, still well describing the deformation induced in the lattice. The equations entering the APFC model, similarly to PFC, can be solved with simple numerical approaches.
For example, using a uniform grid, Smirman \etal \cite{Smirman2017} studied Moir\'e patterns in graphene films with the largest size system of $19.6\,\mu$m $\times$ $33.9\,\mu$m containing more than 25 billion unit cells (although it should be noted that these patterns contain no defects). When dislocations, grain boundaries, and interfaces appear, i.e. when a significant local variation of amplitudes occurs, more advanced numerical approaches can be considered to optimize the calculations. Indeed, these regions require the finest resolution, while a coarser one, typically much larger than the atomic spacing, can be used elsewhere. Adaptive meshing schemes then allow for simulation of large mesoscopic scales and at the same time completely retaining atomic information. Thus the APFC method allows simulations of atomistic features on continuum scales and should play an important role in understanding complex phenomena with multiscale features.

The rest of the review is organized as follows. Section~\ref{sec:origin} describes the original PFC model and the derivation of the APFC model. 
Section~\ref{sec:numerical-methods} outlines various numerical methods that 
have been developed to solve the APFC on regular and adaptive meshes. This 
is followed by Section~\ref{sec:continuumlimit} that provides a connection 
of the APFC model to traditional models of continuum elasticity and plasticity. 
Section~\ref{sec:limits_extensions} outlines the limitations of the approach and some extensions aimed at overcoming some of these constraints.
Following this is Section~\ref{sec:applications} which describes some 
applications of the model to various physical phenomena.  Finally, some 
conclusions and future outlooks are given in Section~\ref{sec:conclusions}.

\section{From phase-field crystal to the amplitude expansion }
\label{sec:origin}
\subsection{Origin of the phase-field crystal model}
The PFC model was proposed 
phenomenologically \cite{Elder2002,Elder2004} to 
model elasticity and plasticity in crystal structures and can be written in terms of a dimensionless Helmholtz free energy functional, $F$, which 
is given as,
\begin{equation}
F_n = \int \dd\rv \left[
\frac{\DBpar}{2}n^2 + \frac{\Bxpar}{2}n(q^2_0+\Lap)^2n
- \frac{\tpar}{3} n^3
+ \frac{\vpar}{4} n^4
\right],
\label{eq:F_PFC}
\end{equation}
and an equation of motion,
\begin{equation}
\pp{n}{t}=  \Lap \frac{\delta F_n}{\delta n},
\label{eq:dndt}
\end{equation}
where $n$ is related to the atomic number density difference and 
$\DBpar, \Bxpar$, $\tpar$ and $\vpar$ are constants that may 
depend on temperature \cite{elder2007}. Although Eq.~\eqref{eq:F_PFC} 
can be derived \cite{elder2007,Huang2010,vanTeeffelen2009} from the classical 
density functional theory of Ramakrishnan and Yussouf \cite{RY79}, 
the approximations used give rise to poor atomic-scale predictions in most materials  since this  free energy is minimized by an almost 
sinusoidal density fluctuations,  while in metals 
for example $n$ is very sharply peaked Gaussians at each lattice point. 
Nevertheless the periodic nature of the solutions of Eq.~\eqref{eq:F_PFC}, 
which mimic a time average of microscopic atomic density \cite{Tupper_2008} 
and evolves over diffusive time scales \cite{Emmerich2012}, make it useful 
for studying a large 
variety of physical systems such as multi-component polycrystals, liquid crystals, quasi-crystals and colloids as well as a broad class of phenomena including crystal growth and nucleation, heteroepitaxy, pattern formation, dislocation dynamics, grain boundary morphology and motion \cite{Elder2004,Emmerich2012,Berry2014,Backofen14,GRANASY2019}. PFC models have been developed also for less conventional materials and systems such as, for instance, active crystals \cite{Alaimo2016,Alaimo2018,Huang2020,Menzel2013,Menzel2014}, active colloids \cite{Praetorius2015}, and viral capsids \cite{Aland2012}. 

\begin{figure}
\includegraphics[width=\textwidth]{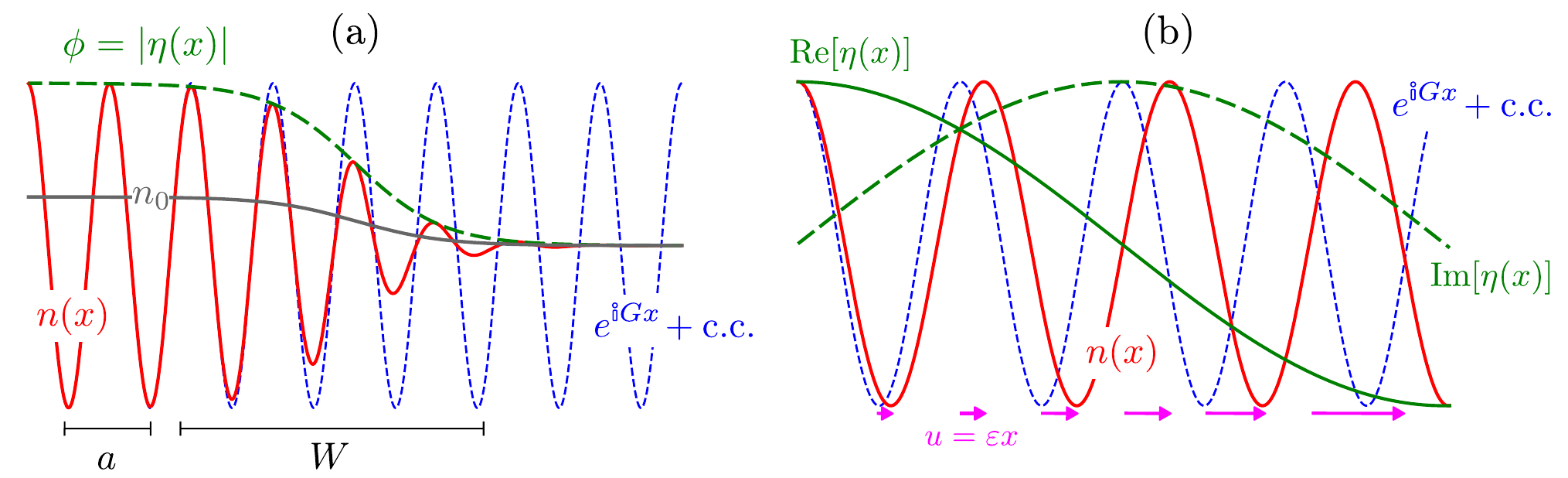}
\caption{(a) Sample (1D) liquid/solid interface, where 
$a$ is the atomic spacing and $W$ is the width of the interface. (b) Sample (1D) deformed lattice by displacement $u=\varepsilon x$.}
\label{fig:figure1}
\end{figure}

\begin{figure}
\includegraphics[width=\textwidth]{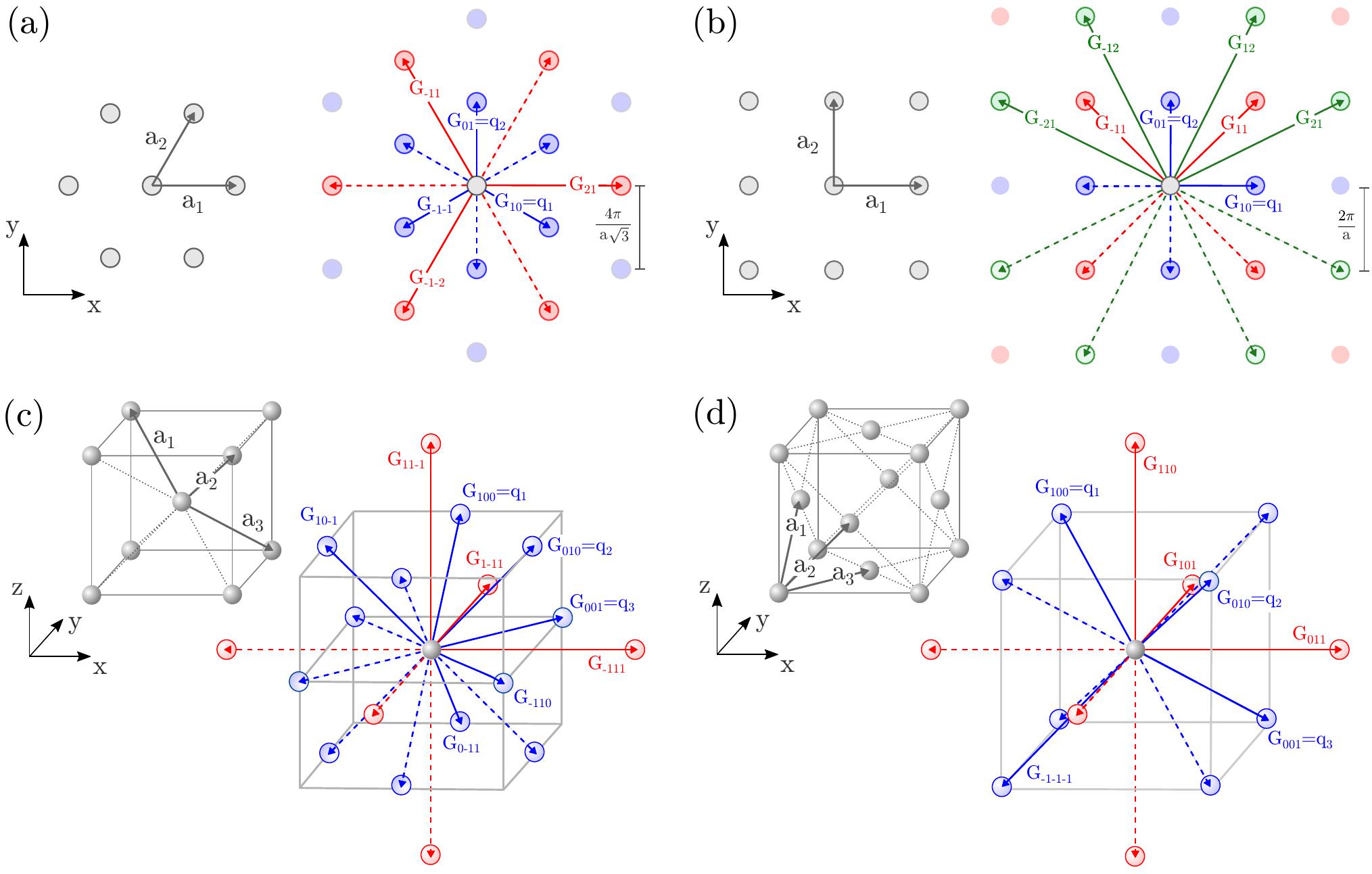}
\caption{Crystal structures (grey) and corresponding shortest reciprocal lattice vectors (colored): (a) triangular, (b) square, (c) body-centered cubic, (d) face-centered cubic. Arrows represent the reciprocal space vectors entering Eq.~\eqref{eq:n} in the one- (blue) and two- (blue and red) mode approximations. For the square lattice the additional reciprocal-space vectors considered in a three-mode approximation involving non-parallel vectors only are also shown (green). Solid arrows indicate an explicit choice of vectors entering Eq.~\eqref{eq:nnn} (as exploited from Sec.~\ref{sec:results_symm} on).}
\label{fig:figure2}
\end{figure}

The fact that the solutions are not sharply peaked means that 
they can be described by a few Fourier components.  
In this 
regard the density is written  in terms of complex amplitudes, $\etai$, as 
follows,
\begin{equation}
n = \no + \si \etai \expF{\I \Gi \cdot \rv},
\label{eq:n}
\end{equation}
where $\no$ is the average density, 
$\Gi=\texttt{h}\qv_1+\texttt{k}\qv_2+\texttt{l}\qv_3$ are reciprocal 
lattice vectors, with $\qv_1=2\pi (\av_2 \times \av_3)/(\av_1\cdot(\av_2 
\times \av_3))$ and cyclic permutations of (1,2,3) the principal reciprocal-lattice vectors, 
and $\av_j$ the vectors defining the primitive cell of the crystal lattice \cite{ashcroft1976solid}. Note that the summation goes over 
both negative and positive $\Gi$'s with $\eta_{-(\hkl)}=\etai^*$ such that $n$ is a real field. In two dimensions, one may define $\Gi$ as above with $\texttt{l}=0$, $\mathbf{q}_i=2\pi\mathcal{R}\mathbf{a}_j/(\mathbf{a}_i\cdot (\mathcal{R}\mathbf{a}_j))$ for $i\neq j$ and $\mathcal{R}$ a 90$^\circ$ rotational matrix (clockwise or anti-clockwise). All these definitions satisfy the condition $\mathbf{a}_i \cdot \mathbf{q}_j=2\pi\delta_{ij}$. Two illustrations of the quantities entering Eq.~\eqref{eq:n} in 1D are shown in Fig.~\ref{fig:figure1}, namely corresponding to a solid-liquid interface and a uniformly strained 1D crystal. Since PFC type models produce smooth solutions it is a good approximation to use the fewest number of complex amplitudes that are needed for any given crystal symmetry (see also Fig.~\ref{fig:figure2}). For example, only six 
$\etai$ (so three independent $\etai$) are needed for a 2D triangular lattice (more explicit examples are given in Sec.~\ref{sec:results_symm}). $\Gi$  entering approximations with the smallest number of modes are shown in Fig.~\ref{fig:figure2}. As discussed in the next section the goal of the APFC model is to derive equations of motion for the amplitudes.

\subsection{Derivation}
\label{sec:derivation}

    There are various methods for deriving the amplitude expansion 
from the original PFC model.  Essentially, it requires a separation 
of length scales by assuming that the complex amplitudes vary 
on length scales much larger than the atomic spacing.  In general 
this is the same assumption of all phase field models which require 
that interfaces or domain walls make a smooth transition from 
one phase to another.  This is illustrated in Fig.~\ref{fig:figure1} for a one dimensional liquid/solid interface for a system of atomic spacing 
$a$ and interface width $W$.   The ``phase field limit" is such that 
$a/W \ll 1$.  For instance, for a two-dimensional triangular lattice it can be 
shown \cite{Galenko2015} that in the limit that $\no=0$ and the complex amplitudes are real and identical (i.e., $\etai=\phi$, for all 
$\hkl$), they are described by traveling wave solutions (with velocity $V$) of the form,
\begin{equation}
\phi = A \bigg[1-\tanh\bigg(\frac{x-Vt}{W}\bigg)\bigg],
\label{eq:tanh}
\end{equation}
where $W$ is the width of the liquid/solid front which can be 
written \cite{Galenko2015} as
\begin{equation}
W= \frac{W^{m}}{1+\sqrt{1-(8/9)\DBpar/\DBpar_{\rm ls}}},
\end{equation}
where $\DBpar_{\rm ls}=8\tpar^2/135\vpar$ is the value of $\DBpar$ at 
liquid/solid coexistence and $W^{m}$ is the maximum value 
of $W$ and is given by
\begin{equation}
W^{m} =  2q_0\sqrt{30 \vpar \Bxpar}/\tpar.
\end{equation}
For $\DBpar > 9/8 \DBpar_{\rm ls}$ no traveling 
wave solution exists as the solid is linearly unstable.
Thus the phase field limit occurs when $\Bxpar \rightarrow \infty$ and 
as such $1/\Bxpar$ can be used as a small parameter 
in a multi-scale calculation.  In light of this, it is convenient 
to make the following rescaling, $\epsilon=-\DBpar/\Bxpar$,  
$\bar{n}=n(\vpar/\Bxpar)^{1/2}$, $\bar{F}=F\vpar/(\Bxpar)^2$, so that Eq.~\eqref{eq:F_PFC} can 
be written 
\begin{equation}
\bar{F} = \int \dd \rv \left[
-\frac{\epsilon}{2} \bar{n}^2 + \frac{1}{2}\bar{n}(q_0^2+\nabla^2)^2\bar{n} - \frac{\tau}{3} \bar{n}^3 
+ \frac{1}{4}\bar{n}^4
\right],
\end{equation}
where $\tau=\tpar/\sqrt{\vpar\Bxpar}$. Now the limit $\Bxpar\rightarrow \infty$ corresponds 
to $\epsilon\rightarrow 0$.
  
Goldenfeld and co-workers \cite{Goldenfeld2005,Athreya2006} report that to 
obtain rotationally invariant equations using multiple-scales analysis requires 
going to sixth order perturbations, which is an extremely tedious task, as to lowest 
order the resulting equations are not rotationally invariant.
However, they have shown that this analysis gives the same result using a simpler renormalization 
group calculation. Other works addressed refinement and assessment of the general renormalization 
group approach \cite{Shiwa2011,Oono2012}.

To grasp the essence of the calculations without using these more 
rigorous methods, Athreya \etal \cite{Athreya2006} developed a 
method that was coined
\textit{``quick and dirty"} that essentially obtains the same result 
in the $W \rightarrow \infty$ limit. The basic idea is to 
assume that the amplitudes are constant on atomic length scales, i.e., 
\begin{equation}
\int_{\rm u.c.} d\rv\, f(\etai) \expF{\I\qv\cdot \rv} \approx
f(\etai) \int_{\rm u.c.} d\rv\,  \expF{\I\qv\cdot \rv},
\label{eq:qd}
\end{equation}
where $\int_{\rm u.c.}$ is an integration over a unit cell and $\qv$ is a 
sum over various $\Gi$. 
Since $\qv$ is periodic in the unit cell, Eq.~\eqref{eq:qd} is zero unless 
$\qv=0$.  This is a considerable simplification that reduces the number of terms that enter the free energy.  For example, consider a term 
\begin{equation}\begin{split}
\int d\rv\, n^2 =
\int d\rv \bigg[ \no^2 + 2\no \bigg(\sum_{\hkl}\etai 
\expF{\I\Gi\cdot\rv} \bigg)+
 \bigg(\sum_{\hkl}\etai  \expF{\I\Gi\cdot\rv}\bigg) 
 \bigg(\sum_{\texttt{h}'\texttt{k}'\texttt{l}'}\eta_{\texttt{h}'\texttt{k}'\texttt{l}'} \expF{\I\mathbf{G}_{\texttt{h}'\texttt{k}'\texttt{l}'}\cdot\rv}\bigg) 
 \bigg].
 \end{split}
\label{eq:nsq}
\end{equation}
Only the first and last term for $\hkl=-(\texttt{h}'\texttt{k}'\texttt{l}')$ give non-zero contributions  
using approximation Eq.~\eqref{eq:qd}, since they do not contain terms multiplied by a periodic function.  Thus, in this 
approximation, Eq.~\eqref{eq:nsq} reduces to 
\begin{equation}
\int d\rv\, n^2 \approx  \int d\rv \bigg[ \no^2 
+ \sum_{\hkl}|\etai|^2\bigg].
\end{equation}
As discussed in the next section, contributions that arise from higher order polynomial terms will depend on the specific crystal symmetry under consideration. 
Terms containing the $\Lap$ operator are treated similarly noting that, assuming constant or slowly varying $\no$, 
\begin{equation}
\Lap n = \sum_{\hkl}\expF{\I\Gi\cdot\rv} \underbrace{(\Lap+2\I\Gi\cdot \vec{\nabla}-|\Gi|^2)}_{\mathcal{L}_{\hkl}}\etai. 
\label{eq:LapG}
\end{equation}
Thus the Laplacian operator transforms as 
$\Lap \rightarrow \mathcal{L}_{\hkl}$. 
While the effective operator on the right hand side of 
Eq.~\eqref{eq:LapG} appears to be anisotropic (due to the specific 
direction of the $\Gi$'s), it can be shown that the free energy is 
independent of the orientation of the pattern formed in $n$ \cite{Provatas2010}.  With these steps an energy functional which depends on amplitudes, $F_\eta$, can be derived (see also Sec.~\ref{sec:formulas}).

The dynamics of $\etai$ approximating \eqref{eq:dndt} can be obtained by multiplying Eq.~\eqref{eq:dndt} by $\expF{-\I\Gi\cdot \rv}$ and integrating over a 
unit cell, i.e., 
\begin{equation}
\int_{\rm u.c.} \frac{d\rv}{V}\, \expF{-\I\Gi\cdot \rv}\, 
\pp{n}{t} \approx \pp{\etai}{t},
\label{eq:cg_dynamics}
\end{equation}
where $V$ is the volume of a unit cell, which may be written as
\footnote{The functional derivative $\delta F/\delta z^*$
is computed treating $z$ and $z^*$ as independent variables.}
\begin{equation}
\pp{\etai}{t} = \mathcal{L}_{\hkl}\frac{\delta F_\eta}{\delta \etai^*} = (\nabla^2+ \I\Gi\cdot\Grad-
|\Gi|^2)\frac{\delta F_\eta}{\delta \etai^*}
\approx -|\Gi|^2\frac{\delta F_\eta}{\delta \etai^*},
\label{eq:detadt}
\end{equation}
where the long-wavelength limit has been used in the last approximation. It is interesting to note that the equation of motion for the amplitudes are non-conserved, implying that an initial liquid (crystal) can completely transform in a crystal (liquid) locally.  

Nevertheless the density is a conserved quantity in a closed system and it is often important 
in liquid solid transitions since in liquid/solid coexistence the 
liquid and solid have different densities.  In addition, the process 
of dislocation climb involves the mass (or vacancy) diffusion.
In the original derivation of the APFC \cite{Goldenfeld2005,Athreya2006} the 
average density was assumed to be constant.  The first inclusion of a spatially 
dependent density was reported by Yeon \etal \cite{Yeon2010}.  In this work 
$\no$ was assumed to vary on the same length scales as the complex amplitudes and Eq.~\eqref{eq:n} should read
\begin{equation}
n({\bf r},t) = \no({\bf r},t) + \sum_{\hkl} \etai({\bf r},t) 
\expF{\I\Gi\cdot {\bf r}}.
\end{equation}
Unfortunately, using the so-called 
\textit{``quick and dirty"} method leads to an equation of 
motion for $\no$ (and free energy) which contains terms like 
$(1+\nabla^2)^2n$ and then implies that crystal state can be obtained 
from constant amplitudes or by a periodically varying $\no$ (which of
course violates the assumption the $\no$ varies on the same length 
scales as the amplitudes).  To overcome this difficulty several simpler models were proposed, which were shown to 
incorporate interfacial energy associated with the 
density difference at liquid/solid front as well as the 
well known Gibbs-Thomson effect \cite{Yeon2010}. The model can be written 
\begin{equation}
\begin{split}
{\cal F}=&\int {\rm d} \mathbf{r} \bigg[ \frac{\DBpar}{2}\no^2-\frac{\tpar}{3}\no^3 + \frac{\vpar}{4}\no^4 +\frac{1}{2}\left(\DBpar -2\tpar \no +3\vpar\no^2\right) \bigg(\sum_{\hkl} \etai \expF{\I \Gi \cdot \mathbf{r}} \bigg)^2 \\
& -\frac{1}{3}\left(\tpar-3\no\right)\bigg(\sum_{\hkl} \etai \expF{\I \Gi \cdot \mathbf{r}} \bigg)^3 +\frac{\vpar}{4}\bigg(\sum_{\hkl} \etai \expF{\I\Gi\cdot \mathbf{r}}\bigg)^4 +\frac{\Bxpar}{2}\sum_{\hkl} (|\mathcal{L}_{\hkl}+q_0^2) \etai|^2\bigg],
\label{eq:F_density_amplitude}
\end{split}
\end{equation}
with dynamics
\begin{equation}
\pp{\etai}{t}=-|\Gi|^2\frac{\delta {\cal F}}{\delta \etai^*},  \ \  \ \ \ 
\pp{\no}{t}=\Lap\frac{\delta {\cal F}}{\delta \no}.
\label{eq:dt_eta_no}
\end{equation}
The specific terms that emerge when averaged over a unit cell are discussed in 
the following section.
This approach is also discussed in Huang \etal \cite{Huang2010}.  If the amplitudes are 
assumed to be real (which eliminates the possibility of elastic and plastic phenomena) this 
reduces to Model C in the Hohenberg/Halperin \cite{HohHal} classification scheme that can  be 
 used to study phenomena such as directional 
 solidification  \cite{Grossmann93}
 or eutectic 
 solidification \cite{Drolet00,Elder94}. Heinonen \etal \cite{HeinonenPRL2016} use a similar free energy functional, but also incorporate momentum through the Navier Stokes equation and add the corresponding 
convective  term to the dynamics of $\etai$ and $\no$. This has the advantage of including 
faster relaxation of elastic fields as discussed in Sec.~\ref{sec:mechanical-equilibrium}.

\subsection{Formulas for amplitude equations} 
\label{sec:formulas}

Let's consider the free energy Eq.~\eqref{eq:F_PFC} with constant average 
density $\no$ and for the sake of simplicity the generic parameters $A=\Bxpar$, 
$B=\DBpar-2\tpar\no+3\vpar\no^2$, $C=-(\tpar+3\no)$, $D=\vpar$, $E=\DBpar\no^2/2-\tpar\no^3/3 + \vpar\no^4/4$. 
The amplitude expansion is based on the approximation of $n$ as from Eq.~\eqref{eq:n} with a finite set of $M$ vectors $\Gi$, reproducing a specific crystal symmetry. This equation, exploiting that $\eta_{-(\hkl)}=\eta_{\hkl}^*$, is here rewritten as 
\begin{equation}
n=\no+\sum_{m=1}^M \eta_{m} \expF{\I \Gv_m \cdot \rv }+\cc
    \label{eq:nnn}
\end{equation}
where for simplicity $\Gi$ is given a single subscript $m$ and $\cc$ is the complex conjugate, highlighting the minimal set of amplitudes to be considered to approximate $n$.
The free energy and the evolution law for the amplitudes can be obtained by exploiting the coarse-graining procedure introduced in Sec.~\ref{sec:derivation}, i.e. by integration over the unit cell of the phase-field crystal energy density \eqref{eq:F_PFC}, with $n$ expressed through its amplitude expansion, Eq.~\eqref{eq:nnn} \cite{MajaniemiPRB2007,Majaniemi2009,Chan2009,Provatas2010,Ofori-Opoku2013}.

To provide a general form of the free energy, consider separately the different powers of $n$ entering Eq.~\eqref{eq:F_PFC}, namely $n^k( \{\eta_m \},\{ \eta_m^*\})\rightarrow \zeta_k$. After averaging over a unit 
cell the following results emerge,

\begin{equation}
\begin{split}
\zeta_2\ = 
\ 2\ &{\textstyle \sum^M_{m}}\ |\eta_m|^2 = \Phi,
\\
\zeta_3\ =
\ \bigg[ 3&{\textstyle \sum_{n > m}^M}\ \big\{
\mathcal{K}_{2m+n}\eta_m^2\eta_n + \mathcal{K}_{m+2n}\eta_m\eta_n^2 + \mathcal{K}_{-2m+n}{\eta_m^*}^2\eta_n + \mathcal{K}_{-m+2n}\eta_m^*\eta_n^2 \big\} \\
+6&{\textstyle \sum^M_{o > n > m}} \big\{
\Kd_{m+n+o}\eta_m\eta_n\eta_o + \Kd_{-m+n+o}\eta_m^*\eta_n\eta_o + \Kd_{m-n+o}\eta_m\eta_n^*\eta_o \\ 
&\ \qquad \quad + \Kd_{m+n-o}\eta_m\eta_n\eta_o^* \big\}
+ \text{c.c.} \bigg],
\\
\zeta_4\ = \ 6 \ &{\textstyle \sum^M_m}\  |\eta_m|^4+
    24 {\textstyle \sum^M_{n>m}} 
     |\eta_m|^2|\eta_n|^2 \\  
    +\bigg[4&{\textstyle\sum_{n>m}^M} \big\{ \Kd_{3m+n}\eta_m^3\eta_n+\Kd_{-3m+n}{\eta_m^*}^3\eta_n  +\Kd_{-m+3n}\eta_m^*\eta_n^3+\Kd_{m+3n}\eta_m\eta_n^3\big\}\\
    +
    12 &{\textstyle\sum^M_{o>n>m}} \big\{ \Kd_{2m+n+o}\eta_m^2\eta_n\eta_o +\Kd_{m+2n+o} \eta_m\eta_n^2\eta_o + 
    \Kd_{m+n+2o} \eta_m\eta_n\eta_o^2 
    \\ &\ \qquad \quad +  \Kd_{-2m+n+o}{\eta_m^* }^2\eta_n\eta_o +
    \Kd_{-m+2n+o} \eta_m^*\eta_n^2\eta_o + 
    \Kd_{-m+n+2o} \eta_m^*\eta_n\eta_o^2 
    \\ &\ \qquad \quad +
    \Kd_{2m-n+o} \eta_m^2\eta_n^*\eta_o +
    \Kd_{m-2n+o} \eta_m {\eta_n^*}^2 \eta_o 
    \Kd_{m-n+2o} \eta_m\eta_n^*\eta_o^2 
    \\ &\ \qquad \quad +
    \Kd_{2m+n-o} \eta_m^2\eta_n\eta_o^* +
    \Kd_{m+2n-o} \eta_m\eta_n^2\eta_o^* + %
    \Kd_{m+n-2o} \eta_m\eta_n{ \eta_o^* }^2 \big\} 
    \\  + %
     24 &{\textstyle \sum^M_{p>o>n>m}} \big\{ %
    \Kd_{-m+n+o+p}\eta_m^*\eta_n\eta_o\eta_p %
    +\Kd_{m-n+o+p}\eta_m\eta_n^*\eta_o\eta_p
    \\ &\ \qquad \qquad +
    \Kd_{m+n-o+p}\eta_m\eta_n\eta_o^*\eta_p
    +\Kd_{m+n+o-p}\eta_m\eta_n\eta_o\eta_p^*
    \\ &\ \qquad \qquad 
    +\Kd_{-m-n+o+p}\eta_m^*\eta_n^*\eta_o\eta_p
    +\Kd_{-m+n-o+p} \eta_m^*\eta_n\eta_o^*\eta_p
    \\ &\ \qquad \qquad 
    +
    \Kd_{-m+n+o-p} \eta_m^*\eta_n\eta_o\eta_p^* +
    \Kd_{m+n+o+p} \eta_m\eta_n\eta_o\eta_p \big\} +\cc \bigg],
 \end{split}
\label{eq:gen_energy}
\end{equation}
with
\begin{equation}
\begin{split}
\mathcal{K}_{im+jn+ko+lp}&=\begin{cases}1& 
\text{if} \ \
|i\Gv_m + j\Gv_n +k\Gv_o+l\Gv_p|=0  \\ 
0 & 
\text{if} \ \ 
|i\Gv_m + j\Gv_n +k\Gv_o+l\Gv_p|\neq 0
\end{cases},  
\end{split}
\end{equation}
and neglecting terms including a factor $\mathcal{K}_{im+in}$ with $i=\pm 1,\pm 2$ which would appear in $\zeta_2$ and $\zeta_4$ as $\Gv_m$ with the same lengths are never parallel (or antiparallel), so $\mathcal{K}_{im+in}=0$. Notice that terms as in the first sum in $\zeta_3$ or the third sum of $\zeta_4$ contributes if considering modes with two or three times the length of others, respectively (e.g. $\Gv_{10}$ and $\Gv_{20}$ in Fig.~\ref{fig:figure2}(b)).

For a one-mode approximation of $n$ through Eq.~\eqref{eq:nnn}, i.e. by considering the shortest $\Gv_m$, and transformation \eqref{eq:LapG}, the excess term becomes
\begin{equation}
\begin{split}
    \int_{\rm u.c.} d{\bf r}\ n(1+\Lap)^2n  = \sum^M_m 2|(1+\mathcal{L}_m) \eta_m|^2
    \stackrel{|\Gv_m|=1}{=} \sum^M_m 2|\mathcal{G}_m \eta_m|^2,
    \end{split}
    \label{eq:diffop}
\end{equation}
with $\mathcal{G}_m=\Lap+2\I \Gv_m\cdot\Grad$ and $\mathcal{L}_m=\mathcal{G}_m-|\Gv_m|^2$. In the one mode approximation, the 
length scales can always be re-parametrized such that $|\Gv_m|=1$.

Interestingly the term $\zeta_2=\Phi$ does not depend on the lattice symmetry, 
while $\zeta_4$ can be written $\zeta_4 =6\sum_m^M |\eta_m|^4 +24\sum^M_{n>m} |\eta_m|^2|\eta_n|^2 
+\zeta_4^{\rm s} = 3\Phi^2 - 6 \sum_m^M |\eta_m|^4+\zeta_4^{\rm s}$, where 
$\zeta_4^{\rm s}$ depends on lattice symmetry. Therefore, 
the free energy as function of amplitudes may be written
\begin{equation}
\begin{split}
F_\eta&=\int_{\Omega} \dd\rv \bigg[ \frac{A}{2}\sum_{m}^M 2|\mathcal{G}_m \eta_m|^2+
\frac{B}{2}\zeta_2+\frac{C}{3}\zeta_3+\frac{D}{4}\zeta_4 + E \bigg] \\
&=\int_{\Omega} \dd\rv \bigg[\frac{B}{2}\Phi+\frac{3D}{4}\Phi^2 +\sum_{m}^M
\left ( A |\mathcal{G}_m \eta_m|^2-\frac{3D}{2}|\eta_m|^4 \right ) + f^{\rm s}(\{\eta_m\},\{\eta_m^*\}) + E \bigg] ,
\label{eq:F_APFC}
\end{split}
\end{equation}
with $f^{\rm s}(\{\eta_m\},\{\eta_m^*\})=\frac{C}{3}\zeta_3+\frac{D}{4}\zeta_4^{\rm s}$.

The dynamics of the amplitudes, based on the PFC formulation in Eq.~\eqref{eq:dndt} and according to transformation \eqref{eq:cg_dynamics} are  given by 
\begin{equation}
    \frac{\partial \eta_m}{\partial t}=\mathcal{L}_m\frac{\delta F_\eta}{\delta \eta_m^*}
    \approx-|\Gv_m|^2\bigg[A\mathcal{G}_m^2\eta_m+B\eta_m+ 3D(\Phi - |\eta_m|^2)\eta_m+\underbrace{\frac{C}{3}\frac{\partial \zeta_3}{\partial \eta_m^*}+\frac{D}{4}\frac{\partial \zeta_4^{\rm s}}{\partial \eta_m^*}}_{\partial f^{\rm s}/\partial \eta_m^*}\bigg],
    \label{eq:ev_apfc}
\end{equation}
where $\mathcal{L}_m \approx -|\Gv_m|^2$ as in Eq.~\eqref{eq:detadt}, and, from Eq.~\eqref{eq:gen_energy},
\begin{equation}
\begin{split}
    \frac{1}{3}\frac{\partial \zeta_3}{\partial \eta_m^*}\ =\ &{\textstyle\sum_{n \neq m}^M} \big\{
2\mathcal{K}_{-2m-n}\eta_m^*\eta_n^* +
 2\mathcal{K}_{-2m+n}\eta_m^*\eta_n + \mathcal{K}_{-m-2n}{\eta_n^*}^2 + \mathcal{K}_{-m+2n}\eta_n^2 \big\} \\
    +2&{\textstyle \sum^M_{o>n\neq m}}\big\{\Kd_{-m-n-o}\eta_n^*\eta_o^* +\Kd_{-m+n+o}\eta_n\eta_o+\Kd_{-m+n-o}\eta_n\eta_o^*+\Kd_{-m-n+o}\eta_n^*\eta_o \big\},
    \\
    \frac{1}{4}\frac{\partial \zeta_4^{\rm s}}{\partial \eta_m^*}\ =\ &{\textstyle \sum_{n\neq m}^M}  \big\{3\Kd_{-3m-n}{\eta_m^*}^2\eta_n^*+3\Kd_{-3m+n}{\eta_m^*}^2\eta_n+\Kd_{-m+3n}\eta_n^3+\Kd_{-m-3n}{\eta_n^*}^3 \big\}
    \\
    +3&{\textstyle \sum^M_{o>n\neq m}} \big\{ 2\Kd_{-2m-n-o}\eta_m^*\eta_n^*\eta_o^*
      +\Kd_{-m-2n-o} {\eta_n^*}^2\eta_o^* +
      \Kd_{-m-n-2o} \eta_n^*{\eta_o^*}^2 
     \\& \qquad \qquad +2\Kd_{-2m+n+o} \eta_m^*\eta_n\eta_o + 
        \Kd_{-m+2n+o} \eta_n^2\eta_o + 
        \Kd_{-m+n+2o} \eta_n\eta_o^2 
    \\& \qquad \qquad +2\Kd_{-2m+n-o} \eta_m^*\eta_n\eta_o^*
    +\Kd_{-m+2n-o} \eta_n^2\eta_o^* +
    \Kd_{-m+n-2o} \eta_n{\eta_o^*}^2 
    \\& \qquad \qquad +2\Kd_{-2m-n+o} \eta_m^*\eta_n^*\eta_o
    +\Kd_{-m-2n+o} {\eta_n^*}^2\eta_o 
    +\Kd_{-m-n+2o} \eta_n^*\eta_o^2
    \big\}\\
    +6&{\textstyle \sum^M_{p>o>n\neq m}} \big\{
    \Kd_{-m+n+o+p}\eta_n\eta_o\eta_p+
    \Kd_{-m+n-o-p}\eta_n\eta_o^*\eta_p^*+
    \Kd_{-m-n+o-p}\eta_n^*\eta_o\eta_p^*
    \\& \qquad \qquad +
    \Kd_{-m-n-o+p}\eta_n^*\eta_o^*\eta_p+
    \Kd_{-m-n+o+p}\eta_n^*\eta_o\eta_p+
    \Kd_{-m+n-o+p}\eta_n\eta_o^*\eta_p
    \\
       & \qquad \qquad +\Kd_{-m+n+o-p}\eta_n\eta_o\eta_p^*+
     \Kd_{-m-n-o-p}\eta_n^*\eta_o^*\eta_p^* \big\},
\end{split}
\label{eq:gen_dynamic}
\end{equation}

\subsubsection{Multi-mode approximations.}
\label{sec:multiM}

To model some crystal lattices, more than one mode is required in Eq.~\eqref{eq:nnn}, i.e. more length scales are set through the choice of the reciprocal space vectors. In this case, $\zeta_m$ reads as reported above, but the excess term takes different forms. However, it may be reduced to Eq.~\eqref{eq:diffop} through approximation \cite{ElderPRE2010,SalvalaglioPRL2021}. For two lengths, $R_1=2\pi/k^{\rm eq}_1$ and $R_2=2\pi/k^{\rm eq}_2$, corresponding to different lengths in the reciprocal space $k^{\rm eq}_1=1$ and $k^{\rm eq}_2=\alpha k^{\rm eq}_1$, with $\alpha \neq 1=k^{\rm eq}_2/k^{\rm eq}_1=R_1/R_2$, the term including the differential operator in the dynamic would read \cite{ElderPRE2010}
\begin{equation}
 (1+R_1^2\Lap)^2(1+R_2^2\Lap)^2n \rightarrow \sum^M_m \alpha^{-4}(1+\mathcal{L}_m)^2 (\alpha^2+\mathcal{L}_m)^2\eta_m=\sum^M_m \mathcal{D}_m \eta_m,
 \label{eq:term1}
\end{equation}
with
\begin{equation}
\mathcal{D}_m=\begin{cases} 
\alpha^{-4}(\mathcal{G}_m)^2 ( \alpha^2-1+\mathcal{G}_m)^2= \frac{(\alpha^2-1)^2}{\alpha^4} (\mathcal{G}_m)^2  \big(1+\frac{\mathcal{G}_m}{\alpha^2-1}\big)^2 & \text{if} \ |\Gv_m|=k^{\rm eq}_1=1 \\ 
\alpha^{-4}(1-\alpha^2+\mathcal{G}_m)^2 (\mathcal{G}_m)^2= \frac{(1-\alpha^2)^2}{\alpha^4} (\mathcal{G}_m)^2  \big(1-\frac{\mathcal{G}_m}{\alpha^2-1}\big)^2 & \text{if} \ |\Gv_m|=k^{\rm eq}_2=\alpha 
\end{cases}
\end{equation}
and lengths have been scaled such that $\mathbf{x}\rightarrow \mathbf{x}/R_1$.
If $2|\mathcal{G}_m\eta_m| \ll |(\alpha^2-1)\eta_m|$,
\begin{equation}
\mathcal{D}_m\eta_m\approx \frac{(\alpha^2-1)^2}{\alpha^4} \mathcal{G}_m^2 \eta_m.
\label{eq:twolen}
\end{equation}
Therefore, the coefficient $A$ can be rescaled by a factor $\alpha^4 / (\alpha^2-1)^2$ and the same energy term as for the one mode approximation can be used. This result may be generalized for a lattice 
having $N_\ell$ different length scales $R_\ell=2\pi/k^{\rm eq}_\ell$ 
and $k^{\rm eq}_\ell/k^{\rm eq}_1=\alpha_\ell$  (noting $k_1^{\rm eq}=1$). Eq.~\eqref{eq:term1} would read 
\begin{equation}
\prod_\ell^{N_\ell}(1+R_\ell^2\Lap)^2n\rightarrow \sum^M_m \bigg[ \prod_\ell^{N_\ell} (1+\alpha_\ell^{-2}\mathcal{L}_m)^2\bigg] \eta_m =\sum^M_m \mathcal{M}_m \eta_m.
\end{equation}
If, $\forall \ell$,  $2|\mathcal{G}_m\eta_m| \ll |(\alpha^2_\ell-|\Gv_m|^2)|\eta_m$, one may write
\begin{equation}
\mathcal{M}_m\eta_m
\approx \left[ |\Gv_m|^{-4}\prod_{\substack{\ell \\ \alpha_\ell \neq |\Gv_m|}}^{N_\ell}\bigg(\frac{\alpha_\ell^2-|\Gv_m|^2}{\alpha_\ell^2}\bigg)^2\right] \mathcal{G}_m^2\eta_m=\Gamma_m \mathcal{G}_m^2\eta_m,
\label{eq:manylen}
\end{equation}
that for $N_l=2$, $\alpha_1=1$ and $\alpha_2=\alpha$ reduces to Eq.~\eqref{eq:twolen}. Then, 
under this approximation, $\sum^M_m \mathcal{M}_m \eta_m = \sum^M_m \Gamma_m 
\mathcal{G}_m^2 \eta_m$.
Notice that in the presence of more than two modes, the coefficient of $\mathcal{G}_m^2$ cannot be taken outside the sum so it cannot be included in the coefficient $A$ through rescaling as in Eq.~\eqref{eq:twolen}.

\subsubsection{Results for specific lattice symmetries.}
\label{sec:results_symm}

Implementations of the APFC equations may be performed in a general fashion by considering Eqs.~\eqref{eq:gen_energy} and \eqref{eq:gen_dynamic}. This delivers a general framework suitable for changes in lattice symmetries and the number of modes used (eventually also different symmetries at once, see also Sec.~\ref{sec:multi-phase-systems}). However, the specific equations corresponding to given lattice symmetries through the choice of reciprocal lattice vectors may be useful for analytic calculations and ad-hoc implementations. In the following, $f^{\rm s} \equiv f^{\rm s}({\{\eta_m\},\{\eta_m^*\}})$ are reported for selected crystal symmetries used in literature, with the length of shortest reciprocal space vectors normalized to $1$ (see, e.g.,  \cite{ElderPRE2010,XuPRB2016,Salvalaglio2017,AnkudinovPRE2020} and Fig.~\ref{fig:figure2}). 

\noindent \textit{Triangular (TRI) symmetry} (2D), one-mode approximation, $N=3$:
\begin{gather*}
\Gv_1^{\rm TRI}= \begin{bmatrix}
           -\sqrt{3}/2 \\
           -1/2 \\
         \end{bmatrix}, 
         \quad 
         \Gv_2^{\rm TRI}=\begin{bmatrix}
          0\\
          1 \\
         \end{bmatrix}, 
         \quad 
\Gv_3^{\rm TRI}= \begin{bmatrix}
           \sqrt{3}/2 \\
           -1/2 \\
         \end{bmatrix},
         \nonumber
         \end{gather*} 
\begin{equation}
f^{\rm tri,1}=2C(\eta_1\eta_2\eta_3+\eta_1^*\eta_2^*\eta_3^*).
\label{eq:energyterm_tri}
\end{equation}
\textit{Triangular (TRI) symmetry} (2D), two-mode approximation, $N=6$:  
 \begin{gather*}
 \Gv_1^{\rm TRI},\quad \Gv_2^{\rm TRI},\quad \Gv_3^{\rm TRI},\quad \Gv_4^{\rm TRI}=\Gv_1^{\rm TRI}-\Gv_2^{\rm TRI},\\ \quad \Gv_5^{\rm TRI}=\Gv_2^{\rm TRI}-\Gv_3^{\rm TRI},\quad \Gv_6^{\rm TRI}= \Gv_3^{\rm TRI}-\Gv_1^{\rm TRI}, \nonumber
 \end{gather*}
\begin{equation}
\begin{split}
f^{\rm tri,2}=&2C(\eta_1\eta_2\eta_3+\eta_1^*\eta_2\eta_4+\eta_1\eta_3^*\eta_6+\eta_2^*\eta_3\eta_5+\eta_4\eta_5\eta_6)\\
&+3D(\eta_1\eta_2^2\eta_5^*+\eta_1^2\eta_2\eta_6+\eta_1^2\eta_3\eta_4^*+\eta_1\eta_3^2\eta_5+\eta_2^2\eta_3\eta_4+\eta_2\eta_3^2\eta_6^*) \\
&+6D(\eta_1\eta_2^*\eta_5\eta_6+
\eta_1^*\eta_3\eta_4\eta_5+\eta_2\eta_3^*\eta_4\eta_6)
+\cc
\end{split}
\label{eq:energyterm_tri2}
\end{equation}
\textit{Square (SQ) symmetry} (2D), two-mode approximation, $N=4$:
\begin{gather*}
\Gv_1^{\rm SQ}= \begin{bmatrix}
           1 \\
           0 \\
         \end{bmatrix},
         \quad 
\Gv_2^{\rm SQ}=\begin{bmatrix}
          0\\
          1 \\
         \end{bmatrix},
         \quad 
\Gv_3^{\rm SQ}= \begin{bmatrix}
           1 \\
           1 \\
         \end{bmatrix},
         \quad 
\Gv_4^{\rm SQ}= \begin{bmatrix}
           -1 \\
           1 \\
           \end{bmatrix},
         \nonumber
         \end{gather*} 
\begin{equation}
f^{\rm sq,2}= 2C(\eta_1\eta_2\eta_3^*+\eta_1\eta_2^*\eta_4) + 3 D(\eta_1^2\eta_3^{*}\eta_4+
\eta_2^2\eta_3^*\eta_4^*) + {\rm c.c.}
\label{eq:energyterm_sq4}
\end{equation}
\textit{Square (SQ) symmetry} (2D), three-mode approximation, $N=8$: 
\begin{gather*}
\Gv_1^{\rm SQ},\ \ 
\Gv_2^{\rm SQ},\ \ 
\Gv_3^{\rm SQ},\ \ 
\Gv_4^{\rm SQ},\ \ 
\Gv_5^{\rm SQ}= \begin{bmatrix}
           2 \\
           1 \\
         \end{bmatrix},
         \ \  
\Gv_6^{\rm SQ}=\begin{bmatrix}
          -2\\
          1 \\
         \end{bmatrix},
         \ \ 
\Gv_7^{\rm SQ}= \begin{bmatrix}
           1 \\
           2 \\
         \end{bmatrix},
         \ \  
\Gv_8^{\rm SQ}= \begin{bmatrix}
           -1 \\
           2 \\
           \end{bmatrix},
\nonumber
\end{gather*}
\begin{equation}
\begin{split}
f^{\rm sq,3}=& 2C(\eta_1  \eta_2  \eta_3^*+
\eta_1  \eta_2^* \eta_4+
\eta_1  \eta_3  \eta_5^*+
\eta_1  \eta_4^* \eta_6+
\eta_2  \eta_3  \eta_7^*+
\eta_2  \eta_4  \eta_8^*+
\eta_3  \eta_6  \eta_8^*  
\\
& +
\eta_4  \eta_5  \eta_7^*) +3D(\eta_1^2 \eta_2\eta_5^*+
\eta_1^2 \eta_2^{*}\eta_6+
\eta_ 1 \eta_ 2 ^2 \eta_ 7 ^{*}+
\eta_ 1^{*} \eta_ 2^2 \eta_8^{*}+
\eta_1^2 \eta_3^{*}\eta_4+
\eta_1^{*} \eta_ 3^2 \eta_7^{*}\\&+
\eta_ 1 \eta_ 4 ^2 \eta_ 8 ^{*}+
\eta_1^2 \eta_7^{*}\eta_8+
\eta_2^2 \eta_3^{*}\eta_4^{*}+
\eta_2^{*} \eta_3^2 \eta_5^{*}+
\eta_2^{*} \eta_4^2 \eta_6^{*}+
\eta_2^2 \eta_5^{*}\eta_6^{*})+
6D(\eta_1 ^* \eta_ 2  \eta_ 5  \eta_ 7 ^*\\
&+
\eta_1  \eta_ 2  \eta_ 6  \eta_ 8 ^*+
\eta_ 1  \eta_ 3  \eta_ 4  \eta_ 7 ^*+
\eta_ 1 ^* \eta_ 3  \eta_ 4  \eta_ 8 ^*+
\eta_ 1  \eta_ 5  \eta_ 6  \eta_ 7 ^* +
\eta_ 1 ^* \eta_ 5  \eta_ 6  \eta_ 8 ^*+
\eta_ 2 ^* \eta_ 3^*  \eta_ 4 \eta_5 \\&+
\eta_ 2 ^* \eta_ 3  \eta_ 4 ^* \eta_ 6+
\eta_ 2 ^* \eta_ 5  \eta_ 7 ^* \eta_ 8+
\eta_ 2 ^* \eta_ 6  \eta_ 7  \eta_ 8 ^*+
\eta_ 3 ^* \eta_ 4^*  \eta_5  \eta_ 6+
\eta_ 3 ^* \eta_ 4  \eta_ 7  \eta_ 8 ^*)
+ {\rm c.c.}
 \end{split}
\label{eq:energyterm_sq8}
\end{equation}
\textit{Body Centered Cubic (BCC) symmetry} (3D), one-mode approximation, $N=6$: 
\begin{gather*}
\frac{\Gv_1^{\rm BCC}}{G_0^{\rm BCC}}= \begin{bmatrix}
           0 \\
           1 \\
           1 \\
         \end{bmatrix},
         \quad 
\frac{\Gv_2^{\rm BCC}}{G_0^{\rm BCC}}= \begin{bmatrix}
           1 \\
           0 \\
           1 \\
         \end{bmatrix},
         \quad
\frac{\Gv_3^{\rm BCC}}{G_0^{\rm BCC}}= \begin{bmatrix}
           1 \\
           1 \\
           0 \\
         \end{bmatrix},
         \quad
\frac{\Gv_4^{\rm BCC}}{G_0^{\rm BCC}}= \begin{bmatrix}
           0 \\
           1 \\
           -1 \\
         \end{bmatrix},
         \nonumber \\
\frac{\Gv_5^{\rm BCC}}{G_0^{\rm BCC}}= \begin{bmatrix}
           1 \\
           -1 \\
           0 \\
         \end{bmatrix},
         \quad
\frac{\Gv_6^{\rm BCC}}{G_0^{\rm BCC}}= \begin{bmatrix}
           -1 \\
           0 \\
           1 \\
         \end{bmatrix}, \quad G_0^{\rm BCC}=\frac{\sqrt{2}}{2}
         \nonumber
         \end{gather*} 
\begin{equation}
\begin{split}
f^{\rm BCC,1} = &2C(
\eta_1\eta_2^*\eta_5+
\eta_1^*\eta_3 \eta_6+
\eta_2\eta_3^*\eta_4+ 
\eta_4\eta_5\eta_6)\\ 
&+6D(
\eta_1^*\eta_2\eta_4\eta_6+
\eta_1\eta_3^*\eta_4\eta_5+
\eta_2^*\eta_3\eta_5\eta_6)+\cc
\end{split}
\label{eq:energyterm_BCC6}
\end{equation}
\textit{Body Centered Cubic (BCC) symmetry} (3D), two-mode approximation, $N=9$
\begin{gather*}
\Gv_1^{\rm BCC},\ \ 
\Gv_2^{\rm BCC},\ \ 
\Gv_3^{\rm BCC},\ \ 
\Gv_4^{\rm BCC},\ \ 
\Gv_5^{\rm BCC},\ \ 
\Gv_6^{\rm BCC},\nonumber \\
\frac{\Gv_7^{\rm BCC}}{G_0^{\rm BCC}}= \begin{bmatrix}
           2 \\
           0 \\
           0 \\
         \end{bmatrix},
         \quad 
\frac{\Gv_8^{\rm BCC}}{G_0^{\rm BCC}}= \begin{bmatrix}
           0 \\
           2 \\
           0 \\
         \end{bmatrix},
         \quad
\frac{\Gv_9^{\rm BCC}}{G_0^{\rm BCC}}= \begin{bmatrix}
           0 \\
           0 \\
           2 \\
         \end{bmatrix},
         \nonumber
         \end{gather*} 
\begin{equation}
\begin{split}
f^{\rm BCC,2} = &2C(
\eta_1 \eta_2^* \eta_5+
\eta_1^* \eta_3 \eta_6+
\eta_1 \eta_4 \eta_8^*+
\eta_1^* \eta_4 \eta_9+
\eta_2 \eta_3^* \eta_4+
\eta_2^* \eta_6 \eta_7+
\eta_2 \eta_6 \eta_9^*+
\\
&\eta_3 \eta_5 \eta_7^*+
\eta_3^* \eta_5 \eta_8+
\eta_4\eta_5\eta_6)
+ 3D(\eta_1^2 \eta_8^{*}\eta_9^{*}+
\eta_2^{2} \eta_7^{*}\eta_9^{*}+
\eta_3^{2} \eta_7^{*}\eta_8^{*}
+\eta_4^2 \eta_8^{*}\eta_9
\\
&+\eta_5^2 \eta_7^{*}\eta_8+
\eta_6^2 \eta_7\eta_9^*) 
+ 6D (
\eta_1^*\eta_2\eta_3\eta_7^*+
\eta_1^*\eta_2\eta_3^*\eta_8+
\eta_1^*\eta_2^*\eta_3\eta_9+
\eta_1^*\eta_2\eta_4\eta_6
\\
&+\eta_1\eta_3^*\eta_4\eta_5+
\eta_1^*\eta_5^*\eta_6\eta_7+
\eta_1^*\eta_5\eta_6\eta_8+
\eta_1\eta_5\eta_6\eta_9^*+
\eta_2^*\eta_3\eta_5\eta_6+
\eta_2\eta_4\eta_5\eta_7^*
\\
&+\eta_2^*\eta_4^*\eta_5\eta_8+
\eta_2^*\eta_4\eta_5\eta_9+
\eta_3^*\eta_4\eta_6\eta_7+
\eta_3\eta_4\eta_6\eta_8^*+
\eta_3^*\eta_4\eta_6^*\eta_9)+\cc
\end{split}
\label{eq:energyterm_BCC9}
\end{equation}
\textit{Face Centered Cubic (FCC) symmetry} (3D), two-mode approximation, $N=7$: 
\begin{gather*}
\frac{\Gv_1^{\rm FCC}}{G_0^{\rm FCC}}= \begin{bmatrix}
           -1 \\
           1 \\
           1 \\
         \end{bmatrix},
         \quad 
\frac{\Gv_2^{\rm FCC}}{G_0^{\rm FCC}}= \begin{bmatrix}
           1 \\
           -1 \\
           1 \\
         \end{bmatrix},
         \quad
\frac{\Gv_3^{\rm FCC}}{G_0^{\rm FCC}}= \begin{bmatrix}
           1 \\
           1 \\
           -1 \\
         \end{bmatrix},
         \quad
\frac{\Gv_4^{\rm FCC}}{G_0^{\rm FCC}}= \begin{bmatrix}
           -1 \\
           -1 \\
           -1 \\
         \end{bmatrix},
         \nonumber \\
\frac{\Gv_5^{\rm FCC}}{G_0^{\rm FCC}}= \begin{bmatrix}
           2 \\
           0 \\
           0 \\
         \end{bmatrix},
         \quad
\frac{\Gv_6^{\rm FCC}}{G_0^{\rm FCC}}= \begin{bmatrix}
           0 \\
           2 \\
           0 \\
         \end{bmatrix}, \quad
\frac{\Gv_7^{\rm FCC}}{G_0^{\rm FCC}}= \begin{bmatrix}
           0 \\
           0 \\
           2 \\
         \end{bmatrix},\quad G_0^{\rm FCC}=\frac{\sqrt{3}}{3}
         \nonumber
         \end{gather*} 
\begin{equation}
\begin{split}
f^{\rm FCC,2} =
&2C(\eta_1\eta_2\eta_7^*+\eta_1\eta_3\eta_6^*+\eta_1\eta_4\eta_5+\eta_2\eta_3\eta_5^*+\eta_2\eta_4\eta_6+
\eta_3\eta_4\eta_7) \\& + 6D (\eta_1\eta_2\eta_3\eta_4
+\eta_1^*\eta_2 \eta_5^* \eta_6  + \eta_1^*\eta_3\eta_5^* \eta_7 +
\eta_1^*\eta_4\eta_6\eta_7 \\ 
	      &+ \eta_2^*\eta_3\eta_6^*\eta_7+\eta_2^*\eta_4\eta_5\eta_7 + \eta_3^*\eta_4\eta_5\eta_6)+\cc \\
\end{split}
\label{eq:ffFCC}
\end{equation}
Other symmetries may be considered, provided that the proper set of the reciprocal space vectors are known and that the encoded symmetry corresponds to a global energy minimum for some parameters (see Sec.~\ref{sec:stability}). Alternatively, stability of phases/symmetries may be enforced with the APFC formulation outlined in Sec.~\ref{sec:different-formulations}.

\subsubsection{Stability of phases.} 
\label{sec:stability}
In a relaxed, bulk crystal, real and constant amplitudes $\phi$ may be computed by energy minimization. For instance, for one-mode approximations and $\no=0$, one gets the energy 
\begin{equation}
    F[\phi] = \int_\Omega h(\phi)\dd \rv = \int_\Omega \bigg[M B \phi^2 + 3DM\left(M - \frac{1}{2}\right)\phi^4 + \frac{C}{3}\zeta_3(\phi)+\frac{D}{4}\zeta_4^{\rm s}(\phi)\bigg] \dd \rv.
    \label{eq:Frealamp}
\end{equation}
Letting $\zeta_3=p\phi^3$ and $\zeta_4^{\rm s}=q\phi^4$ where $p$ and $q$ where are integers, and minimizing the free energy given in  
Eq.~\eqref{eq:gen_energy}, with respect to $\phi$ 
($\delta F[\phi]/{\delta \phi} = \partial h[\phi]/{\partial \phi} = 0$) gives the solutions,
\begin{equation}
    \phi_{1,2}=\frac{-pC \pm \sqrt{(pC)^2-8MB D (12M^2-6M+q)}}{2D(12M^2-6M+q)},
    \label{eq:phireal}
\end{equation}
with $\pm$ the solution for $C\lessgtr 0$. For instance, for a triangular symmetry described by a one mode approximation (see Fig.~\ref{fig:figure2}) where $M=3, p=12, q=0$, gives $\phi_{1,2}=(-C\pm \sqrt{C^2-15BD})/15D$. Similarly, for a BCC lattice described by a one mode approximation (see Fig.~\ref{fig:figure2}) where $M=6$, $p=48$, $q=144$ the result is 
$\phi_{1,2}=(-2C\pm \sqrt{4C^2-45BD})/45D$. Real solutions of Eq.~\eqref{eq:phireal} exist if $(pC)^2>8MB D (12M^2-6M+q)$. Moreover, the general stability of the solid phase described by a real amplitude $\phi_{1,2}$ can be assessed by evaluating the condition $F[\phi_{1,2}] < F[0]$. Notice that, $F[0]$ is trivially $0$ from Eq.~\eqref{eq:Frealamp}, but it may have different values for $\no\neq0$ as a non-zero average density would enter explicitly the energy \eqref{eq:Frealamp} and modifies the value of the real amplitudes at equilibrium (see e.g. Ref.~\cite{ElderPRE2010}).
Phase diagrams can then be devised generally for both PFC and APFC approaches \cite{ElderPRE2010,Mkhonta2013} by evaluating the relative stability of different phases described by $\phi$. Generally, for a given set of parameters $C$ and $D$, liquid phase results favored for values of $B$ smaller than a critical value $B^{\rm c}$. This parameter phenomenologically encodes the role of the temperature. $|B-B^{\rm c}|$ is often referred to as quenching depth. Notice that $B^{\rm c}=0$ for $C=0$.

\begin{figure}
\includegraphics[width=\textwidth]{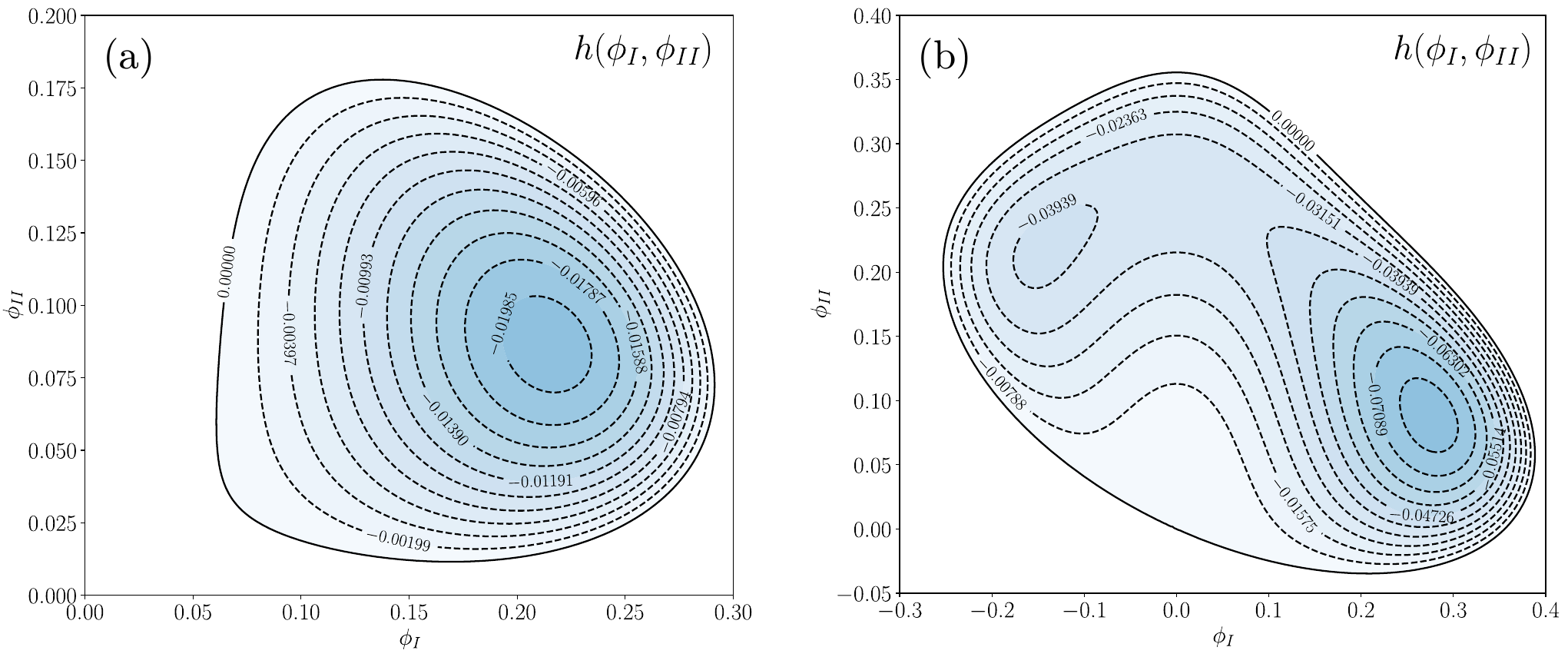}
\caption{$h(\phi_{\rm I},\phi_{\rm II})$ as obtained for a two-mode approximation of the triangular symmetry with $C=-2.0$ and $D=1.0$ at two quenching depths: (a) $B=0.3$, (b) $B=0.0$. Dashed lines show representative isolines for negative values of $h(\phi_{\rm I},\phi_{\rm II})$.}
\label{fig:figure3}
\end{figure}

When considering approximations with more modes, different values of $\phi$ should be considered for every set of amplitudes corresponding to different lengths of $\mathbf{G}_m$. Typically this task should be addressed numerically. Consider an approximation with $K$ equal to the number of the modes of different length (under approximations introduced in Sec.~\ref{sec:multiM}). In this case the following function must 
be minimized,
\begin{equation}
h[\{\phi_k\}] = \sum_{k=1}^K\bigg[ B M_k \phi_k^2 - \frac{3}{2}DM_k\phi_k^4\bigg] + 3D\left(\sum_{k=1}^K M_k\phi_k^2 \right)^2+ \frac{C}{3}\zeta_3(\{\phi_k\})+\frac{D}{4}\zeta_4^{\rm s}(\{\phi_k\}).
\end{equation}
with $M_k$ the number of reciprocal space vectors for each considered mode (the solid arrows in Fig.~\ref{fig:figure2}). For instance, for the three-mode approximation of a cubic lattice in Fig.~\ref{fig:figure2}, we would have $M_1=2$, $M_2=2$ and $M_3=4$. $\zeta_3(\{\phi_k\})$, $\zeta_4^{\rm s}(\{\phi_k\})$ are the symmetry-dependent polynomials resulting by substituting $\eta_j$ with the amplitude associated to the length of the reciprocal space vector they correspond to. To introduce an explicit example, consider the two mode approximation of the triangular symmetry (see Fig.~\ref{fig:figure2}(a)), i.e. $\{\phi_k\}=[\phi_I,\phi_{\rm II}]$, $M_{\rm I}=M_{\rm II}=3$, and $\zeta_3(\phi_{\rm I},\phi_{\rm II})$, $\zeta_4^{\rm s}(\phi_{\rm I},\phi_{\rm II})$ the polynomial resulting by setting $\eta_j=\phi_I$ for $j=1,2,3$ and $\eta_j=\phi_{\rm II}$ for $j=4,5,6$ in Eq.~\eqref{eq:energyterm_tri2}. Plots of $h(\phi_{\rm I},\phi_{\rm II})$ for selected parameters ($C=-2.0$ and $D=1.0$) are shown in Fig.~\ref{fig:figure3}. At a value $B=0.3$ (Fig.~\ref{fig:figure3}(a)), relatively close to the solid-liquid phase transition, the free energy has a single minimum corresponding to $\phi_{\rm I} \approx 0.274$ and $\phi_{\rm II}\approx 0.087$. By increasing the quenching depths, the global minimum shifts to $\phi_I \approx 0.215$ and $\phi_{\rm II} \approx 0.086$ for $B=0.0$. Moreover, another relative minimum appears (see  Fig.~\ref{fig:figure3}(b)), which corresponds to a graphene-like phase. Some extended discussions on all the possible phases which can be described in two dimensions with combination of more modes can be found in Ref.~\cite{Mkhonta2013}.
 
\subsection{Amplitude XPFC} 
\label{sec:different-formulations}

A formulation based on the the so-called structural PFC (XPFC) \cite{Greenwood2010,xtal2}, describing more detailed features and phenomena in crystalline systems such as, e.g. multicomponent systems, structural transformations, anisotropies, and extended defects  \cite{Greenwood2011binary,Berry2012,Ofori-Opoku2013}, has been proposed in Ref.~\cite{Ofori-Opoku2013}. In a dimensionless form, the XPFC free energy $F_{\rm X}$ reads
\begin{equation}
F_X = \int \dd\rv \left[
F_{\rm ex}+
\frac{n^2}{2} 
- P\frac{n^3}{3}
+ Q\frac{n^4}{3} 
\right], \qquad F_{\rm ex}=-\frac{n(\rv)}{2}\int \dd \rv' X_2(|\rv-\rv'|)n(\rv'),
\label{eq:F_XPFC}
\end{equation}
where $P$ and $Q$ are parameters and $X_2(|\rv-\rv'|)$ is the direct two-point correlation function at the reference density $\no$. In this approach, this function is typically expressed in the reciprocal space, $\hat{X}_2(|\kv|)$. Following \cite{Ofori-Opoku2013}, it may be expressed as an envelope of Gaussian peaks associated with different modes of the periodic density or, in other words, to a family of planes of a crystal structure,  \cite{xtal2}
\begin{equation}
    \hat{X}_{2,j}=\expF{-\frac{1}{2w_j^2}(k-{k}_j)^2-\frac{\sigma^2}{p_ja_j} k_j^2},
\end{equation}
where $w_j$ controls the elastic and surface energies (the width of the $j$-th Gaussian peak), $\sigma$ is an effective temperature parameter \cite{Alster17}, $p_j$ and $a_j$ are the planar and atomic densities associated with the family of planes corresponding to the $j$-th mode, respectively, while $k_j$ is the inverse of the interplanar spacing for the $j$-th family of planes. Then, by assuming an amplitude formulation and volume average as in Sects~\ref{sec:derivation}--\ref{sec:formulas},  the polynomial in $n$ that enters $F_X$ leads to terms similar to the energy in 
Eq.~\eqref{eq:F_density_amplitude} except for the excess term which becomes \cite{Ofori-Opoku2013}  \begin{equation}
\begin{split}
    F_{\rm ex, \eta}=\int \dd\rv
    \bigg[
    \sum_m^M&-\frac{\eta_m^*}{2} \mathcal{F}^{-1}\bigg\{ \hat{X}_{2}(|\kv+\mathbf{G}_m|)\hat{\eta}_m(\kv)\bigg\}
    \\
    &-\frac{\no}{2}\mathcal{F}^{-1}\bigg\{\hat{\xi}_V(\kv)\hat{X}_{2}(|\kv|)\hat{n}_{\rm o}(\kv)  \bigg\} +\cc\bigg], 
\end{split}
\label{eq:XAPFC}
\end{equation}
where the hat symbol denotes the Fourier transform, $\mathcal{F}^{-1}$ the inverse Fourier transform, and $\hat{\xi}_V$ an averaging (convolution) kernel in Fourier space that restricts the wave number to small values, approximately approaching the extension of the first Brillouin zone, which filters out spatial variations smaller than the lattice spacing. Interestingly, this model has been proposed with an ansatz for the amplitude expansion encoding different (two) lattice symmetries (see Sec.~\ref{sec:multi-phase-systems}). This ansatz is expected to work with other forms of the energy and it consists just of a different formulation for Eq.~\eqref{eq:nnn} leading to results that may be formulated in terms of the equations reported in Sec.~\ref{sec:formulas}.

\section{Numerical methods}
\label{sec:numerical-methods}

In this section, two standard methods (finite difference 
and spectral) for solving first order in time 
partial differential equations that are applicable to 
APFC models are described. Following this, a finite element approach 
for solving APFC models is outlined and the description of a mesh refinement algorithm is reported.

\subsection{Finite differences}
\label{sec:fd}

In general there are many methods for solving an equations of the form
\begin{equation}
\pp{\psi}{t} = H(\psi),
\end{equation}
where $H(\psi)$ is a  function of $\psi$. To solve it numerically it is useful to first consider  
integrating the equation over time from $t$ to $t+\Delta t$ to obtain, 
\begin{equation}
\psi(t+\Delta t) = \psi(t)+ \int_t^{t+\Delta t} \dd t' \, H(\psi).
\end{equation}
The main question is how to approximate the integral in the above equation. 
In explicit methods only prior knowledge of $\psi$ and its derivatives are 
used, i.e., 
\begin{equation}
\psi(t+\Delta t) = \psi(t)+ \int_t^{t+\Delta t} \dd t' \, 
\left[ H(t) 
+ \left.\pp{H}{t'}\right|_t t' 
+\frac{1}{2!} \left.\pp{^2H}{t'^2}\right|_t t'^2  + \cdots \right].
\label{eq:numnum}
\end{equation}
where $H(t)=H(\psi(t))$.
The simplest method, Euler's method, just retains the first term 
in the expansions, i.e., 
\begin{equation}
\psi(t+\Delta t) = \psi(t)+ \Delta t H(t).
\label{eq:Euler}
\end{equation}
This approach must be supplemented by methods to evaluate 
spatial gradients in $H$, which in (A)PFC type models are 
typically even order derivatives, i.e., $\Lap, \nabla^4, \dots$.  
Often these are evaluated using a central difference formula. For instance, in two dimensions with a 5-points stencil (\textit{quincunx}), the Laplacian is given by 
\begin{equation}
\Lap f = \frac{f(i+1,j)+f(i,j+1)+f(i-1,j)+f(i,j-1)-4f(i,j)}{\Delta s^2},
\label{eq:cdiff}
\end{equation}
where $(x,y)=(i\Delta s,j\Delta s$). Eq.~\eqref{eq:cdiff}, in conjunction with Eq.~\eqref{eq:Euler}, is quite simple to implement for numerical integrations. Moreover, it is easy to incorporate different boundary conditions.  However, the time step $\Delta t$ is limited by the grid spacing due to stability constraints, typically 
\begin{equation}
\Delta t < \alpha \Delta s^{-k},
\end{equation}
where $k$ is the highest order spatial derivative (i.e., $k=6$ for 
the PFC equation) and $\alpha$ is a constant that is model specific.  
If $\Delta t$ is too large, the solution very rapidly diverges
(a pitchfork instability).  The specifics of the origin of this 
instability are described in detail in Ref.~\cite{Provatas2010}.
It is possible to slightly reduce 
this instability by including next nearest neighbours as done 
by Oono and Puri \cite{Oono87}.
This limitation is quite severe in PFC and APFC models as $k=6$ 
in the former case and $k=4$ in the latter.  
This instability can be avoided using semi-implicit approaches 
that are typically done in Fourier space.
However, implicit or more generally semi-implicit approaches may be exploited, 
evaluating terms in the integrals in Eq.~\eqref{eq:numnum} within the range
$[t,t+\Delta t]$, to have more stable numerical schemes (see also Sec.~\ref{sec:fem}). Also, finite difference approaches may be combined with spatial adaptivity which may allow for efficient simulations (see also Sect.~\ref{sec:mesh-adaptivity}). A few examples of APFC numerical simulations performed with finite differences can be found, e.g., in Refs.~\cite{Goldenfeld2005,AthreyaPRE2007,Yeon2010,Bercic2018,Bercic2020,Geslin2015,Guan16}. Alternatively, the instability mentioned above can be avoided using spectral methods, as discussed in the next section.

\subsection{Fourier spectral method}
\label{sec:spectral}

Spectral methods solve differential equations treating variables as a sum of basis functions with coefficients to be computed, i.e., through a global representation. The so-called Fourier spectral method exploits the Fourier transform, typically in its discrete formulation for numerical integrations (therefore often referred to as pseudo-spectral, Fourier method). This method is particularly suited for periodic boundary conditions. A key feature of this approach is that, in the Fourier space, differential operators become algebraic expression of the wave vector, e.g. $\nabla^2 \psi(t) \rightarrow -|k|^2 \widehat \psi_k(t)$, where $\widehat \psi_k$ is the (discrete) Fourier transform of $\psi$. No finite difference approximations are then required if solving for $\widehat\psi_k(t)$, and $\psi(t)$ may be then obtained through a (discrete) inverse Fourier transform. Moreover, efficient algorithms exist to compute $\widehat{\psi}_k$ from $\psi$ and vice-versa, namely exploiting the Fast Fourier Transform (FFT) algorithm \cite{cooley1965algorithm}.  The adaptation of such approaches to phase field modeling in 
materials physics can be found in reference \cite{Chen98}.
This method generally allows for splitting off the linear term in $H$ and solving that part exactly, i.e., 
\begin{equation}
\pp{\psi}{t} = {\cal L} \psi + N(\psi),
\end{equation}
where ${\cal L} $ is a linear operator and $N$ is a non-linear function 
of $\psi$. Indeed, in Fourier space, this would then read
\begin{equation}
\pp{\widehat{\psi}_k}{t} = {\cal L}_k \widehat{\psi}_k + \widehat{N}_k,
\label{eq:ode}
\end{equation}
with $\widehat{N}_k$ 
the Fourier transform of $N(\psi)$ and ${\cal L}_k$ is an algebraic expression of the wave vector. Eq.~\eqref{eq:ode} is an ordinary differential equation with solution
\begin{equation}
\widehat{\psi}_k(t) = \expF{{\cal L}_k t}\widehat{\psi}_k(0)
+\expF{{\cal L}_k t}\int_0^{t} \dd t' \, 
\expF{-{\cal L}_k t'}\widehat{N}_k (t'). 
\label{eq:ff1}
\end{equation}
Typically, the numerical instability in Euler's method 
occurs when ${\cal L}_k$ is the most negative (i.e., at large 
wavevectors). However, in this method, $\expF{{\cal L}_k t}$ is very small in this limit so that instability is completely avoided.  To complete the picture, the non-linear term must be approximated as was done for $H(\psi)$ in the preceding section.  Considering Eq.~\eqref{eq:ff1} for $\widehat{\psi}_{k}(t+\Delta t)$ and approximating (explicitly) $\widehat{N}_k(t') \approx \widehat{N}_k(t)$ gives
\begin{equation}
\begin{split}
\widehat{\psi}_{k}(t+\Delta t) =\  &
\expF{{\cal L}_k \Delta t}\widehat{\psi}_k(t)
+\expF{{\cal L}_k (t+\Delta t)}\int_t^{t+\Delta t} \dd t' \, 
\expF{-{\cal L}_k t'}\widehat{N}_k (t')\\
\approx\ & \expF{{\cal L}_k\Delta t}\widehat{\psi}_k(t)
+\frac{\expF{{\cal L}_k\Delta t}-1}{{\cal L}_k} \widehat{N}_k(t),
\label{eq:spec}
\end{split}
\end{equation}
while other approximations of $\widehat{N}(t')$ may be considered as well. Eq.~\eqref{eq:spec} provides a relatively simple method of updating 
the field $\psi$ at one time step, although it requires Fourier transforms 
of $\psi$ and $N(\psi)$ and an inverse Fourier transform of $\widehat{\psi}_k$ 
per time step. 
While the method eliminates the Euler instability, the free energy will increase if the time step is too large, which should not occur. Nevertheless, depending on the specific model, it is possible to use time steps that are tens or hundreds of times larger than those used in the Euler algorithm.  
For the amplitude expansion, this method is directly applicable as 
the linear pieces of the equations of motion for $\eta_m$ are not 
coupled to any other amplitudes. Representative examples of APFC numerical simulations exploiting the Fourier pseudo-spectral method can be found, e.g., in Refs.~\cite{Hirvonen2016,Smirman2017,Huang2010,Mkhonta2013,Ofori-Opoku2013,XuPRB2016,Huter2017,Spatschek2010,HeinonenPRE2014,Jreidini2021}.

\subsection{Finite element method}
\label{sec:fem}

The Finite Element Method (FEM) emerged as a particularly suitable framework for solving the APFC model's equations \cite{Salvalaglio2017,Praetorius2019,SalvalaglioJMPS2020,Xiaoting2022}, besides being also employed in PFC studies in the first place \cite{backofen07,Gomez2012,Vignal2015,Ruihan2016,Liupeng2020}.
Indeed, it conveniently discretizes partial differential equations (PDEs) while exploiting inhomogeneous and adaptive meshes.

Within FEM, the PDEs are expressed in an integral form (weak form) over their domain of definition ($\Omega$), typically having a rectangular/cubic shape. For the discretization of the resulting equations, a conforming triangulation $\mathcal{T}_{h}$ of the domain $\Omega$ is considered, usually with simplex elements $S\in \mathcal{T}_{h}$ (with characteristic size $h$). In the context of APFC simulations, linear elements have been mostly adopted. This means considering a discrete function space of local polynomial of order 1 ($\mathbb{P}_1$), namely $\mathcal{V}_h^1=\{v\in C(\Omega,\mathbb{R}) : v|_S \in \mathbb{P}_1(S,\mathbb{R}), S \in \mathcal{T}_{\rm h}\}$. A function $y \in \mathcal{V}^1_h$ can be written in terms of a basis expansion $y=\sum_i Y_i \Xi_i$ with real coefficients $Y_i$ and basis $\{\Xi_i\}$ of $\mathcal{V}_h^1$. To solve for complex functions, as $\eta_m$, their real and imaginary part can be considered as two (real) independent unknowns. Alternatively,
complex coefficients with real basis functions may be considered.

The FEM approach which has been used to solve APFC equations as in Eq.~\eqref{eq:ev_apfc}, features a splitting into two second-order equations for $\partial \eta_m / \partial t$ and $\rho_m=\mathcal{G}_m \eta_m$ (with $m=1,...,M$ as in Sec.~\ref{sec:formulas}) \cite{Salvalaglio2017,Praetorius2019}:
\begin{equation}
\begin{split}
    \frac{\partial \eta_m}{\partial t}&=-|\Gv_m|^2\bigg[A\mathcal{G}_m\rho_m+B\eta_m+ 3D(\Phi - |\eta_m|^2)\eta_m+ \frac{\partial f^{\rm s}}{\partial \eta_m^*}\bigg],\\
    \rho_m&=\mathcal{G}_m\eta_m=\Lap \eta_m + 2 \I \Gv_m \cdot \Grad \eta_m.
    \label{eq:splitted}
\end{split}
\end{equation}
This choice is convenient within the APFC framework as it allows the computing of relevant quantities straightforwardly as, e.g., the stress field, which may be rewritten in terms of both $\eta_m$ and $\rho_m$ and their spatial derivatives \cite{SalvalaglioJMPS2020} (see also Sec.~\ref{sec:stress-field}). Moreover, even though it is defined for $\mathcal{G}_m$, $\rho_m$ can be readily be used for computing $\mathcal{L}_m$, for instance when considering multi-mode approximations. From a numerical point of view, the splitting in Eq.~\eqref{eq:splitted} allows exploiting linear elements as only second-order operators appear, which translate to first order operators acting on elements of $\mathcal{V}_h^1$ in the weak form. With $(f,\,g) \coloneqq \int_\Omega f(\rv) g(\rv)\,\dd \rv$ the $L^2(\Omega,\mathbb{R})$ scalar product, and considering the integral form of Eq.~\eqref{eq:splitted}, the problem to solve then reads: for $t\in[0,T]$, find $\eta_m(t)=a_m(t)+\I b_m(t)$ and $\rho_m(t)=c_m(t)+\I d_m(t)$, with $a_m, b_m, c_m, d_m \in \mathcal{V}^1_h$ (implying hereafter their dependence on $t$), such that
\begin{equation}\label{eq:weak_evolution}\begin{split} 
  \left(\frac{\partial a_m}{\partial t},\,v\right) - A|\Gv_m|^2\bigg[ (\Grad c_m,\,\Grad v) + 2(\Gv_m \cdot\Grad d_m,\,v)\bigg] &= \left({\rm Re}[H(\{\eta\})],\,v\right),\ \\
  \left(\frac{\partial b_m}{\partial t},\,v\right) - A|\Gv_m|^2\bigg[ (\Grad d_m,\,\Grad v) - 2(\Gv_m \cdot\Grad c_m,\,v)\bigg] &= \left({\rm Im}[H(\{\eta\})],\,v\right),\ \\
  (c_m,\, v) + (\Grad a_m,\,\Grad v) + 2(\Gv_m\cdot\Grad b_m,\,v) &= 0\ , \\
    (d_m,\, v) + (\Grad b_m,\,\Grad v) - 2(\Gv_m\cdot\Grad a_m,\,v) &= 0\ ,
\end{split}\end{equation}
$\forall v \in \mathcal{V}^1_h$ subject to an initial conditions $\eta_m(0) = \eta_m^0$, and $H(\{\eta\})$=$\partial f^{\rm s} /\partial \eta_m+B\eta_m+3D(\Phi-|\eta_m|^2)\eta_m$. The time derivatives are approximated by $\partial a_m /\partial t=(a_m^{j+1}-a_m^j)/\Delta t_j$ and $\partial b_m /\partial t=(b_m^{j+1}-b_m^j)/\Delta t_j$, with $\Delta t_j=t_{j+1}-t_{j}$ the time step, and $j \in \mathbb{N}_0$ the index labelling time steps. The time discretization is obtained through an implicit-explicit (IMEX) scheme. It consists of evaluating all the linear (nonlinear) terms in Eq.~\eqref{eq:weak_evolution} implicitly (explicitly), i.e. at time $t^{j+1}$ ($t^j$) \cite{Salvalaglio2017,Praetorius2019}, with $a^{j+1}_m$, $b^{j+1}_m$, $c^{j+1}_m$, $d^{j+1}_m$ the unknowns to solve for. Eq.~\eqref{eq:weak_evolution} consists of a set of nonlinear equations due to $H(\{\eta\})$. This term can be generally linearized and handled through iterative approaches as Picard Iterations or the Newton method. A simple but effective approach, which can be exploited for methods introduced in previous sections too, consists of applying a one-iteration Newton method \cite{Salvalaglio2017}, i.e. approximating $H(\eta^{j+1})$ as 
\begin{equation}
    H(\eta^{j+1})=H(\eta^{j})+H'(\eta^{j})(\eta^{j+1}-\eta^{j}).
\end{equation}
To solve Eq.~\eqref{eq:weak_evolution}, basis function expansions of unknowns are considered, e.g. $a^{j+1}_m = \sum_i Y_{m,i}^{j+1} \Xi_i$, with ${Y}_{m,i}^{j+1}$ the coefficients to be computed at the $j$-th timestep (and analogous expressions and coefficients' definition for $b^{j+1}_m, c^{j+1}_m, d^{j+1}_m$). These coefficients are computed by substituting the basis function expansions into Eq.~\eqref{eq:weak_evolution}, setting basis functions as test functions, and solving the resulting system of equations. Notice that $M$ coupled systems \eqref{eq:weak_evolution} must be solved concurrently, with $M$ the number of independent amplitudes according to the considered lattice symmetry and approximation (see Sec.~\ref{sec:formulas}).
Boundary conditions (BC) such as Dirichlet, Neumann, or Periodic BC, may be included as in common FEM approaches. Further discussions and explanations of standard aspects can be found in specialized textbooks.

The FEM approach outlined above proved efficient in handling relatively large systems in both two and three dimensions, in combination with standard direct and iterative solvers within FEM toolboxes like, e.g., AMDiS \cite{Vey2007,WitkowskiACM2015}.
Further improvements may be devised to increase the performances. An example is reported in \cite{Praetorius2019} where the development of a dedicated preconditioner \cite{Praetorius2014,PraetoriusThesis2015} allowing for fast solver convergence has been proposed and exploited for simulations of hundreds of nanometers domains in three dimensions for some materials.

The approach described in this section is also prone to coupling with other equations. Indeed, other variables would share spatial features with amplitudes. Coupling terms could be considered as additional terms entering $\partial{\eta}_m/\partial t$. At the same time, other equations may be discretized readily following the main FEM features described above (linear elements, operator splitting in second-order PDEs, IMEX time discretization). This has been exploited for instance when imposing mechanical equilibrium \cite{SalvalaglioJMPS2020} (see Sec.~\ref{sec:mechanical-equilibrium}), to simulate binary systems \cite{SalvalaglioPRL2021} (see Sec.~\ref{sec:binary-systems}), and to investigate the effect of magnetic field on small-angle grain boundaries \cite{Backofen2022}.

\subsection{Mesh adaptivity}
\label{sec:mesh-adaptivity}

\begin{figure}
\includegraphics[width=\textwidth]{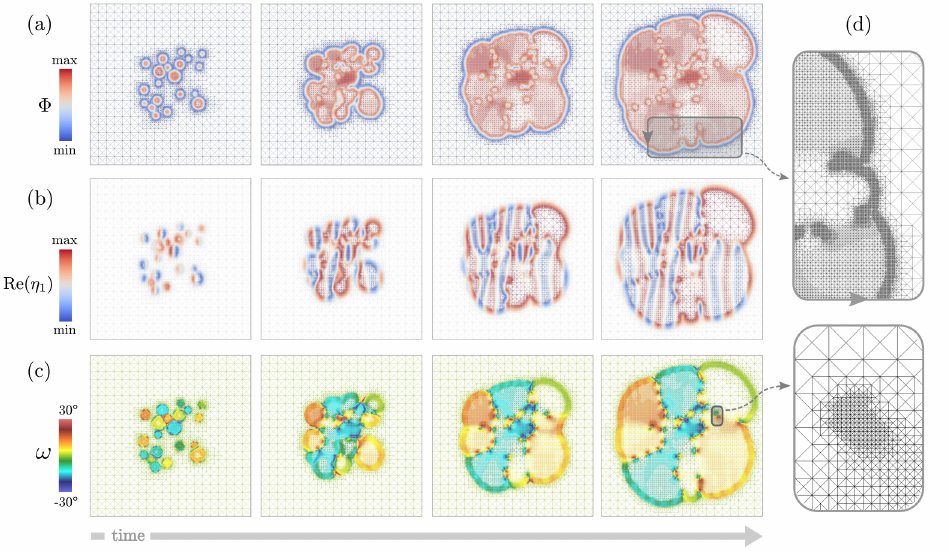}
\caption{Illustration of the growth of 20 crystal seeds (with a triangular lattice, one-mode approximation) having random orientation ranging in $[-15^\circ,15^\circ]$, as obtained by an APFC simulation with an adaptive mesh. The spatial discretization is represented by means of the mesh while colors represent: (a) $\Phi=\sum_m^M|\eta_m|^2$, (b) Re$(\eta_1)$, as indexed in \eqref{eq:energyterm_tri}, (c) local rotation $\omega$ w.r.t the reference crystal set by $\Gv_m$, computed by Eq.~\eqref{eq:strain2D}. (d) Magnification of two regions showing the mesh on a smaller length scale at the solid-liquid interface (top) and at a defect (bottom). Reprinted from \cite{Praetorius2019} $\copyright$ IOP Publishing Ltd. All
rights reserved.} 
\label{fig:figure4}
\end{figure}

Exploiting spatial adaptivity is a convenient strategy for performing efficient simulations with the APFC model \cite{AthreyaPRE2007,Salvalaglio2017,Bercic2018,Praetorius2019}. Indeed, amplitudes are constant for relaxed crystals, oscillate with different periodicity according to the local distortion of the crystal with respect to the reference one (see, e.g., Fig.~\ref{fig:figure1}) and exhibit significant variation at defects and solid-liquid interfaces. Depending on the numerical approach and set of 
equations, one may devise different strategies to set a local refinement, e.g., based on error estimates or indicators. 

An optimized local resolution based on the amplitudes oscillations, which works even for the standard approaches considered so far, has been achieved focusing on phases of the complex amplitudes, $\text{arg}(\eta_m)=\theta_m$. By looking at this quantity, it is possible to determine the wavelength of oscillating amplitudes $\lambda_m$ \cite{Praetorius2019}. 
Then for a good resolution of all the amplitudes, the discretization $h$ should be a fraction of the smallest $\lambda_m$, i.e. $h_{\rm amp}=\min_m{(\lambda_m)}/n$, with $n \geq 10$. 

To use this criterion in practice, the deformation, strain and/or rotation fields must 
be derived  from amplitudes. This will be discussed in detail in the following section (see Sec.~\ref{sec:stress-field}).
In addition to the oscillation of amplitudes, a refinement for the interfaces and defects controlled by $h_{\rm min}$ where $|\Grad \Phi|$ is significantly larger than a relatively small threshold $\varsigma$ and imposed as finest resolution in the mesh is considered \cite{Salvalaglio2017}, while a large discretization bound $h_{\rm max}$ is defined for region where $\Phi \sim 0$ or where $\theta_m \rightarrow 0$ (i.e. for constant amplitudes). Summarizing these concepts, this method ensures a local discretization, $h$, as 
\begin{equation}
  h=\begin{cases}
  h_{\rm min}, & \mbox{if } |\Grad \Phi| \geq \varsigma \\
  \min(\hspace{1pt}\max(h_{\rm amp},h_{\rm min})\hspace{1pt},\hspace{1pt} h_{\rm max} \hspace{1pt}), & \mbox{if } \Phi > 0 \mbox{ and } |\Grad \Phi| < \varsigma \\
  h_{\rm max},& \mbox{elsewhere.} 
  \end{cases}
  \label{eq:meshcriterion}
\end{equation}
This approach has been exploited together with the FEM approach outlined in Sec.~\ref{sec:fem}, in particular within the FEM toolbox AMDiS \cite{Vey2007,WitkowskiACM2015}. However, it is expected to work with any real-space method readily. Further optimization of the mesh refinement can be achieved by a polar representation \cite{AthreyaPRE2007,Bercic2018} which involves, however, some changes in the amplitude equations, the coupling with additional fields, and other technical details to be considered. An examples of an APFC simulation performed with the adaptive refinement strategy here outlined is given in Fig.~\ref{fig:figure4}.

\section{Continuum limit: elasticity and plasticity}
\label{sec:continuumlimit}

\subsection{Elasticity} 
\label{sec:linear-elasticity}

The elastic properties in the amplitude expansion 
arise from the term 
$A \sum_m^M \Gamma_m|\mathcal{G}_m\eta_m|^2$
(see Eq.~\eqref{eq:manylen}). Indeed, all the other terms 
in the free energy do not give rise to gradients in the phase 
of the amplitudes and 
as such do not contribute to the elastic energy.
To obtain the consequences 
of this term it is useful to consider  deformations ($\uv\equiv \uv(\rv)$) from 
a perfect lattice, i.e., 
\begin{equation}
\eta_m = \phi_m \expF{-\I \theta_{m}},
\label{eq:smalld}
\end{equation}
where $\theta_{m}\equiv {\bf G}_m \cdot \uv$ 
and $\phi_m$ is weakly dependent on $\uv$ (see a 1D illustration 
in Fig.~\ref{fig:figure1}(b)).  This leads to 
\begin{equation}
\begin{split}
\mathcal{G}_m \eta_m
&= \phi_m\,\expF{-\I \mathbf{G}_m \cdot \uv} \left(-\I \Lap \theta_m
-|\Grad \theta_m|^2+2\mathbf{G}_m\cdot\Grad\theta_m\right) \\
&\approx \phi_m\, \expF{-\I \mathbf{G}_m \cdot \uv} \left(
-|\Grad \theta_m|^2+2\mathbf{G}_m\cdot\Grad\theta_m\right),
\end{split}
\label{eq:elas_full}
\end{equation}
where in the last line higher order gradients in $\uv$ have 
been neglected. So that
\begin{equation}
\begin{split}
\sum_m^M \Gamma_m |\mathcal{G}_m \eta_m|^2 &
= 4 \sum_m^M \Gamma_m \phi_m^2 G_i^m G_j^m G_k^m G_l^m \left(
u_{ij} u_{kl} - u_{ij}u_{ko}u_{lo}+\frac{1}{4}u_{io}u_{jo}u_{kp}u_{lp} \right),
\label{eq:elas1}
\end{split}
\end{equation}
where $u_{ij}\equiv \partial u_i/\partial x_j$, $G_i^m$ is the $i$-th component of $\Gv_m$ and the Einstein summation convention is used. 
Eq.~\eqref{eq:elas1} contains linear and non-linear terms.
In terms of the non-linear Eulerian-Almanasi strain measure 
(${\bf U}$) 
\cite{Chan2009,Huter2017} with elements
\footnote{The strain measure ${\bf U}$ belongs to the general class of strain (material, Lagrangian) called Seth-Hill tensors $\varepsilon_n=(1/n)(\mathbf{C}^n-\mathbb{1})$, with $\mathbf{C}=\mathbf{F}^T\mathbf{F}$, $F_{ij}=\partial x_i/\partial X_j$ the deformation gradient and $\mathbf{x}$ and $\mathbf{X}$ the spatial (eulerian) and material (lagrangian) coordinates respectively, such that $d\mathbf{x}=\mathbf{F}d\mathbf{X}$ and $d\mathbf{X}=\mathbf{F}^{-1}d\mathbf{x}$ \cite{seth1961generalized,hill1968constitutive,hill1970constitutive,bruhns2015multiplicative,Neff2016}. ${\bf U}$ corresponds to $\varepsilon_{-1}$. This definition mixes a Lagrangian tensor due to the dependence on $\mathbf{F}^T\mathbf{F}$ (an Eulerian tensor would depend on $\mathbf{F}\mathbf{F}^T$), with an Eulerian strain measure $\mathbb{1}-\mathbf{F}^{-1}$ (a Lagrangian strain measure would depend on $\mathbf{F}-\mathbb{1}$), see also Ref.~\cite{Huter2017}.}, 
\begin{equation}
U_{ij}=\frac{1}{2}\left(u_{ij}+u_{ji}-u_{ik}u_{jk}\right),
\end{equation}
Eq.~\eqref{eq:elas1} can be written as
\begin{equation}
\sum_m^M \Gamma_m |\mathcal{G}_m \eta_m|^2 = 4\sum_m^M \Gamma_m \phi_m^2
G_i^m G_j^m  G_k^m G_l^m U_{ij}U_{kl}.
\end{equation}
The elastic part of the free energy is then 
\begin{equation}
F_{\rm elas} =\frac{1}{2}\int d\rv \ [\sigma_{ij}U_{ij}]=  4A \int d\mathbf{r}
\bigg[\sum_m^M \Gamma_m \phi_m^2 G_i^m G_j^m  G_k^m G_l^m U_{ij}U_{kl}\bigg].
\label{eq:Felas}
\end{equation}
The components of the stress tensor defined as 
\begin{equation}
\sigma_{ij}=\mathcal{C}_{ijkl}U_{kl},
\label{eq:hooke}
\end{equation}
where $\mathcal{C}_{ijkl}$ is the elastic modulus tensor \cite{LLEL} are then given by 
\begin{equation}
\mathcal{C}_{ijkl}= 8A\sum_m^M \Gamma_m\phi_m^2 G_i^m G_j^m  G_k^m G_l^m.
\label{eq:lambda}
\end{equation}
Thus Eq.~\eqref{eq:lambda} provides a general formula for 
the elastic moduli for arbitrary crystal symmetry. Some 
specific examples are given below.

\subsubsection*{Examples:}
\label{sec:ElasEx}
For a free energy with a single mode, i.e., containing the term  
$n(1+\Lap)^2n/2$, 2D triangular and 3D BCC structures minimize 
the free energy in certain parameter ranges.  At a minimum 
these systems can be described by modes with the same 
length scale and thus $\Gamma_m=1$  and 
$\phi_m=\phi$, $\forall m$. Following the definition of $\mathbf{G}_m$ as in Sec.~\ref{sec:results_symm} for these symmetries (one-mode approximation), Eq.~\eqref{eq:Felas} gives
\begin{equation}
\begin{split}
F^{\rm TRI}_{\rm elas}=&A\phi^2 \int \dd\rv \bigg[
\frac{9}{2} \sum_i U_{ii}^2+3U_{xx}U_{yy}+6U_{xy}^2
\bigg], \\
F_{\rm elas}^{\rm BCC} =&A\phi^2 \int \dd\rv \bigg[ 
4 \sum_i U_{ii}^2+4 \sum_{i,j>i} U_{ii}U_{jj}+8\sum_{i,j>i}U_{ij}^2
\bigg].
\end{split}
\label{eq:BCCelas}
\end{equation}
For the FCC symmetry in the two-mode approximations (see Sec.~\ref{sec:results_symm}), $\Gamma_m=1/16$\footnote{A factor of $1/9$ appears 
in Ref.~\cite{AnkudinovPRE2020} as a different scaling was 
employed.}, $\forall m$.  This gives 
\begin{equation}
F_{\rm elas}^{\rm FCC}=\frac{A}{9}\int \dd\rv \bigg[
(\phi^2+4\psi^2)\sum_iU_{ii}^2+2\phi^2\sum_{i,j>i}U_{ii}U_{jj} 
+4\phi^2 \sum_{i,j>i} U_{ij}^2
\bigg],
\label{eq:FCCelas}
\end{equation}
where $\eta_m = \phi \expF{-\I\theta_m}$ for $i=1,...,4$ and 
$\eta_m = \psi \expF{-\I\theta_m}$ for $i=5,...,7$.

One of the difficulties in parameterizing PFC models is that 
the ratio of the elastic moduli cannot be changed in the one 
mode triangular and BCC cases. However, it is interesting to 
note that in the FCC case, the ratio of the elastic moduli 
depends on $\psi$, which in principle can be tuned. It 
suggests that adding more length scales will allow for more 
tuneability in the models as shown in XPFC models \cite{xtal2}.   
However, it is important to note 
that if the added vectors have the same symmetry as the 
original ones this will not change the ratios. 

\subsection{Strain and stress field from the amplitudes}
\label{sec:stress-field}

When examining the results of APFC simulations, it is useful to
develop methods to extract the strain and stress fields directly 
from the complex amplitudes. As shown by 
Salvalaglio \etal \cite{SalvalaglioNPJ2019} the displacement field, 
$\uv$ that enters continuum elasticity field can be extracted  directly from the phase of the amplitudes ($\theta_m$). In two dimensions (2D), inverting Eq.~\eqref{eq:smalld}, the expression is
\begin{equation}
u^{\rm 2D}_i=-\frac{\epsilon_{ij}}{\hat{\mathbf{p}}\cdot(\mathbf{G}_l\times\mathbf{G}_m)}\big[
G_j^m \theta_l-G_j^l\theta_m
\big],
\label{eq:st1}
\end{equation}
with $(i,j)=(x,y)$ and cyclic permutations, $\epsilon_{ij}$ is 
the 2D Levi-Civita symbol, $l$ and $m$ label two different amplitudes, $\hat{\mathbf{p}}=\hat{\mathbf{x}}\times{\hat{\mathbf{y}}}$ the normal vector of the xy-plane and $\theta_m=\arg({\eta_m})=\arctan \left[ \text{Im} (\eta_m)/ \text{Re}
(\eta_m)  \right]$.  In three dimensions (3D) it can be shown that 
\begin{equation}\begin{split}
u_i^{\rm 3D} =- \frac{1}{\mathbf{G}_n\cdot(\mathbf{G}_m\times \mathbf{G}_l)}\big[&
\theta_l(G_k^mG_j^n-G_j^mG_k^n)+\theta_m(G_k^nG_j^l-G_j^nG_k^l) \\
&+\theta_n(G_k^lG_j^m-G_j^lG_k^m)
\big].
\end{split} \end{equation}
with $(i,j,k)=(x,y,z)$ and cyclic permutations, and $l$, $m$, $n$, labelling three different amplitudes. These quantities are discontinuous. However the component of the (linear) strain tensor $\mathbf{U}^{\rm L}$ become expressions of $\partial \theta_m / \partial
x_i$ with 
\begin{equation} \frac{\partial \theta_m}{\partial
x_i}= \frac{1}{|\eta_m|^2} \left[ \frac{\partial \text{Im} (\eta_m) }{\partial
x_i}\text{Re} (\eta_m) - \frac{\partial \text{Re} (\eta_m) }{\partial x_i}
\text{Im} (\eta_m) \right] ,
\label{eq:derphase}
\end{equation}
which is continuous almost everywhere in the solid phase, with a singularity for vanishing amplitudes in correspondence of phase singularities, e.g., at the cores of defects. Then, with a regularization for these amplitudes (see also Sec.~\ref{sec:comparison-continuum-elasticity}), elastic field can be readily computed and conveniently exploited. In two dimensions, for $\mathbf{U}^{\rm L}$ and the rotation field $\omega=\nabla \times \mathbf{u}$ we then get
\begin{equation}
\begin{split}
U_{xx}^{\rm L}&=-\frac{1}{\hat{\mathbf{p}}\cdot(\mathbf{G}_l \times \mathbf{G}_m)} \left(G_y^m \frac{\partial
\theta_l}{\partial x} - G_y^l \frac{\partial \theta_m}{\partial x}\right), \\
U_{yy}^{\rm L}&=- \frac{1}{\hat{\mathbf{p}}\cdot(\mathbf{G}_l \times \mathbf{G}_m)} \left(G_x^l \frac{\partial
\theta_m}{\partial y}  - G_x^m \frac{\partial \theta_l}{\partial y}\right), \\
U_{xy}^{\rm L}&=-\frac{1}{2\hat{\mathbf{p}}\cdot(\mathbf{G}_l \times \mathbf{G}_m)} \left(G_y^m \frac{\partial
\theta_l}{\partial y} - G_y^l \frac{\partial \theta_m}{\partial y} + G_x^l
\frac{\partial \theta_m}{\partial x}  - G_x^m \frac{\partial
\theta_l}{\partial x} \right), \\
\omega&=-\frac{1}{2\hat{\mathbf{p}}\cdot(\mathbf{G}_l \times \mathbf{G}_m)} \left(G_y^m \frac{\partial \theta_l}{\partial y}
- G_y^l \frac{\partial \theta_m}{\partial y} - G_x^l \frac{\partial
\theta_m}{\partial x}  + G_x^m \frac{\partial \theta_l}{\partial x}\right).
\end{split}
\label{eq:strain2D}
\end{equation}
Explicit expressions for 3D strain and rotation fields can be found in Ref.~\cite{SalvalaglioNPJ2019}. The stress field can then be computed through the Hooke's law \eqref{eq:hooke}.

In 2018 Skaugen, Angheluta and Vi\~nals  
\cite{SkaugenPRB2018}
derived an expression for the stress tensor, $\sigma_{ij}$ from the 
density field using the standard definition of $\sigma_{ij}$, i.e, 
\begin{equation}
\sigma_{ij} = \frac{\delta \Delta F}{\delta (\partial_i u_{j})},
\label{eq:sigmapfc}
\end{equation}
where $\Delta F = F(n(\rv+\uv))-F(n(\rv))$ and $\uv$ is 
the displacement field.  This gives 
\begin{equation}
\sigma_{ij}= [\partial_i {\cal L}n] \partial_j n
-[{\cal L}n](\partial_{ij}n)+P\delta_{ij},
\end{equation}
where $P=f-n(\delta F/\delta n)$ is a pressure term summing up to the mechanical stress, with $f$ the integrand in Eq.~\eqref{eq:F_PFC}, the second term arising when considering mass-conserving deformations \cite{Skogvoll_stress2021}, and ${\cal L}\equiv 1+\nabla^2$. In terms of amplitudes, integrating over
the a unit cell with $n$ expressed via Eq.~\eqref{eq:nnn} and neglecting the pressure terms gives \cite{SalvalaglioJMPS2020}
\begin{equation}
\begin{split}
    \sigma_{ij}=\sum_{m}^M&
    \bigg\{ \big[(\partial_i + \I G_i^m)(\nabla^2 + 2\I\mathbf{G}_m\cdot \nabla)\eta_m \big] \big[(\partial_j - \I G_j^m)\eta_m^*\big] \\
    &- \big[(\nabla^2 + 2\I\mathbf{G}_m\cdot \nabla) \eta_m\big] \big[(\partial_i - \I G_i^m)(\partial_j - \I G_j^m)\eta_m^*+\cc\big] 
    \bigg\},
    \label{eq:sigmapsi1}
\end{split}
\end{equation}
for one-mode approximations, while it can be generalized for more modes accounting for the full ${\cal L}_m$ operators (see Eq.~\eqref{eq:diffop}).

\subsection{Plasticity and defect dynamics}
\label{sec:plasticity} 

As seen in previous sections, the amplitude formalism can describe the 
elastic behavior of crystals as encoded in the PFC model. Moreover, 
by focusing on singularities in the corresponding phases, the motion of defects may be connected to the evolution amplitudes \cite{SkaugenPRL2018,SkaugenPRB2018,SalvalaglioPRL2021,Skogvoll2021}.

A dislocation in a crystalline lattice corresponds to a 
discontinuity in the phase $\theta_m$. 
At the same time, a dislocation with Burgers vector $\bv$ is defined by $\oint d\uv = \bv$ \cite{anderson2017}, thus it can be shown that $\oint \dd\theta_m = -\Gv_m \cdot \bv = -2\pi s_m$, where $s_m$ is the winding number. As discussed in Ref.~\cite{SkaugenPRB2018}, a vortex solution for amplitudes at dislocation cores may be assumed, that reads $\eta_m \propto x - \I s_m y $ with $s_m = \pm 1$.
The Burgers vector distribution of a dislocation can be defined as a localized (vectorial) topological charge $\bv\delta(\rv-\rv_0)$ with $\rv_0$ the nominal position of the dislocation core, assumed pointwise from a continuous point of view. By extension, the Burgers vector density can be defined to be $\BBv(\rv)=\sum_{d=1}^D \bv^d \delta(\rv-\rv^{d}_0)$, with $d$ indexing the dislocations and $D$ their total number. To connect this quantity to amplitudes, note that the position of the core is where the amplitudes go to zero. Therefore, following the theoretical framework reported in \cite{Mazenko1997,Mazenko2001,Angheluta2012}, a change of coordinates from the canonical one to the amplitudes' components can be considered. Namely, for point dislocations in two dimensions, or straight dislocations in three dimensions, one gets 
\begin{equation}
\BBv(\rv)=- \beta \sum_m^M \Gv_m D_m \delta(\eta_m),
\quad  D_m = \frac{\epsilon_{ij}}{2\I}\partial_i
\eta^*_m \partial_j \eta_m,
\label{eq:coord-change}
\end{equation}
with $D_m$ the Jacobian determinant of the coordinates' transformation, $\beta \equiv \beta_k=2\pi /\sum_m^M (G^m_k)^2$ as $\beta_x=\beta_y=\beta_z$ (as can be verified explicitly with $\Gv_m$ defined in Sec.~\ref{sec:results_symm}), $\epsilon_{ij}$ is the Levi-Civita symbol, delta functions transforming as $D_m \delta(\eta_m)=-(2\pi)^{-1}\sum_d^D (\qv_m \cdot \bv^d)\delta(\rv-\rv^d_0)$ \cite{Mazenko1997,Mazenko2001,SkaugenPRB2018}, and implying the Einstein
summation convention. Aiming at the velocity of dislocations, the dynamics of $\BBv(\rv)$ is considered. Exploiting that the determinant fields $D_m$ have conserved currents \cite{Angheluta2012}, $\partial D_m / \partial t = - \partial_i J_i^m$, with
\begin{equation}
J^m_i=\epsilon_{ij}\text{Im}\left(\pp{\eta_m}{t}\partial_j\eta_m^*\right),
\label{eq:JJJ}
\end{equation}
and that a similar continuity equation holds true for $\delta(\eta_m)$, from Eq.~\eqref{eq:coord-change} the equation of motion for $B_i$ may be written,
\begin{equation}
\begin{split}
\frac{\partial B_i }{\partial t} =& - \partial_j {\cal J}_{ij}= -\partial_j\bigg[\beta \sum_m^M G_i^m J_j^m \delta(\eta_m)\bigg] \\
=&\ \partial_j\bigg[ \frac{\beta}{2\pi}\sum_m^M G_i^m J_j^m\sum_d^D\frac{\Gv_m \cdot \bv^d}{D_m}\delta(\rv-\rv_0^d)\bigg],
\label{eq:dBdt}
\end{split}
\end{equation}
where the last term was obtained by transforming back the delta function to spatial coordinates.
For dislocations moving at a velocity $\vv^d$, it also follows that ${\cal J}_{ij}=\sum_d^D b_i^d v_j^d \delta(\rv-\rv^d_0)$. Therefore, by equating this latter expression with the corresponding quantity in Eq.~\eqref{eq:dBdt}, the dislocation velocity can be related to the evolution of amplitudes as
\begin{equation}
\mathbf{v}^d = \frac{\beta}{2\pi} 
\sum_m^M  \frac{(\Gv_m \cdot \bv^d)^2}
{|\bv^d|^2}\frac{\mathbf{J}_m}{D_m}.
\label{eq:vel_amp}
\end{equation}
At the dislocation core, a few simplifications may be considered. For the amplitudes which are zero at the dislocation core,
\begin{equation}
\pp{\eta_m}{t} 
= -|\Gv_m|^2 A \Gamma_m {\cal G}_m^2 \eta_m \approx -\I 8A\Gamma_m|\Gv_m|^2
 (\Gv_m \cdot \Grad \phi_m)
\left(\Gv_m \cdot \Grad \theta_m
\right) \expF{\I \theta_m},
\label{eq:detadtvel}
\end{equation}
while others do not contribute to Eq.~\eqref{eq:vel_amp}. The latter term in Eq.~\eqref{eq:detadtvel} is obtained by imposing again a form for amplitudes as in Eq.~\eqref{eq:smalld} and retaining the lowest order only in $\phi_m$ and $\theta_m$. Combing all the equations reported above gives
\begin{equation}
v_i^d = \frac{8 \beta A b_j^d}{|\bv^d|^2} 
\epsilon_{ik} \sum_m^M \Gamma_m|\Gv_m|^2 G_j^m G_k^m G_l^m G_p^m U_{lp}.
\label{eq:vj2}
\end{equation}
where $U_{ij}=(\partial_i u_j + \partial_j u_i)/2$. This equation is consistent with the classical Peach-Koehler force
\cite{anderson2017}. 
For the case of a 2D triangular lattice or a 3D BCC crystal where
it is possible to construct the lattice by retaining only one mode of
the lowest order (with $|\Gv_m|=1$, $\Gamma_m=1$), the velocity takes the form
\begin{equation}
v_i^d = M \epsilon_{ij}\left( \sigma_{jk} b_k^d \right),
\label{eq:vj4}
\end{equation}
with M a mobility factor.

With this formalism, the dynamic of defects may be obtained once $\partial{\eta_m}/\partial{t}$ are known. This applies independently to the specific contributions affecting the dynamics of amplitudes. See, for instance, an application to binary systems in Sec.~\ref{sec:binary-systems}. The equations presented here apply for point dislocations in two dimensions or straight dislocations in three dimensions. A generalization to curved dislocations in three dimensions has been recently introduced in Ref.~\cite{Skogvoll2021}.

\subsection{Comparisons with elasticity theories}
\label{sec:comparison-continuum-elasticity}

As noted in previous sections, the APFC model may be employed to the study elasticity and plasticity in crystalline systems. A few prototypical cases have been investigated, delivering direct comparisons with predictions from other theories \cite{Huter2016,SalvalaglioNPJ2019,SalvalaglioJMPS2020}. Of particular note is the comparison with continuum elasticity results, as the coarse-grained nature of APFC may deliver advanced/improved continuum approaches.

\begin{figure}
\includegraphics[width=\textwidth]{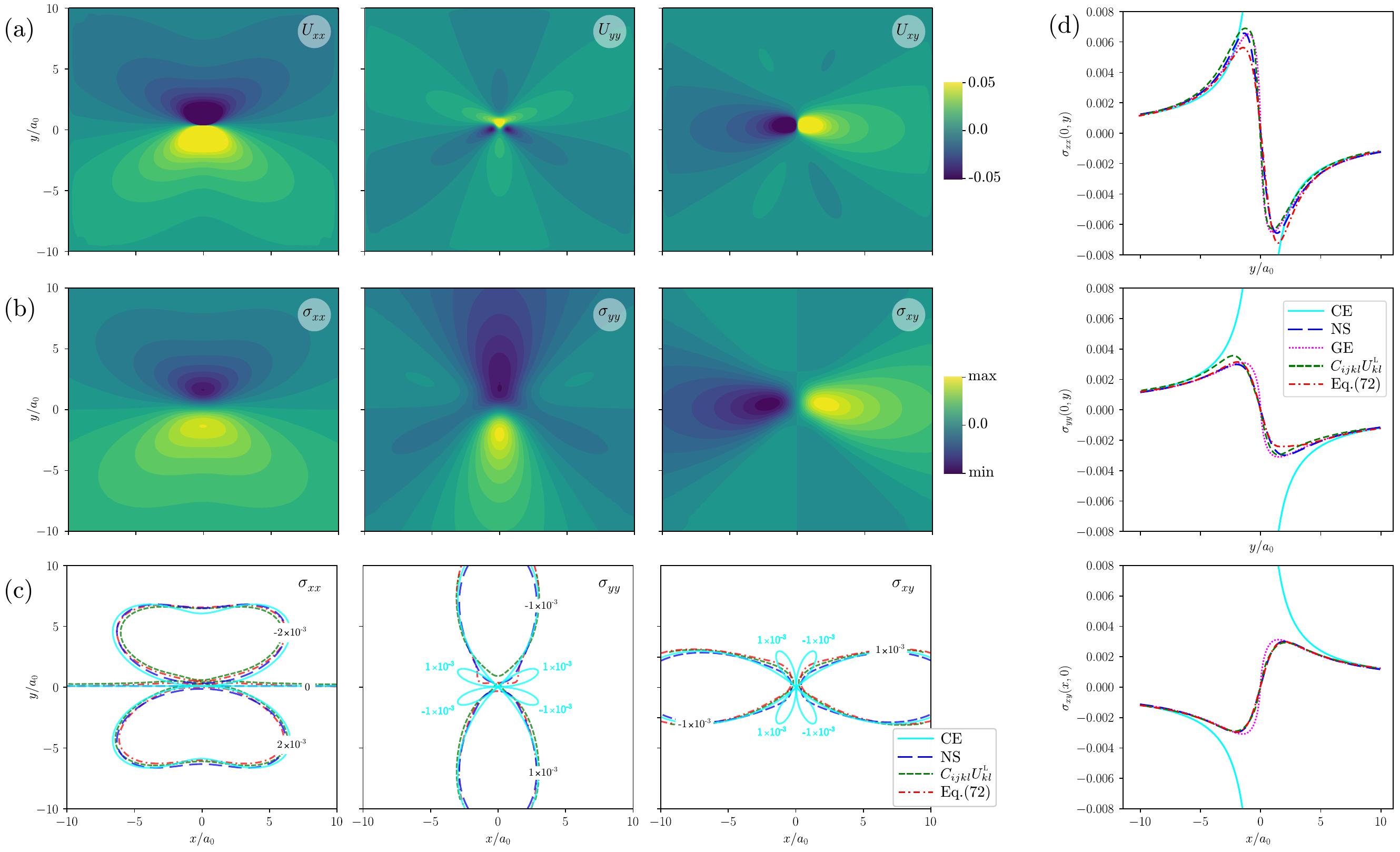}
\caption{Elastic field induced by an edge dislocation in a 2D triangular lattice (one-mode approximation) with $b=a_0=4\pi/\sqrt{3}$. Parameters for the considered APFC simulation: $A=0.98$, $B=0.044$, $C=-0.5$, $D=1/3$. (a) Strain field $U_{ij}^{\rm L}$ from Eqs.~\eqref{eq:strain2D} with $l=1$ and $m=3$, i.e. the amplitudes with singular phases. (b) Stress field from Eqs.~\eqref{eq:sigmapsi1}. (c) Comparison of representative isolines of the 2D stress fields obtained by different methods and continuum theories: Eq.~\eqref{eq:hooke} with $U_{ij}^{\rm L}$ as in panel (a), stress fields from panel (b), classical continuum elasticity from Eq.~\eqref{eq:stressWC} with $c=0$ (CE), non-singular field theory from Eq.~\eqref{eq:stressWC} with $c=a_0$ (NS). 
(d) Comparison of stress fields as in panel (c) along a line crossing the defect core, including also the stress field from the strain gradient formulation of Eq.~\eqref{eq:stressGE} (GE).}
\label{fig:figure5}
\end{figure}

A representative case is the elastic field generated by dislocations at mechanical equilibrium, which is well known in the continuum (linear) elasticity for isotropic media \cite{anderson2017,Cai2006}. In the APFC model, configurations with dislocations in prescribed positions may be obtained with different approaches. The phase of amplitudes $\sigma_{\hkl}$ can be initialized with singularities as discussed in Sec.~\ref{sec:plasticity} at given positions and then the APFC model is used 
to minimize the free energy. By restricting the description to 2D crystals for the sake of simplicity, a convenient approach consists of setting phases $\theta_{m}=-\Gv_m\cdot \uv^{\rm dislo}$ with
\begin{equation}
    \begin{split}
        u_x^{\rm dislo}&= \frac{b}{2\pi} \bigg[ \arctan{\left(\frac{y}{x}\right)} +\frac{xy}{2(1-\nu)(x^2+y^2)} \bigg],\\
        u_y^{\rm dislo}&= -\frac{b}{2\pi} \bigg[ \frac{(1-2\nu)}{4(1-\nu)}\log{\left( x^2+y^2 \right)}+\frac{x^2-y^2}{4 (1-\nu) (x^2+y^2)} \bigg],\\
    \end{split}
    \label{eq:udislo}
\end{equation}
the displacement field of an edge dislocation having Burgers vector $\bv=b\hat{\mathbf{x}}$ and $\nu$ the Poisson's ratio \cite{anderson2017}. Alternatively, an initial strain that induces the formation of dislocations can be considered. For instance, a pair of dislocations having the Burgers vector $\pm \bv^{\rm edge}$ is obtained by defining layers with initial deformation $\uv=[Dx,0]$ with $D = \pm b/L$ and allowing the system to relax \cite{Salvalaglio2017}. Dislocations move when Peach-Koehler force is 
finite assuming no barriers exist (see Sec.~\ref{sec:barriers}). As discussed in Sec.~\ref{sec:mechanical-equilibrium}, for dynamical configurations, corrections are needed to account for mechanical equilibrium within (A)PFC. Special cases are the configurations where defects do not move, and relaxation given by dynamical equations effectively approaches mechanical equilibrium. These may be represented, for example, by equally spaced arrays of dislocations along $\hat{\mathbf{x}}$ and $\hat{\mathbf{y}}$ with alternating Burgers vectors, i.e., a ``grid" where four defects with the same Burgers vectors surround another one with opposite Burgers vector. It is worth mentioning that a single dislocation, in the absence of external stress, would be in principle stationery too (as the Peach-Koehler force is zero). Still, its elastic field would inherently interact with the boundaries of any finite simulation domain as it is long-range, with energy dependent on the system size and diverging for an infinite medium. A possible solution would be studying a single dislocation in a finite crystal \cite{SalvalaglioJMPS2020}, which, however, is expected to induce changes in the elastic field \cite{anderson2017,Head_1953,Marzegalli2013}.

Fig.~\ref{fig:figure5} shows the elastic field of a dislocation belonging to a two dimensional grid with alternating Burgers vector along $\hat{\mathbf{x}}$ and $\hat{\mathbf{y}}$. Both strain components resulting from computing Eqs.~\eqref{eq:strain2D} (Fig.~\ref{fig:figure5}(a)) and stress components from Eq.~\eqref{eq:sigmapsi1} (Fig.~\ref{fig:figure5}(b)) are shown. These fields agree well with the field expected in classical continuum elasticity \cite{anderson2017}. The elastic field obtained from Eqs.~\eqref{eq:strain2D} is to some extent easier to compute as it involves only the first derivatives of amplitudes. Still, they are singular at the core of vanishing amplitudes, here regularized by setting to $1/(|\eta_m|^2+\delta)$ with a small $\delta$ as prefactor in Eq.~\eqref{eq:derphase}. On the other hand, the elastic field from Eq.~\eqref{eq:sigmapsi1} does not require such a numerical regularization. This approach involves higher-order derivatives than Eq.~\eqref{eq:derphase}, which can be handled efficiently when combined with a proper splitting of the APFC equations (see also Sec.~\ref{sec:numerical-methods}).

More insights are given in Fig.~\ref{fig:figure5}(c) and Fig.~\ref{fig:figure5}(d). Therein, a comparison of the stress field components obtained with different continuum theories for representative isolines (panel c) and along lines crossing the defect core (panel d) is reported. In particular, it shows 
the stress fields components computed from the APFC simulation, namely Eq.~\eqref{eq:sigmapsi1} and Eq.~\eqref{eq:hooke} with $U_{ij}^{\rm L}$ from Eq.~\eqref{eq:strain2D} with $\phi^2=\sum_{m=1}^3 |\eta_m|^2 /3$. These fields are compared with the non-singular isotropic theory (NS) reported by Wei Cai \etal in Ref.~\cite{Cai2006}, 
\begin{equation}
        \frac{\sigma_{xx}^{\rm NS}}{\sigma_0}= -\frac{y (3c^2+3x^2+y^2)}{(c^2+x^2+y^2)^2}, \ \
        \frac{\sigma_{yy}^{\rm NS}}{\sigma_0}= - \frac{y (c^2-x^2+y^2)}{(c^2+x^2+y^2)^2}, \ \
        \frac{\sigma_{xy}^{\rm NS}}{\sigma_0}= \frac{x (c^2+x^2-y^2)}{(c^2+x^2+y^2)^2},
    \label{eq:stressWC}
\end{equation}
and $\sigma_{zz}^{\rm NS}=\nu(\sigma_{xx}^{\rm NS}+\sigma_{yy}^{\rm NS})$, with $\sigma_0=E b_x/(4\pi (1-\nu)^2)$, $E$ the Young modulus, $\nu$ the Poisson ratio, and $c$ a parameter controlling the extension of the core-regularization ($c=0$ reduces to classical continuum elasticity (CE) formulations $\sigma^{\rm CE}$ \cite{anderson2017}). The triangular symmetry considered here, which results isotropic, and under the plane strain condition, gives $\upmu=\uplambda=3\phi^2$ while $E=\upmu(3\uplambda+2\upmu)/(\uplambda+\upmu)=(5/2)\phi^2$, and $\nu=\uplambda/(2\uplambda+2\upmu)=1/4$ 
\footnote{Plane strain setting corresponds to have $U_{zz}=U_{xz}=U_{yz}=0$ given by $u_z=0$, and $\sigma_{zz}=\nu(\sigma_{xx}+\sigma_{yy})$ (entering, e.g., Eq.~\eqref{eq:stressWC} and \eqref{eq:stressGE}). It leads to the expressions for $\nu$ and $E$ in the text. The alternative is the plane stress setting where $\sigma_{zz}=0$ and thus $U_{zz} \neq 0$ and $u_z \neq 0$). It leads to $E=4\upmu(\uplambda+\upmu)/(\uplambda+2\upmu)=(8/3)\phi^2$, and $\nu=\uplambda/(\uplambda+2\upmu)=1/3$.}. 
Another comparison with continuum elasticity is provided with a regularized formulation of the stress emerging in the framework of strain-gradient elasticity (Helmholtz type) \cite{Lazar2005,Lazar2017}
\begin{equation}
    \begin{split}
    \frac{\sigma_{xx}^{\rm GE}}{\sigma_0}&=-\frac{y}{r^4}\bigg[(y^2+3x^2)+\frac{4\ell^2}{r^2}(y^2-3x^2)-2y^2\frac{r}{\ell}K_1(r/\ell)-2(y^2-3x^2)K_2(r/\ell)\bigg], \\
    \frac{\sigma_{yy}^{\rm GE}}{\sigma_0}&=-\frac{y}{r^4}\bigg[(y^2-x^2)-\frac{4\ell^2}{r^2}(y^2-3x^2)-2x^2\frac{r}{\ell}K_1(r/\ell)+2(y^2-3x^2)K_2(r/\ell)\bigg],\\
    \frac{\sigma_{xy}^{\rm GE}}{\sigma_0}&=\frac{x}{r^4}\bigg[(x^2-y^2)-\frac{4\ell^2}{r^2}(x^2-3y^2)-2y^2\frac{r}{\ell}K_1(r/\ell)+2(x^2-3y^2)K_2(r/\ell)\bigg],\\
    \end{split}
    \label{eq:stressGE}
\end{equation}
and $\sigma_{zz}^{\rm GE}=\nu(\sigma_{xx}^{\rm GE}+\sigma_{yy}^{\rm GE})$, with $K_n(r/\ell)$ the modified Bessel function of the second type, and $\ell$ a characteristic internal length parameter of the material. The elastic field obtained from APFC simulations encodes a smoothing similar to the non-singular theories in Eq.~\eqref{eq:stressWC} and Eq.~\eqref{eq:stressGE}. A good agreement is obtained with $c=2a_0$ and $\ell=a_0$. However, notice that these parameters are expected to vary for different quench depths as they are related to the extension of the core \cite{Cai2006,Lazar2005} and this shrinks with decreasing the temperature. It is worth mentioning that strain gradient terms may be indeed identified in Eq.~\eqref{eq:elas_full}, supporting the qualitative agreement shown in Fig.~\ref{fig:figure5}. For isotropic materials, a more accurate description is actually given by the so-called Mindlin's isotropic first gradient elasticity, which feature two characteristic lengths \cite{Mindlin1964,Mindlin1968,Lazar2018} and may therefore provide descriptions closer to the APFC results. Comparisons for 3D configurations and for rotation fields from Eq.~\eqref{eq:strain2D} can be found in Ref.~\cite{SalvalaglioNPJ2019}.

Another example is offered by a recent APFC formulation \cite{Chockalingam2021} encoding a mechanical deformation not caused by a defect or an external mechanical stress (namely an eigenstrain \cite{Kinoshita1971}). In practice, a spatially dependent $q_0\equiv q(\mathbf{r})$ is set in the free energy \eqref{eq:F_PFC}, such that 
\begin{equation}
    q(\mathbf{r}) = \frac{q_0}{1+\varepsilon^*(\mathbf{r})} = \beta(\mathbf{r}) q_0,
    \label{eq:betaeshelby}
\end{equation}
with $\varepsilon^* = (a(\mathbf{r})-a_{0})/a_{0} = q_0 / q(\mathbf{r})-1$ the eigenstrain encoding a deformation from a lattice parameter $a_{0}$ to a lattice parameter $a(\mathbf{r})$.
When setting $\beta(\mathbf{r})\gtrless 1$ and constant, corresponding to an eigenstrain $\varepsilon^* \lessgtr 0$, within a region embedded in a medium having $\beta(\mathbf{r})=1$ the resulting elastic field matches well with the solution of the Eshelby inclusion problem \cite{eshelby1957determination,eshelby1959elastic,mura2013micromechanics} as shown in \cite{Chockalingam2021}.

\section{Limits and extensions}
\label{sec:limits_extensions}

\subsection{Large tilts: the problem of beats}
\label{sec:beats}

\begin{figure}
\includegraphics[width=\textwidth]{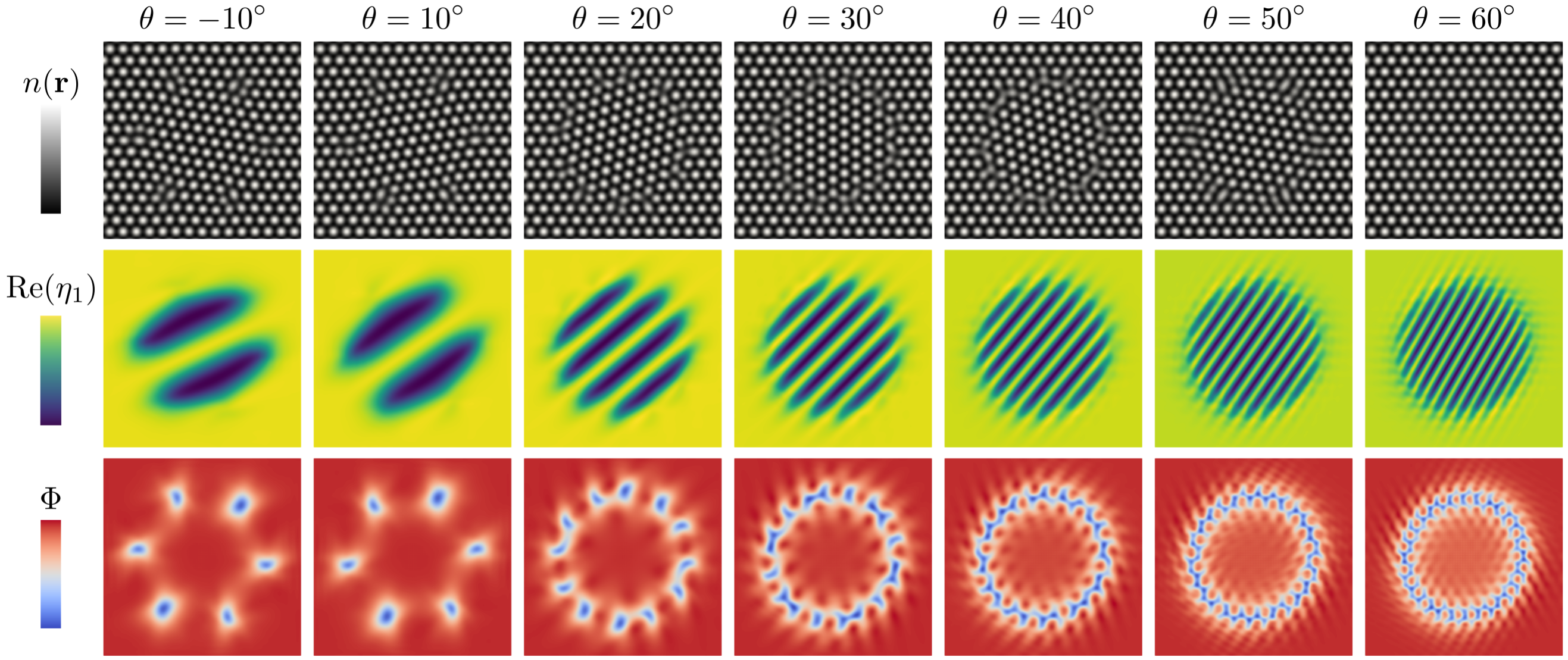}
\caption{APFC description of (small) circular rotated inclusion in a 2D crystal with triangular symmetry (one-mode approximation), for different tilts with respect to the surrounding matrix. Different rows show: the reconstructed density $n(\rv)$, the real part of $\eta_1$ and $\Phi$.}
\label{fig:figure6}
\end{figure}

Complex amplitudes consistently describe deformations, i.e., the energy is rotationally invariant while accounting for elastic energy associated with distortion with respect to the reference state (see Sec.~\ref{sec:linear-elasticity}). However, the larger the rotation with respect to the reference crystal (described by Eq.~\eqref{eq:n} and the choice of $\Gv_m$) is, the shorter (larger) is their wavelength (frequency), resulting in the so-called \textit{problem of beats} \cite{AthreyaPRE2007,Spatschek2010,Huter2017}. 
Indeed, in the presence of a rotation $\Theta$, the density (assuming here zero average), can be written 
\begin{equation}
    n=\sum_m^M \eta_m^\Theta \expF{\I \Gv_m \cdot \rv}=\sum_m^M \phi_m \expF{\I \Gv_m(\Theta)-\Gv_m \cdot \rv} \expF{\I \Gv_m \cdot \rv}=\sum_m^M\phi_m\expF{\I \Delta\Gv_m(\Theta) \cdot \rv} \expF{\I \Gv_m \cdot \rv},
    \label{eq:beats}
\end{equation}
where $G^m_i(\Theta)=G^m_j R_{ij}(\Theta)$ and $R_{ij}(\Theta)$ is the counter-clockwise rotational matrix. Therefore, oscillations of $\eta_m^\Theta$ have a wavelength $2\pi/|\Delta \Gv_m(\Theta)|$. This leads to a crucial two-fold limitation for the APFC model. On one side, the spatial resolution required to discretize the corresponding equations depends on their relative orientation with respect to the reference lattice encoded in $\Gv_m$. For large rotations this results in significant variations of the amplitudes over lengths approaching the lattice spacing, inconsistent with the assumption in their derivation and also requiring mesh sizes approaching the ones required in the PFC model.
On the other side, while the energy of a single crystal remains rotationally invariant, the rotational symmetry of bicrystals is lost, and unphysical grain boundaries are obtained for large relative tilts corresponding to small or no deviations in the density field $n$ (e.g., when rotating a 2D triangular lattice by $\sim 60^\circ$). An illustration of this behavior is reported in Fig.~\ref{fig:figure6}. When increasing the relative rotation of a circular inclusion, the oscillation of amplitudes increases requiring finer mesh as illustrated by $\text{Re}(\eta_1)$. Even though the fields are properly resolved, unphysical grain boundaries appear in $\Phi$ for $\theta \gtrsim 30^\circ$ (e.g., according to symmetry, $\theta=-10^\circ$ and $\theta=50^\circ$ should coincide, as well as $\theta=60^\circ$ should have no defects with a $\Phi$ uniform).

An attempt to overcome this issue followed the first publications on the APFC model and consists of a polar representation of amplitudes \cite{AthreyaPRE2007}. In practice, the complex amplitudes are expressed in terms of the real fields $\phi_m=|\eta_m|$ and $\theta_m={\rm arg}(\eta_m)$. 
The resulting set of equations for $\partial{\phi}_m/\partial t$ and $\partial{\theta}_m/\partial t$ derived from Eq.~\eqref{eq:ev_apfc}, have issues related to the discontinuous nature of $\theta_m$ and that $\phi_m$ vanishes in the liquid phase, in principle requiring robust and structured regularization algorithm. Therefore, further approximations are introduced \cite{AthreyaPRE2007}: i) a hybrid formulation exploiting the aforementioned polar representation only for crystal bulk, i.e. away from defects and interfaces, while solving the equations for the complex amplitudes everywhere else; ii) neglecting third and higher-order spatial derivatives of $\phi_m$ and $\theta_m$ in their dynamics and iii) assuming that gradients in the phase are zero within grains. 
This method has been shown to allow for efficient inhomogenous spatial discretization for numerical methods working in real space.

Recently the same issue has been addressed by exploiting a Cartesian representation of the amplitudes and allowing for local rotation of the basis vector $\Gv_m$ \cite{Bercic2018,Bercic2020}. This model considers a set of locally rotated amplitudes $\tilde \eta_m$ such as $\eta_m=\tilde \eta_m \expF{-\I \Delta\Gv_m(\Theta)\cdot \rv}$. A rotation field $\Theta$ is then computed such that $\eta_m$ have vanishing oscillation, i.e., satisfying the condition
\begin{equation}
    \Grad \tilde{\eta}_m=(\Grad \eta_m)\expF{-\I \Delta\Gv_m(\Theta)\cdot \rv}-\I \eta_m \Delta\Gv_m(\Theta) \expF{-\I \Delta\Gv_m(\Theta)\cdot \rv}=0,
    \end{equation}
    thus
    \begin{equation}
\Delta \Gv_m(\Theta)=\Gv_m(\Theta)-\Gv_m=\frac{\Grad \eta_m}{\I \eta_m}.
\end{equation}
The local rotation field may be explicitly extracted from amplitudes, e.g. exploiting the results reported in \cite{SalvalaglioNPJ2019}. Then, it may be shown \cite{Bercic2018,Bercic2020} that operators defined in the rotated system, $\mathcal{O}^\Theta$, applied to rotated fields, $f^\Theta$, transform as $\mathcal{O}^\Theta f^\Theta=\expF{-\I \Delta\Gv_m(\Theta)\cdot \rv} \mathcal{O} f$, as e.g. ${\partial \eta_m^\Theta}/{\partial  t}=\expF{-\I \Delta\Gv_m(\Theta)\cdot \rv}{\partial \eta_m}/{\partial t}$ or $\mathcal{G}_m^{\Theta}\eta_m^\Theta=\expF{-\I \Delta\Gv_m(\Theta)\cdot \rv}\mathcal{G}_m\eta_m$. The evolution for $\eta^\Theta$ is evaluated while computing $\Gv_m(\Theta)$ everywhere. This approach still requires a proper numerical implementation \cite{Bercic2018}, but has been proved successful in describing crystal structures through the ``rotated" amplitudes avoiding beats due to crystal rotation, exploiting efficient mesh refinement (see Sec.~\ref{sec:mesh-adaptivity}), and matching the dynamics obtained by the original amplitude expansion. Importantly, this approach has also been combined with an algorithm selecting the closest reference crystal for a given local orientation \cite{Bercic2020} which avoids the presence of unphysical grain boundaries, at least in two dimensions for triangular lattices.

\subsection{Elastic relaxation and mechanical equilibrium}
\label{sec:mechanical-equilibrium}

The dynamics of the PFC model and, in turn, its
amplitude expansion approximation, was initially assumed to be overdamped, i.e.~driven by minimization of the corresponding free-energy functional through a gradient flow \cite{Elder2002,Elder2004}. Although this setting can be justified in some circumstances, it constrains the dynamic to diffusive timescales. This may lead to some issues for the description of elastic relaxation, which usually occurs on faster timescales with respect to the diffusive dynamics of the density field. A few investigations addressed these issues, delivering either a framework able to ensure mechanical equilibrium at every time, describing the limit of instantaneous elastic relaxation \cite{HeinonenPRE2014,SkaugenPRL2018,SalvalaglioJMPS2020}, or modeling explicitly elastic excitations \cite{HeinonenPRL2016}. 

In the work of Heinonen \etal~\cite{HeinonenPRE2014,HeinonenThesis}, the amplitudes are expressed similarly to Eq.~\eqref{eq:smalld}, assuming small displacements in $\uv$. Then a formal separation of the timescales of the field $\phi_m$ from the field $\theta_m$, is considered. To ensure mechanical equilibrium, i.e. $\Div \boldsymbol{\sigma}=0$, it is then 
demonstrated to be equivalent to solving 
\begin{equation}
\sum_m^M \Gv_m \frac{d\theta_m}{dt}=-\sum_m^M \Gv_m {\rm Im}\left(\frac{1}{\eta_m} \frac{\delta F_\eta}{\delta \eta_m^*} \right)=-\frac{1}{2}\sum_m^M\Gv_m \frac{\delta F_\eta}{\delta \theta_m}=0,
\label{eq:ime}
\end{equation}
at every step after solving for $\partial \eta_m / \partial t$. In \cite{HeinonenPRE2014}, a factor $\phi_m^{-2}$ appears in the second-last term in \eqref{eq:ime}. However, as discussed in \cite{HeinonenThesis}, this expression allows for a more formal connection to the displacement $\mathbf{u}$. Moreover, equilibrating Eq.~\eqref{eq:ime} would corresponds to a real energy minimization problem.

A different approach, which computes the mechanical equilibrium deformation from the incompatible one, fully accounting for the singular distortion of defects as conveyed by $n$ and/or $\eta_m$ has been proposed in Ref.~\cite{SkaugenPRL2018} for PFC and then translated to APFC in Ref.~\cite{SalvalaglioJMPS2020}. Therein, the smooth distortion $u_i^\delta$ required to fulfill mechanical equilibrium is determined, and then the amplitudes are corrected as $\eta_m^{\rm m.e.}=\eta_m \expF{-i\Gv_m \cdot \uv^\delta}$. In brief, the smooth stress, ${\sigma}^{\delta}_{ij}$, to be added to the stress field computed from the amplitudes, $\sigma^\eta_{ij}$ (see also Sec.~\ref{sec:stress-field}), to satisfy mechanical equilibrium is obtained through the Airy Function ($\chi$) formalism:
\begin{equation}
\begin{split}
\sigma^{\delta}_{ij}=&\sigma^{\rm m.e.}_{ij}-\sigma^\eta_{ij}=\epsilon_{ik}\epsilon_{jl}\partial_{kl}\chi-\sigma_{ij}^\eta,\\
(1-\nu) \Grad^4 \chi =&2\mu \epsilon_{ij}\partial_i B_j(\mathbf{r})= (\epsilon_{ik}\epsilon_{jl} \partial_{ij}\sigma_{kl}^\eta-\nu \Lap \sigma_{kk}^\eta),
\end{split}
\label{eq:sdelta}
\end{equation}
where $ \mathbf{B}(\rv)$ the Burgers vector density, and $\nu$, $\uplambda$ and $\upmu$ as in Sec.~\ref{sec:comparison-continuum-elasticity}, while $\uv^\delta$ is then computed exploiting a Helmholtz decomposition into curl- and divergence-free parts, 
\begin{equation}
    u_i^\delta=\partial_i \varphi +\epsilon_{ij} \partial_j \alpha, \qquad \Lap \varphi = {\rm{Tr}}(\mathbf{U}^\delta), \qquad \Grad^4 \alpha = -2\epsilon_{ij}\partial_{ik} U_{jk}^\delta.
    \label{eq:udelta}
\end{equation}
Once $u_i^\delta$ is calculated, correction to the 
amplitudes can be imposed.
This approach has been shown to work well in two dimensions for isotropic materials, while its generalization to three dimensions is non-trivial due to the Airy function formalism. A more general method to correct $n$ by computing $\mathbf{u}^{\delta}$ in three dimensions has been recently proposed in Ref \cite{Skogvoll2021} for PFC, and it is expected to work for the APFC model.

In Ref.~\cite{HeinonenPRL2016}, a model accounting explicitly for elastic relaxation has been considered by coupling the mesoscale description of the microscopic structure of the materials achieved by amplitudes to a hydrodynamic velocity field. It recovers the instantaneous relaxation as a limit of the model. It consists of describing the crystal lattice through $\eta_m$ and a slowly varying density field, $\no$, via the energy \eqref{eq:F_density_amplitude}. The evolution laws are then derived accounting for mass density and momentum density conservation and read
\begin{equation}
\begin{split}
    \no\frac{ D\vv }{ Dt}=&-\no \Grad \frac{\delta \mathcal{F}}{\delta \no}-\sum_m^M\left[\eta_m^* \mathcal{Q}_m \frac{\delta \mathcal{F}}{\delta \eta_m^*} + \cc \right]+\mu_{\rm S}\Lap \vv + (\mu_{\rm B}-\mu_{\rm S})\Grad(\Div \vv),\\
    \frac{\partial \no}{\partial t}=&-\Div (\no \vv) + \mu_{n} \Lap \frac{\delta \mathcal{F}}{\delta \no} + \frac{1}{2} \mu_{n} \Lap (|\vv|^2),\\
    \frac{\partial\eta_m}{\partial t}=&-\mathcal{Q}_m\cdot(\eta_m \vv)-\mu_\eta |\Gv_m|^2\frac{\delta \mathcal{F}}{\delta \eta_m^*},
\end{split}
\label{eq:hydro2}
\end{equation}
with $\vv$ the velocity field, $D\vv / Dt=\partial \vv/\partial t +\vv \cdot \Grad \vv$, $\mathcal{Q}_m=\Grad + \I \Gv_m$, and $\mu_\eta$, $\mu_n$, $\mu_{\rm B}$, $\mu_{\rm S}$ are parameters. Previous attempts to include fast time scales in the dynamics introduced an explicit second order time derivative in the equation of motion for the PFC mass density field \cite{StefanovicPRL2006,GalenkoPRE2009}. This approach gives rise to short wavelength oscillations accelerating relaxation processes, but fails to describe large scale vibrations \cite{MajaniemiPRB2007}.  The model described by Eq.~\eqref{eq:hydro2} gives the correct long wavelength 
elastic wave dispersion relationship $(\omega \sim k)$.

\begin{figure}
\includegraphics[width=\textwidth]{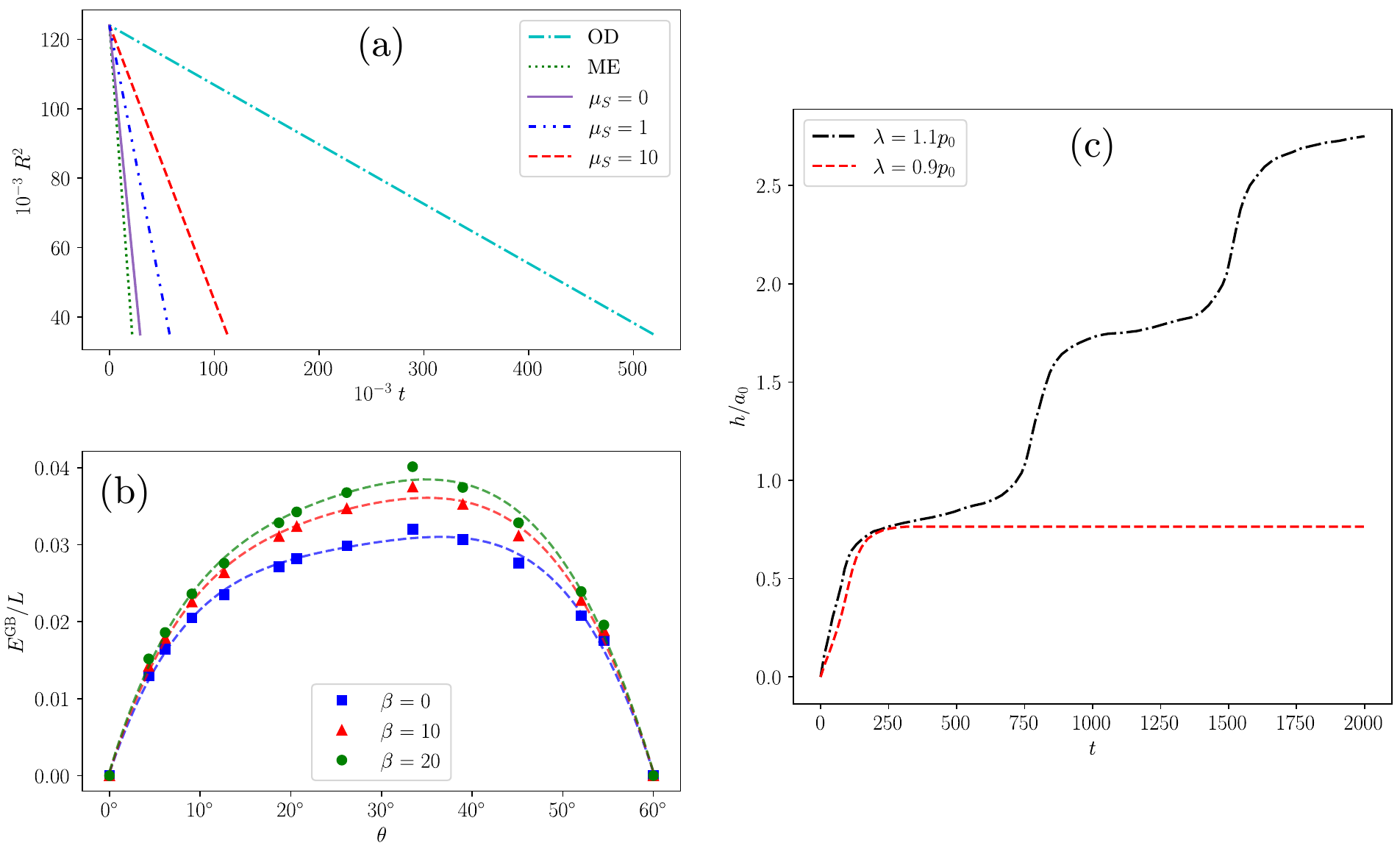}
\caption{Representative results for extensions of the APFC model. (a) Shrinkage of a circular small-angle grain boundary (2D, triangular lattice) in terms of its radius $R(t)$ with the model illustrated in Eq.~\eqref{eq:hydro2} (for different $\mu_S$), instantaneous mechanical equilibrium (ME) as from Eq.~\eqref{eq:ime}, and classical (overdamped) APFC dynamics (OD). Reconstructed from Ref.~\cite{HeinonenPRL2016}. (b) (Symmetric) grain-boundary energy per unit length $E^{\rm GB}/L$ (2D, triangular bicrystal) as a function of the tilt angle $\theta$ for different $\beta$ values in Eq.~\eqref{eq:coreenergy}. Reconstructed from Ref.~ \cite{Salvalaglio2017}. (c) Sample growth of a one dimensional front for 
two driving forces $\lambda$.  Reconstructed from Ref.~\cite{Huang2013}.}
\label{fig:figure7}
\end{figure}

A key test case for all the approaches reported in this section is the shrinkage of rotated grains (see Fig.~\ref{fig:figure7}). Their results consistently show a faster dynamic in the limit of instantaneous mechanical equilibrium \cite{HeinonenPRE2014,SkaugenPRB2018,SalvalaglioJMPS2020} while tuning of parameters in the model reported in Eq.~\eqref{eq:hydro2} allows for the investigation of intermediate regimes \cite{HeinonenPRL2016}.

\subsection{Control of interface and defect energy}

The original APFC (or PFC) model contains a small set of parameters which limits quantitative fitting to match experimental measures or theoretical calculations. In Ref.~\cite{Salvalaglio2017}, it has been shown that the addition of a single term to the free energy functional can be used to control the solid-liquid interface and defect energies in a well-controlled fashion, without affecting the crystal structure. Exploiting the information conveyed by $\Phi=2\sum_m^M |\eta_m|^2$, which is a measure of the crystalline order, and in analogy with the gradient term of order parameters in interfacial free energies \cite{Cahn1958}, an additional energy contribution can be phenomenologically introduced in Eq.~\eqref{eq:F_APFC}, reading 
\begin{equation}
F_{\beta}= \int_\Omega  \frac{\beta}{4}|\Grad\Phi|^2 \dd\rv,
\label{eq:coreenergy}
\end{equation}
where $\beta$ is a free parameter. This leads to an additional term to Eq.~\eqref{eq:ev_apfc} as
\begin{equation}
\frac{\delta F_{\beta}}{\delta \eta_m^*} = -\beta\eta_m \Lap \Phi.
\label{eq:evolcoreenergy}
\end{equation}
For small $\beta$, this additional contribution is found to change the interface and defect energy linearly with $\beta$, while deviations are observed for large values. Fig.~\ref{fig:figure7}(b) shows the tuning of symmetric tilt grain boundary energies by $\beta$ due to the local change in the defect-core energies \cite{Salvalaglio2017}. Notice that, due to the issues discussed in Sec.~\ref{sec:beats}, it is not possible to compute the whole range of $\theta$ only by increasing the relative angle (see also \cite{Hirvonen2016}). In this case, energy values for theta $\gtrless 30^\circ$ are obtained with two different simulation settings. The framework reported in \cite{Bercic2020} would allow addressing these calculations without considering such different settings.

It is worth mentioning that formulations allowing for tunable energies at 
defects and interfaces similar to the one discussed here can be devised 
from microscopic length scales exploiting smoothing kernels in 
Fourier space \cite{Kocher2015,Guo2016}.

\subsection{Lack of barriers}
\label{sec:barriers}

In the derivation of the amplitude equations it was implicitly assumed that 
the atomic- and meso-scales (interface widths, etc.) completely decouple. It 
appears that this approximation eliminates barriers for defect or grain 
boundary motion. Huang has shown that incorporating the first-order coupling of the atomic and mesoscales leads to interface 
pinning \cite{Huang2013}. Consider multiplying the equation of 
motion by $\expF{-\I \qv \cdot \rv}$ and integrating over a unit 
cell while keeping terms previously assumed to be zero.  This leads to 
additional terms in Eq.~\eqref{eq:ev_apfc}.
For instance, for a triangular lattice:
\begin{equation}\begin{split}
    \frac{\partial \eta_m}{\partial t}=&\mathcal{L}_m 
    \frac{\delta F_\eta}{\delta \eta_m^*}\approx
    -|\Gv_m|^2\bigg[A\mathcal{G}_m^2\eta_m+B\eta_m+ 3D(\Phi - 
    |\eta_m|^2))\eta_m+\frac{\partial f^{\rm s}}{\partial \eta_m^*}  \\ &
    +\frac{1}{A_{\rm u.c.}}\int_{\rm u.c.} 
    \dd \rv^{\,\prime} f_{p_1}\expF{-\I q_{\rm o} y^\prime} 
    + (\cdots) \bigg],
    \end{split}
    \label{eq:Cev_apfc}
\end{equation}
where $A_{\rm u.c.}$ is the area of a unit cell  and
\begin{equation}
f_{p_1}=3q_{\rm o}^2\left[(6\no+2C)\eta_1\eta_2^*+3v(\eta_1^2\eta_3+\eta_2^{*^2}\eta_3^*)
\right],
\end{equation}
with $(\cdots)$ implying six other similar terms that contain 
a $\expF{-\I\qv \cdot \rv'}$ term (see reference \cite{Huang2013} for 
details).  The last term(s) in Eq.~\eqref{eq:Cev_apfc} implicitly 
couple atomic ($\expF{-\I q_0 y^\prime}$) and slow scales ($\eta_m$) 
terms.  The equation for the average density becomes 
\begin{equation}\begin{split}
    \frac{\partial \no}{\partial t}=& 
    \Lap \frac{\delta F_\eta}{\delta \no}
    -\frac{1}{A_{\rm u.c.}}\int_{\rm u.c.} 
 \dd \rv^{\,\prime} 
        f^*_{p_1}\expF{-\I q_0(\sqrt{3}/2x^\prime+3/2y^\prime)}
    + (\cdots). 
    \end{split}
    \label{eq:Cev_no}
\end{equation}

To understand the consequences of this coupling, Huang 
derived an equation of motion for a liquid/solid 
front moving in the $y$ direction with slow variations 
in the $x$ direction using the projection operator method of 
Elder \etal \cite{Elder2001}.
In this method a coordinate transformation from $(x,y)$ 
to $(u,s)$ is made where $u$ is a coordinate normal to 
the interface position and $s$ is parallel.  
Equation (\ref{eq:Cev_apfc}) (in the limit $\mathcal{L}_m\approx -|{\bf G}_m|^2=-1$) 
is multiplied by $\partial \eta_m/\partial u$ and 
Eq.~\eqref{eq:Cev_no} by $\partial \no/\partial u$ and 
integrated over $u$ in the  inner region.  In the outer 
regime the Equations (\ref{eq:Cev_apfc}) and (\ref{eq:Cev_no}) 
are linearized around a liquid state and then solved using 
Green's functions.  The  inner and outer solutions are then 
matched such that the chemical potential is continuous across 
the interface.

One main result of these calculations is the equation 
for the interface normal velocity, $v_n$, given by 
\begin{equation}
c_0 v_n = \lambda -\gamma\kappa -p_0\sin(q_0 h +\phi),
\label{eq:cv}
\end{equation}
where $c_0$ is the kinetic coefficient, $\lambda \propto 
\Delta n_0^0 \delta \mu(0,s)$, $\Delta n_0^0$ is the difference 
in liquid/solid density, $\delta\mu(0,s)$ is the chemical 
potential difference from equilibrium along the interface, $\gamma$ is the surface tension, 
$\kappa$ is the curvature, $p_0$ is the pinning strength, 
$h$ is the distance from the front and $\phi$ is the phase.
Expressions for each of these terms is given in 
Huang \cite{Huang2013}.  This equation coupled with 
mass diffusion in the outer regions ($\eta_m$ at equilibrium 
liquid values) and the usually matching condition 
$v_n\Delta n_0^0=
\partial \delta \mu/\partial u|_{0^-}
-\partial \delta \mu/\partial u|_{0^+}$ constitutes a 
free boundary problem.

If gradients in $h$ are assumed to be small, Eq.~(\ref{eq:cv}) reduces to
\begin{equation}
c_0 \pp{h}{t}=\lambda+\gamma\pp{^2h}{x^2}+ \frac{\lambda}{2}
\left(\pp{h}{x} \right)^2 -p_0 \sin(q_0 h +\phi).
\label{eq:hkp}
\end{equation}
In the limit of non-conserved dynamics (fixed $\lambda$) this is 
a driven sine-Gordon equation introduced by Hwa \etal \cite{Hwa1991} 
to study, when thermal fluctuations are included, the interface 
roughening during crystal growth. Huang showed that the pinning 
term can lead to step by step growth of the interface as is 
observed in experiments and even completely arresting the growth 
if the driving force ($\lambda$) is too small, as illustrated in 
Fig.~\ref{fig:figure7}(c).  It is also shown that the pinning strength 
increases as temperature (controlled by $B=\DBpar$) or 
the elastic moduli (controlled by $A=\Bxpar$) are lowered as both have 
the effect of decreasing the width of the liquid/solid domain wall.
Later, Huang \cite{Huang2016} extended this work to a binary system 
with a eutectic phase diagram and derived more general expressions 
for the surface energy and barrier strength as a function of 
concentration, temperature, and crystallographic 
orientation of the liquid/solid front.

\section{Applications}
\label{sec:applications}

\subsection{Solid-liquid interfaces and the phase field limit}
\label{sec:solid-liquid-interfaces}

Solid-liquid interfaces are regions where $n$ may vary over length scales larger than the atomic spacing. Therefore, the APFC model may be exploited to focus on these regions while neglecting the fine details at the atomic scale elsewhere \cite{kundin2014bridging}. Real amplitudes have been first considered to address the modeling of solid-liquid interfaces in the seminal works by Khachaturyan \cite{Khachaturyan1983,Khachaturyan1996}. Therein, the order parameters resemble the ones entering classical phase-field approaches \cite{chen2002phase,boettinger2002phase,Steinbach2009,Provatas2010} and they may be linked to atomistic descriptions. They can be used, for instance, to account for bridging-scale descriptions of elasticity effects by means of additional contributions as, e.g., in the presence of precipitates, alloys, or point defects.\cite{Chen1991,Bugaev2002,Tewary2004,Varvenne2012,Varvenne2017}. However, this approach does not directly encode rotational invariance and elasticity associated with the deformations of the crystal lattice.

In Refs.~\cite{Galenko2015,Nizovtseva2017,AnkudinovPRE2020}, traveling waves characterized by the ansatz \eqref{eq:tanh} have been shown to describe the solid-liquid interfaces within PFC quite well near melting. Real amplitudes result in a classical phase-field model. Indeed, it is shown that a general form for the free energy can be obtained by considering real amplitudes,
\begin{equation}
    F_\phi=\int_\Omega \dd \rv \big[\texttt{a}\phi^2+\texttt{b}\phi^3+\texttt{c}\phi^4+\texttt{d}|\Grad \phi|^2\big],
\end{equation}
where the parameters $\texttt{a}$, $\texttt{b}$, $\texttt{c}$, $\texttt{d}$ depend on the lattice symmetry and the number of modes considered. Different crystalline cubic lattices, and their effect on growth dynamics are still retained \cite{AnkudinovPRE2020}. 
In addition, the framework is consistent with atomistic simulations and 
can be used for matching parameters to specific materials.

In Refs.~\cite{kundin2014bridging,Ofori-Opoku2018} similar underlying ideas led to a phase-field model connecting anisotropic surface energy and corresponding Wulff shapes to the lattice symmetry of various crystals through the choice of reciprocal lattice vectors. The model remarkably encodes a regularization term leading to corner rounding of faceted shapes similarly to diffuse interface theories \cite{Wheeler2006,Tor2009,Sal2015b}. Amplitudes are assumed to be real, but they are still considered separate variables. In the notation adopted in this review from Eq.~\eqref{eq:F_APFC}, and assuming zero average density, this gives 
\begin{equation}
    F_{\phi_m}=\int_\Omega \dd \rv \bigg[\sum_m^M\bigg( A[\nabla^2\phi_m]^2 + 4A[\mathbf{G}_m\cdot \nabla \phi_m]^2 -\frac{3D}{2}\phi_m^4\bigg) + \frac{B}{2}\Phi+\frac{3D}{4}\Phi^2 + f^{\rm s}(\{\phi_m\}) \bigg],
    \label{eq:anisoPF}
\end{equation}
with $\Phi=2\sum_m^M \phi_m^2$ and $f^{\rm s}(\{\phi_m\})$ the polynomial as in Sec.~\ref{sec:formulas} but as function of the real amplitudes only. Eq.~\eqref{eq:anisoPF} is similar to Ginzburg-Landau free energies entering multi-order-parameter phase-field models. The higher-order gradient contribution $[\nabla^2\phi_m]^2$ enforces the rounding of corners appearing among facets. A coefficient may be also introduced to tune its influence \cite{Ofori-Opoku2018}.

\subsection{Grain growth with dislocation networks and small-angle grain boundaries}
\label{sec:small-angle-GBs-3D}

\begin{figure}
\includegraphics[width=\textwidth]{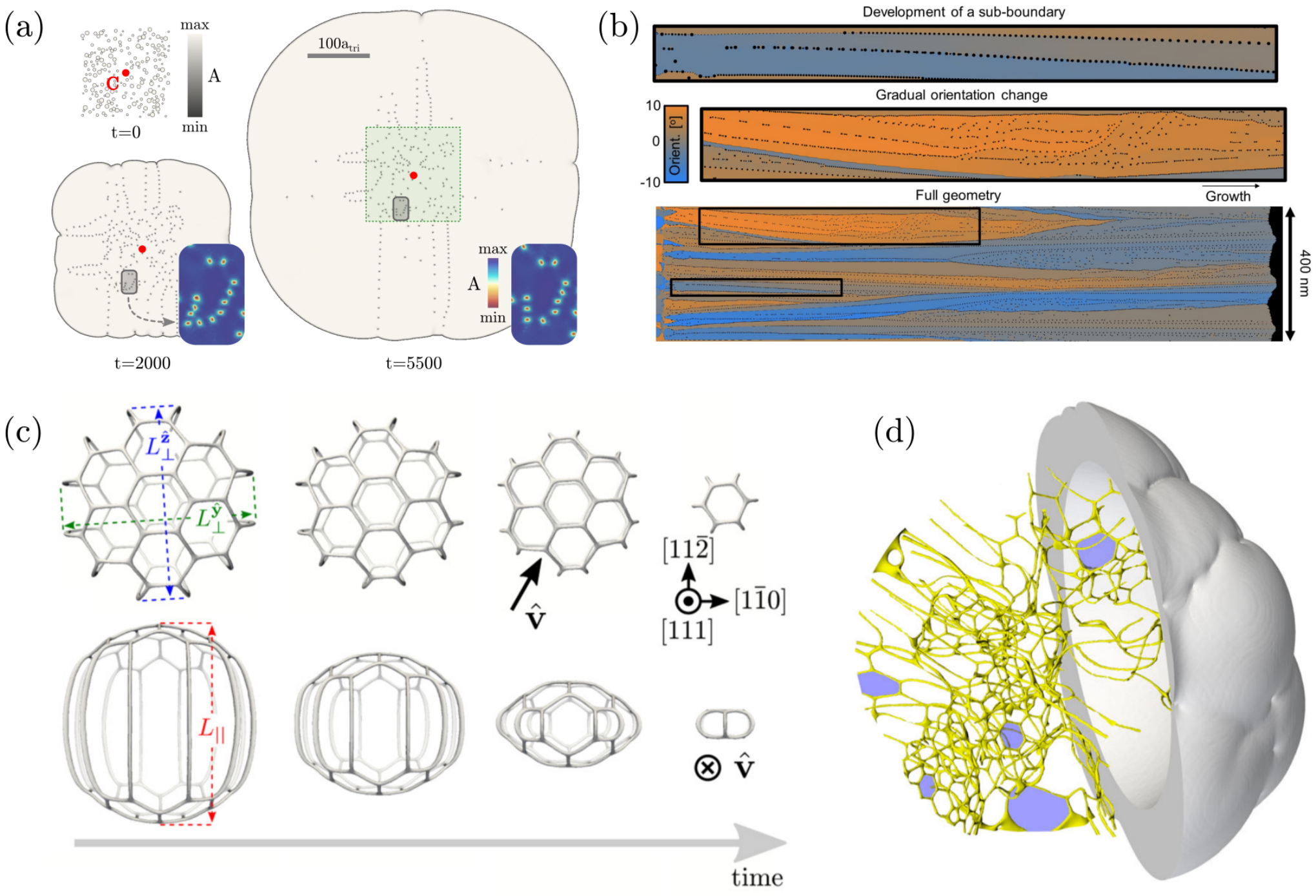}
\caption{Examples of crystal growth and defects networks as obtained by APFC simulations. (a) Growth of $200$ seeds with orientations ranging in $(-15^\circ,15^\circ)$, forming straight sub-boundaries at later stages in a growing polycrystal. Reprinted from \cite{Praetorius2019} $\copyright$ IOP Publishing Ltd. All
rights reserved. (b) Sub-boundaries and orientational gradients in thin aluminium films by APFC. Reprinted  from \cite{Jreidini2021}, under a creative commons attribution (CC BY) license. (c) Evolution of the defect network forming between an FCC crystal and spherical inclusion with the same structure tilted by $5^\circ$ about the [111] direction. Views aligned (top) and perpendicular (bottom) to the rotation axis are shown (see also the orientation of $\hat{\mathbf{v}}$). The network shrinks anisotropically with $\dot{L}_{||}>\dot{L}^{\hat{\mathbf{y}}}\sim\dot{L}^{\hat{\mathbf{x}}}$. Reprinted with permission from \cite{Salvalaglio2018} $\copyright$ (2018) by the
American Physical Society. (d) Network forming after the growth and impingement of thirty crystals with random tilt $\theta \in (-10^\circ,10^\circ)$ about the $[111]$ direction. Defects (yellow network) are shown within a spherical region at the center of the growing polycrystal. Adapted from \cite{SalvalaglioNPJ2019}, under a creative commons attribution (CC BY) license.}
\label{fig:figure8}
\end{figure}

The PFC model has been exploited to investigate rather small systems due to the atomic-scale resolution. According to the features described in Sec.~\ref{sec:continuumlimit} and \ref{sec:limits_extensions}, the APFC is especially suited to describe systems with small deformation and rotation while including isolated defects such as dislocations. Examples include small-angle GBs in graphene structures \cite{Hirvonen2016}, GBs premelting and shearing in BCC iron \cite{Adland2013}, and the  dynamics of small-angle GBs in general \cite{Huter2017}. In two dimensions, it is possible to examine 
systems on the micrometer scale \cite{Praetorius2019,Smirman2017} (see, e.g., Fig.~\ref{fig:figure8}(a)). A recent, remarkable application at this length scale is the simulation of sub-boundaries formation due to orientational gradients in thin aluminium films \cite{Jreidini2021,Pinomaa2022} (Fig.~\ref{fig:figure8}(b)).

The limitation in size for PFC becomes even more evident in three dimensions, requiring advanced numerical methods to simulate rather small systems \cite{PraetoriusThesis2015,mot3}. The APFC model has been proved powerful in addressing the study of defects in crystalline systems in three dimensions \cite{SalvalaglioNPJ2019,Salvalaglio2018,Praetorius2019}. In particular, small-angle grain boundaries can be well captured and also characterized thanks to the advanced description of elasticity as described in Sec.~\ref{sec:continuumlimit}. Representative cases are the shrinkage of dislocation networks forming at the boundaries between rotated inclusions and unrotated surrounding matrix (see Fig.~\ref{fig:figure8}(c)), also in combination with additional effects (see also Sec.~\ref{sec:binary-systems}), and the growth of slightly misoriented crystal seeds (see Fig.~\ref{fig:figure8}(d)). Interestingly, the shrinkage or rotated inclusions and the resulting dislocation networks have been proposed directly using a classical PFC approach \cite{mot3}. This investigation delivered very similar results to the ones obtained by APFC, as reported for instance in Fig.~\ref{fig:figure8}(c), thus assessing the coarse-graining achieved by the APFC model in an applied case.

The shrinkage of grains is generally associated with their rotation. A fingerprint of this process emerges in APFC, as shown in Ref.~\cite{SalvalaglioNPJ2019} where rotations are tracked thanks to Eq.~\eqref{eq:strain2D}. Therein it is shown that when defects at the boundary of a grain get closer, their deformation fields superpose, increasing the effective orientation of the grain.

\subsection{Binary systems}
\label{sec:binary-systems}
Coarse-grained approaches are often required in multiphase systems and alloys to handle simultaneously the deformation induced in the lattice, the resulting phase separations leading to \textit{Cottrell atmospheres} \cite{Cottrell1949,Cottrell1949_2,CottrellBOOK}, and effects on dislocation motion. The APFC model has been proved powerful in describing these effects at the mesoscale for binary systems, beyond results achieved by focusing on either atomistic or continuum length scales \cite{Zhang_2008,Sills2016,Gu2020,MISHIN2019,Koju2020DirectAM,DarvishiKamachali2020}. Also, it can be used to study these systems comprehensively, without focusing on concentration profiles, stress distribution around dislocations, and the force-velocity curves for defect motion separately. 

The original binary PFC model \cite{elder2007} is formulated in terms of the dimensionless atomic number density variation field and a solute concentration field $\psi$. In the APFC model, the expansion Eq.~\eqref{eq:nnn} is considered and a Vegard's law for the lattice spacing $R=R_0(1+\alpha\psi)$ is assumed with $\alpha$ the solute expansion
coefficient. This results in an energy \cite{ElderPRE2010,SalvalaglioPRL2021}
\begin{equation}
\begin{split}
F_{\alpha\psi}  = F_\eta+\int_\Omega \bigg[&
(\texttt{w}+\texttt{Y}\Phi)\frac{\psi^2}{2}+\frac{\texttt{u}}{4}\psi^4+\frac{\texttt{K}}{2}|\nabla \psi|^2\\ &-2A\alpha\sum_m^M|\mathbf{G}_m|^2\left(\eta_m {\cal G}_m^* \eta_m^*+{\rm
\cc}\right)\psi
\bigg]d\rv,
\label{eq:energy_binary}
\end{split}
\end{equation}
with definitions as in previous sections and $\texttt{w}$, $\texttt{u}$, $\texttt{Y}$, $\texttt{K}$, are additional model 
parameters as described in Ref.~\cite{elder2007}. Dynamics in terms of $\partial{\eta_m}/\partial t$ is then described by Eq.~\eqref{eq:detadt} with energy \eqref{eq:energy_binary} and $\partial{\psi}/\partial t=\nabla^2 \delta F_{\alpha\psi}/\delta \psi$, similarly to \eqref{eq:dt_eta_no}. It can be shown that, given
$\mathbf{G}_m$ the basic wave vectors corresponding to a pure system, 
the equilibrium wave vectors for binary systems read 
$\mathbf{G}_m^{\,\rm eq}=\mathbf{G}_m\sqrt{1-2\alpha\psi}$ \cite{Huang2010}. 

This approach allows the study of solute segregation and migration at grain boundaries, eutectic solidification, and quantum dot formation on nanomembranes 
\cite{Spatschek2010,ElderPRE2010,Elder_2010PFCmembranes,SalvalaglioPRL2021}. A similar approach has been exploited to accurately describe the interactions among grain boundaries and precipitates in two-phase solids \cite{XuPRB2016,Geslin2015}.

By applying the framework illustrated in Sec.~\ref{sec:plasticity} to this model, the velocity of dislocations including effects of the solute segregation has been also derived. By retaining only one mode of
the lowest order (with $|\Gv_m|=1$) and using the expression for $\partial{\eta_m}/\partial t$ for binary systems into Eqs.~\eqref{eq:JJJ}--\eqref{eq:vel_amp} one gets
\begin{equation}
v_i^d = \frac{8 \beta A b_j^d}{|\bv^d|^2} 
\epsilon_{ik} \sum_m^M |\Gv_m|^2 G_j^m G_k^m \left(G_l^m G_p^m U_{lp} - |\Gv_m|^2 \alpha \delta \psi \right).
\label{eq:v_binary}
\end{equation}
Eq.~\eqref{eq:v_binary} is consistent with the classical Peach-Koehler force similarly to Eq.~\eqref{eq:vj2}. 
For the case of a 2D triangular lattice or a 3D BCC crystal, the velocity takes the form
\begin{equation}
v_i^d = M \epsilon_{ij}\left( \sigma_{jk} b_k^d 
- 2A\phi_0^2\alpha \delta \psi b_k^d \sum_m^M G_k^m G_j^m
\right),
\label{eq:vj3}
\end{equation}
with a mobility $M= 2\beta/(\phi_0^2|\mathbf{b}^d|^2)$.
The last term in Eqs.~\eqref{eq:v_binary}-\eqref{eq:vj3} accounts for the
contribution from the compositionally generated stress, as a result of
the compositional strain ($\sim \alpha \psi$) arising from local concentration
variations, i.e. from solute preferential segregation (Cottrell atmospheres)
around defects. 
The stress field may be written as
\begin{equation}
\sigma_{ij} = 8 A U_{kl} \sum_m^M \phi_m^2 G_i^m G_j^m G_k^m G_l^m + \frac{\partial f_{\alpha\psi}}{\partial U_{ij}}.
\label{eq:sigma_jk}
\end{equation}
with 
\begin{equation}
f_{\alpha \psi}=-2A\alpha\sum_m |\mathbf{G}_m|^2
\left(\eta_m {\cal G}_m^*\eta_m^*+{\cc}\right) \psi \approx 8 A\alpha\psi \sum_{m}^M \phi_{m}^2 |\mathbf{G}_{m}|^2  G_i^m G_j^m \partial_j u_i,
\end{equation}
neglecting higher order terms in the last approximation obtained with $\eta_m = \phi_m \expF{-\I\mathbf{G}_m\cdot\uv}$ \cite{SalvalaglioPRL2021}.

Results predicted by these equations are the deflection of dislocation glide paths, the variation of climb speed and direction, and the change or prevention of defect annihilation \cite{SalvalaglioPRL2021}. Simulations exploiting the FEM approach outlined in Sec.~\ref{sec:fem} also enable the advanced description of these effects in three dimensions, in particular for small-angle grain boundaries \cite{SalvalaglioPRL2021}.

\subsection{Multi-phase systems}
\label{sec:multi-phase-systems}

\begin{figure}
\includegraphics[width=\textwidth]{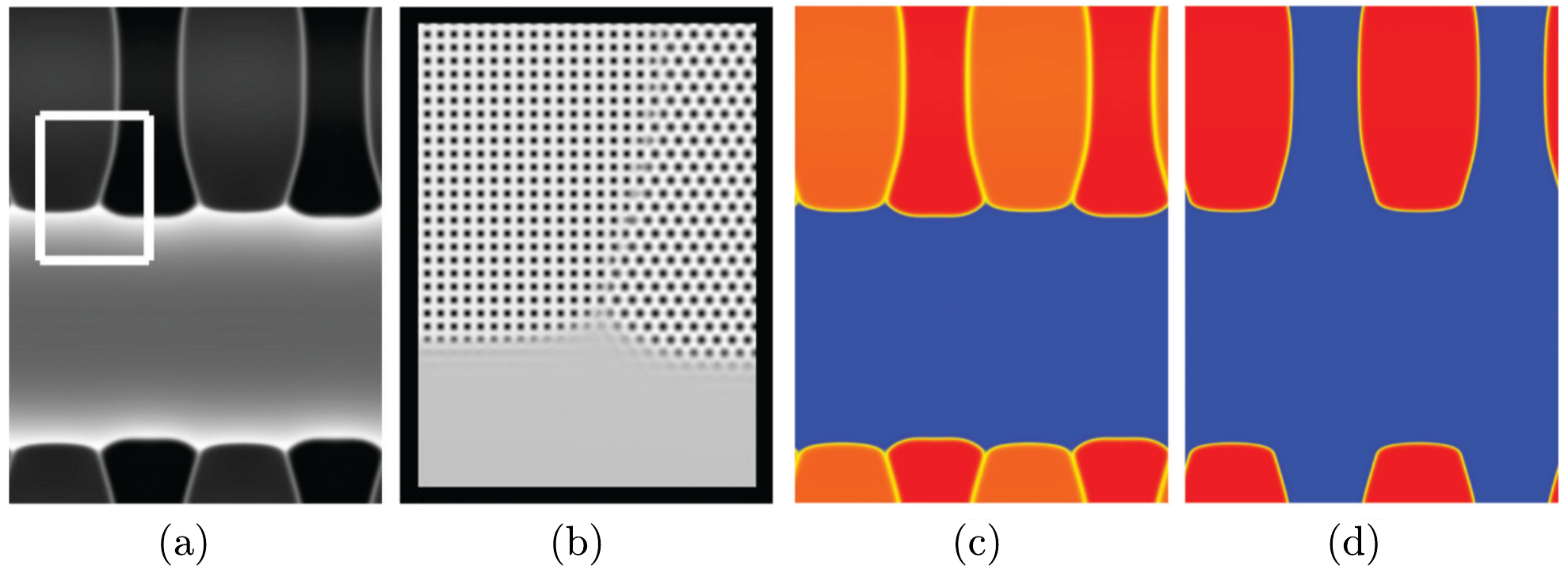}
\caption{Example of peritectic solidification. (a) Average density $\no$ (white to black greyscale). (b) Reconstructed $n$ (white to black greyscale). (c) Magnitude of an amplitude $\eta$, which is nonzero in both solid structures; areas of larger magnitudes are depicted in red and zero magnitudes are blue. (d) Magnitude of amplitude $ \chi$, which is only nonzero in the square phase. Color scheme is the same as in (c). Reprinted with permission from \cite{Ofori-Opoku2013} $\copyright$ (2013) by the
American Physical Society.}
\label{fig:figure9}
\end{figure}

Most of the APFC literature focuses on systems with a single solid phase. In a seminal work by Kubstrup \etal \cite{Kubstrup1996}, studying pinning effects between different phases, namely crystalline systems having triangular/hexagonal and square lattices, a construction has been proposed handling variable phases through a single density expansion. Extending this idea, in Ref.~\cite{Ofori-Opoku2013} an ansatz for the atomic density has been proposed to include more symmetries at once
\begin{equation}
n=\no+\sum_{j}^J\eta_{j}\expF{\I \mathbf{G}_{j}\cdot \rv}+\sum_{m}^M\chi_{m}\expF{\I \mathbf{Q}_{m}\cdot \rv} + \cc
\end{equation}
with $\{\eta_{j}$\} and $\{\chi_{m}$\} representing different set of amplitudes associated to reciprocal lattice vectors $\mathbf{G}_{j}$ and $\mathbf{Q}_{m}$, respectively. These two sets were chosen to account for the first and second modes necessary for reproducing triangular and square symmetry together, namely corresponding to $J=6$ and $M=6$ amplitudes. However, they can be arranged differently among the two sums, and, importantly, a  reduced set of amplitudes can be exploited (see specific choices of $\mathbf{G}_{j}$ and $\mathbf{Q}_{m}$ in Ref.~\cite{Ofori-Opoku2013}). Amplitude equations would simply follow from the general equations reported in Sec.~\ref{sec:formulas}. Simulations performed with this approach, combined with the formulation illustrated in Sec.~\ref{sec:different-formulations} for the excess term, showed the ability to study solidification, coarsening, peritectic growth, and the emergence of the second square phase from grain boundaries and triple junctions in a triangular polycrystalline system. See an example in Fig.~\ref{fig:figure9}. So far, this has been shown only for the lattice symmetry mentioned above in two dimensions. The same applies to extensions of the APFC to account for additional degrees of complexity in the crystal structure, such as for the amplitude expansion of the so-called anisotropic PFC model \cite{kundin2014bridging,Kundin_2017}.

\subsection{Heteroepitaxial growth} 
\label{sec:strained-systems}

An ideal application of the APFC model is heteroepitaxial growth, 
where a substrate provides a single crystallographic basis 
for layers growing on top. In such processes, the 
growing film typically has similar crystal symmetry and
lattice constant. The amplitudes vary on
long length scales for these systems, so a relatively large 
computational grid spacing can be used. 
In this context, the large angle issue discussed in 
Sec.~\ref{sec:beats} is not present.  Therefore, this would be an ideal application 
for using an adaptive mesh since the amplitudes in many
cases vary on very large length scales. To the authors'
knowledge this has not been done to date. Nevertheless, 
even uniform lattices can be used to study relatively 
large systems.

An example application is a single or small number of mismatched layers grown on a substrate. 
The mismatch leads to interesting strain-induced Moir\'e patterns that 
have been observed in experimental systems \cite{Marino2011,Roos2011,Balog2010}.  
In these cases, it is possible to model the film as a 
single two-dimensional layer with amplitudes. To the authors' knowledge, 
the largest APFC simulation of such systems was on the 
study of Moir\'e in graphene films in which the large simulation size was 
$19.6$ $\mu$m $\times$ $34.0$ $\mu$m which corresponded to 
roughly twenty-five billion carbon atoms. Some sample 
works are reviewed in the next subsection. Similarly, the amplitude expansion can also effectively be used to study the growth of many layers in two and three dimensions, i.e. to 
examine the Asaro-Tiller-Grinfeld \cite{Asaro72,Grinfeld86,Srolovitz89} instability and 
the subsequent nucleation of dislocations. This aspect will be also illustrated in the following. This section shows the APFC model in an applied context, reproducing experimental results and outlining general properties of mismatched, multilayered systems.

\subsubsection{Ultrathin films: strain induced ordering}
\label{subsubsec:sio}
\begin{figure}
\includegraphics[width=\textwidth]{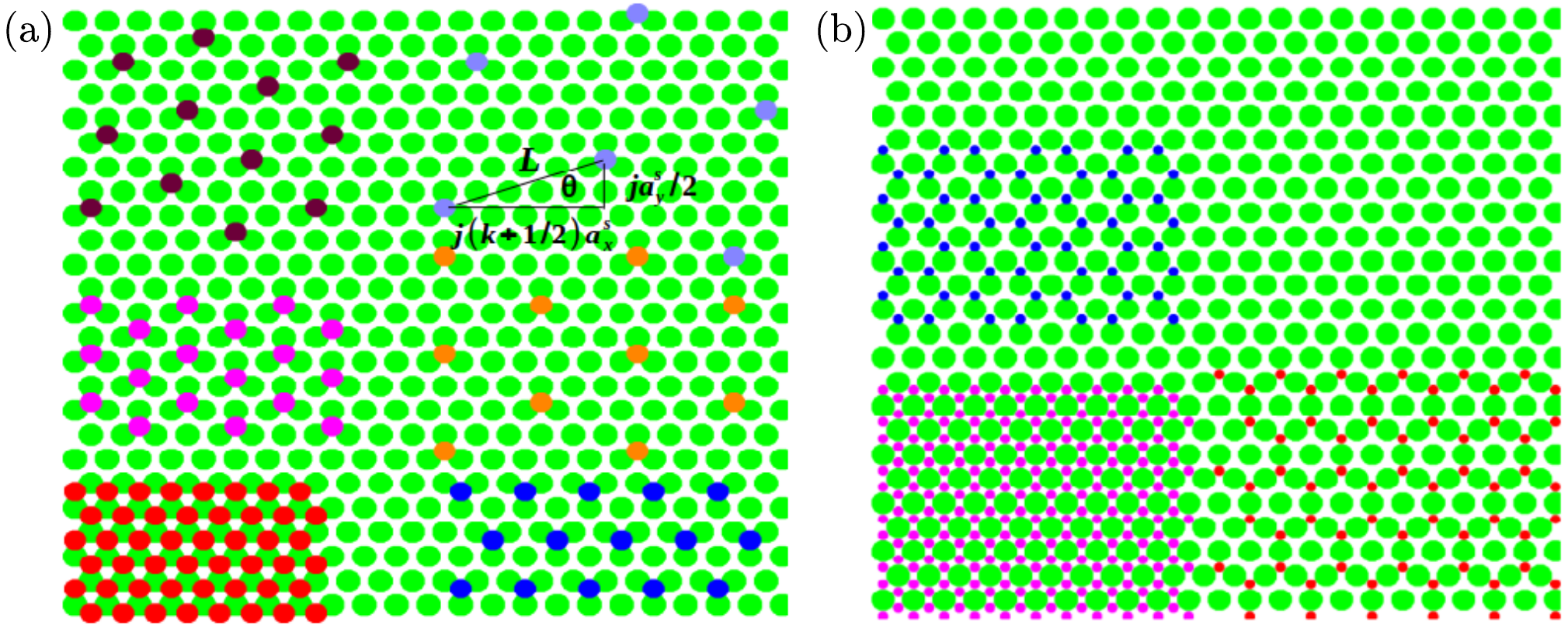}
\caption{Ordering of a triangular (honeycomb) 
lattice  on a substrate with a triangular array of potential 
maxima is depicted as green dots. In (a) the red, blue, pink, orange and purple 
dots correspond to $1\times 1$ (e.g., Cu/Ru(0001) or 
Cu/Pd(111)), $2\times 2$ (e.g., O/N(111)), 
$\sqrt{3}\times \sqrt{3}$ R30$^{\circ}$ (e.g., Xe/graphite), 
$2(\sqrt(3)\times\sqrt{3})$,   $(\sqrt{7}\times\sqrt{7})$ R19.1$^{\circ}$ 
(e.g., S/Pd(111)) and  $(\sqrt{7}\times\sqrt{7})$ R19.1$^{\circ}$ respectively.  In 
(b) the pink, red and blue atoms correspond to $1\times 1$ 
(e.g., graphene/Cu(111)), $2\times 2$ 
and $(\sqrt{3}\times\sqrt{3})$ R30$^{\circ}$.
}
\label{fig:figure10}
\end{figure}

When a monolayer (or several layers) of one material 
are grown on a substrate, the lattice mismatch can 
lead to interesting strain induced patterns \cite{Gunther95,Schmid97} and 
the APFC model is ideally suited to model such patterns 
\cite{elder2017striped,Smirman2017,Elder16,Elder16B,Elder13,Elder12}. 
Their nature depends on the misfit strain, $\vepm = 
( a^{\rm s}-a^{\rm f})/a^{\rm s}$, where $a^{\rm s}$ and $a^{\rm f}$ are the substrate and  film lattice constants, 
the relative crystal symmetry of the layer/substrate system and 
the film/substrate coupling strength. 
For example, when layers of Cu are grown on a Ru(0001) substrate, 
the substrate potential provides a triangular (honeycomb) array of potential 
maxima (minima) for the Cu atoms.  
Since the lattice constants of Cu and Ru(0001) are similar ($\vepm=5.5\%$),
a $1\times 1$ ordering occurs as depicted by the red dots in Fig.~\ref{fig:figure10}(a). 
For larger mismatches other orders can occur as shown in this 
Fig.~\ref{fig:figure10} for the ordering of triangular film on a 
triangular substrate (TT) in (a) and a honeycomb film on a triangular substrate (HT) in (b).  By symmetry a (TT) system is equivalent to a (HH) system 
and a (HT) system is equivalent to a (TH) system.  These patterns can 
be characterized by two integers $(k,j)$ or equivalently a 
length and angle ($L,\theta$) as depicted in Fig.~\ref{fig:figure10}(a).  
The relationship between them is $L=ja^s_x(\sqrt{(2k+1)^2+3})/2$ 
and $\tan\theta=\sqrt{3}/(2k+1)$.

\begin{figure}
\includegraphics[width=\textwidth]{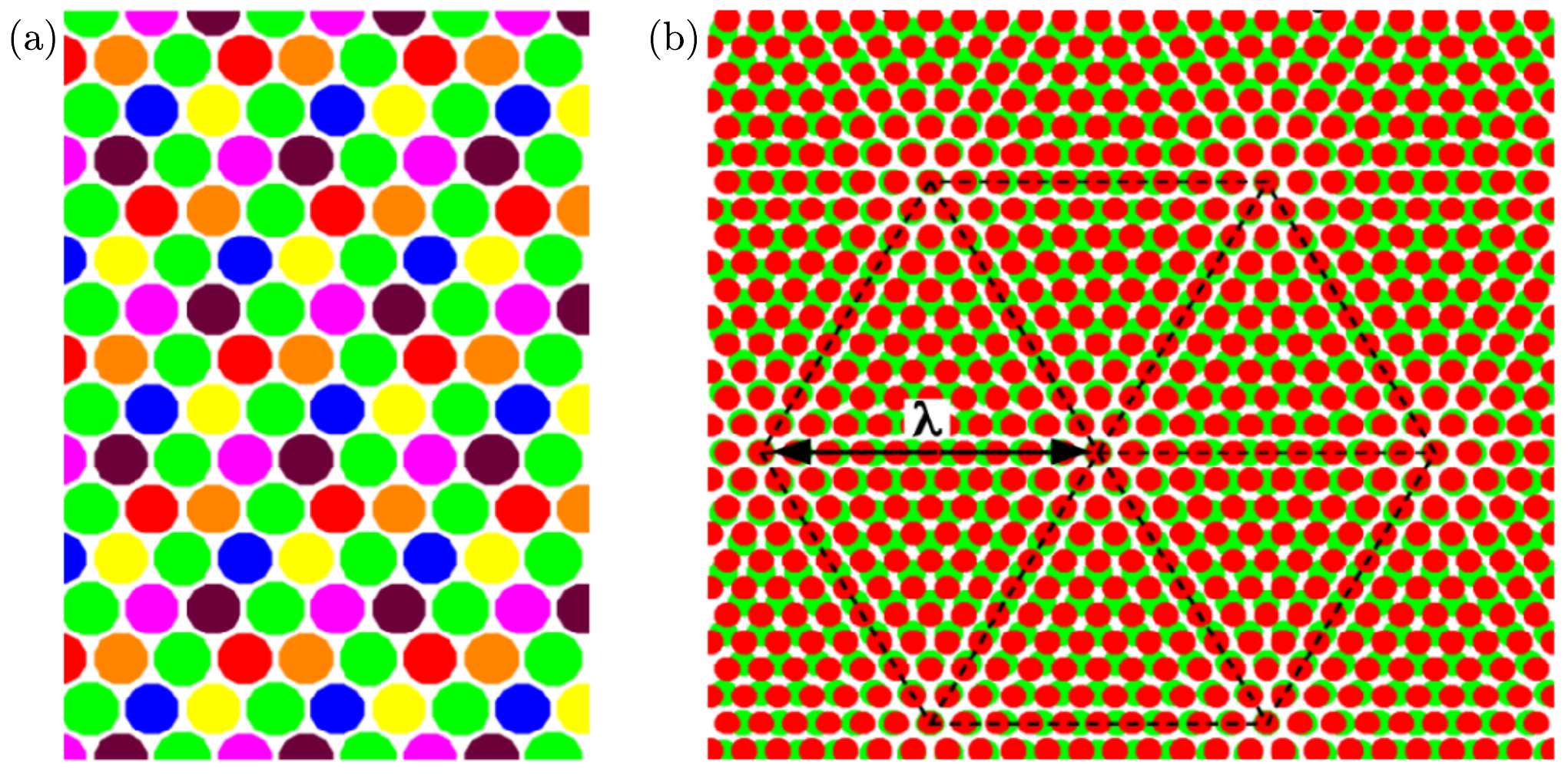}
\caption{(a) Illustration of the six equivalent degenerate sublattices 
for a TT $\sqrt{3}\times\sqrt{3}$ R30$^{\circ}$ system. The green dots 
are potential maxima due to the substrate and the other colored dots 
correspond to the sublattices. (b) Depiction of Moir\'e  pattern 
for a $1\times 1$ system in the limit $V_0=0$.}
\label{fig:figure11}
\end{figure}

In Fig.~\ref{fig:figure10}(a) the $1\times 1$ state could occupy two 
equivalent separate sublattices, while in (b) this state has only one 
sublattice. In general, the degeneracy ($N_{\rm S}=$ number of equivalent 
sublattices) is given by,
\begin{equation}
N_{\rm S}=\frac{j^2}{2}\left(\left(2k+1\right)^2+3\right),
\label{eq:degeneracy}
\end{equation}
for the TT system and half of Eq.~\eqref{eq:degeneracy} for the HT system.  
Figure~\ref{fig:figure11}(a) 
illustrates the different sublattices for a TT $\sqrt{3}\times\sqrt{3}$ R30$^{\circ}$ 
system.

\begin{figure}
\includegraphics[width=\textwidth]{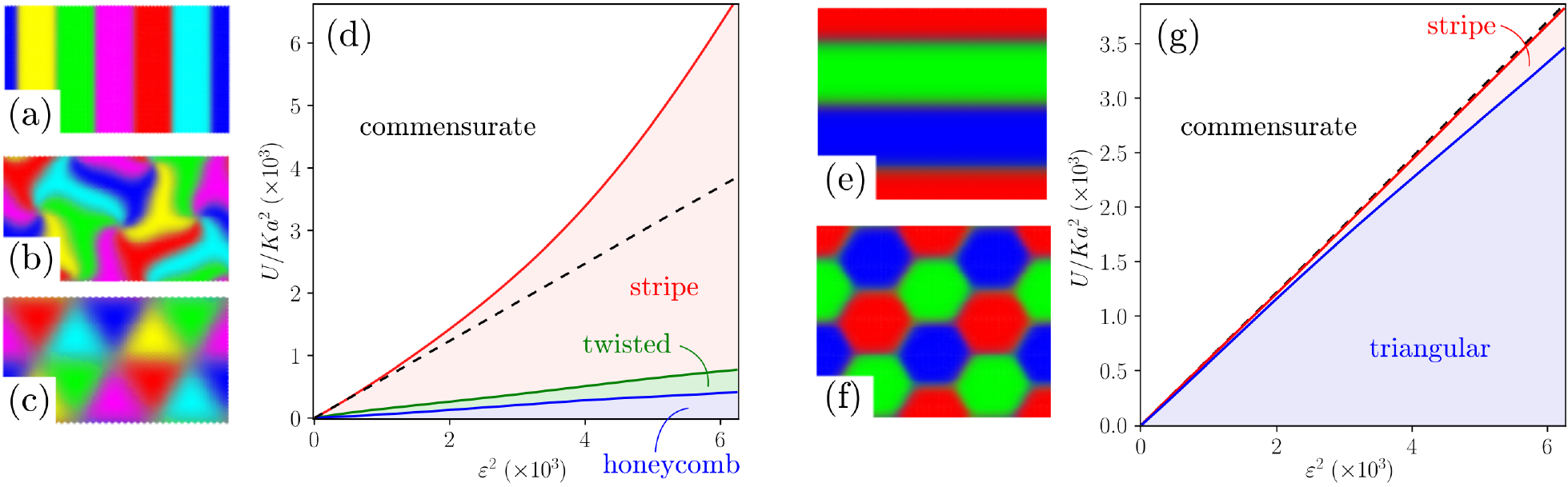}
\caption{Sample patterns and phase diagrams for 
$\sqrt{3}\times\sqrt{3}$ R30$^{\circ}$ system for TT (a)-(d) 
and HT (e)-(g) systems. For the TT system, 
the stripe, twisted honeycomb and honeycomb 
patterns are illustrated in (a), (b) and (c) respectively, 
and the phase diagram is shown in (d).  Stripe and triangular 
patterns for the HT system are shown in (e) and (f) respectively 
and (g) shows the HT phase diagram.
Each color in the patterns corresponds to a different sublattice. 
In (d) and (g) the dashed line is the analytic prediction 
for the stripe/commensurate transition given by Eq.~\eqref{eq:SGpred}.
The figures were reconstructed from 
\cite{elder2017striped}.}
\label{fig:figure12}
\end{figure}

The nature of the patterns that form depend on the degeneracy of sublattices, 
$N$, the mismatch strain, $\vepm$, and the strength of the 
coupling, $V_0$, between the film and substrate. In the limit $V_0=0$, 
a 2D Moir\'e pattern forms in terms of a honeycomb array of commensurate 
regions bounded by a triangular network of domain walls for the TT 
system, with length scale $\lambda = a^{\rm f}/\vepm$. This is illustrated in 
Fig.~\ref{fig:figure11}(b) for a $1\times 1$ system with a 
mismatch consistent with a Cu/Ru(0001).  As $V_0$ increases, the 
commensurate regions increase in size, and the domain walls and 
junctions decrease in size but increase in energy. For the TT 
system, the displacement across a junction is larger than the 
displacement across a domain wall. Thus for the TT system 
at a certain $V_0$ it becomes energetically favorable to eliminate 
the junctions and form stripes. At even larger values of $V_0$ 
the film becomes commensurate with the substrate.   A peculiar 
state in the TT arises for some values of $(V_0,\vepm)$ in between the 
stripe and honeycomb patterns in which the junction energy is lowered 
by twisting the domain walls and moving the junction to a lower 
energy location. Sample patterns for the 
TT system are shown in Figs.~\ref{fig:figure12}(a), (b) and (c).
In the case of the 
$1\times 1$ the junction energy is so high that it can create 
dislocation pairs and lead to zig-zag type patterns \cite{Elder12,Elder13}.

The HT system is considerably different since the domain wall energy 
is higher than the junction energy and of course the symmetry is 
different. At very low $V_0$, a triangular network of commensurate 
regions forms. At a $V_0$ much higher than in the TT case, a 
stripe phase emerges. At a slightly larger $V_0$, the commensurate 
state appears. There appears to be no equivalent twisted state 
in this system.  Sample stripe and triangular patterns are shown 
if Fig.~\ref{fig:figure12}(e) and (f).

To model these patterns within a PFC approach and corresponding APFC 
it useful to consider adding an additional coupling term, $F^{\rm c}$,  
to the free energy  functional given in Eq.~\eqref{eq:F_PFC} of 
the form,
\begin{equation}
F^{\rm c} = \int d{\bf r}\left[ V n^{j(k+1)} \right],
\end{equation}
where
\begin{equation}
V=V_0 \left(\sum_m^M \expF{\I{\bf  G}^{\rm s}_m\cdot{\bf r}} + {\rm c.c}
\right).
\end{equation}
$V_0$ is the coupling strength, 
the summation is over lowest order modes needed to reconstruct 
the symmetry of the substrate and ${\bf G}_m^{\rm s}$ corresponds to 
the reciprocal lattice vectors of the substrate (which will 
have a different magnitude that the film).  
The coupling factor $n^{j(k+1)}$ is 
needed since orders greater that $1\times 1$, a coupling $Vn$ 
would give no contribution in the amplitude expansion, since $V$ and 
$n$ would have different lattice spacings.  In principle, higher 
order harmonics of $V$ (or $n$) could be included, even though this would lead more computational expensive models.  In the amplitude expansion 
this term leads to a coupling term  $F^{\rm c}_\eta$, of the form
\begin{equation}
F^{\rm c}_\eta = V_0 D_{kj} \left(\left\{
\left[\left(\eta_1^*\right)^k\eta_2\right]^j + {\rm cyclic \ permutations}
\right\} + {\rm c.c.}
\right).
\end{equation}
where $D_{kj}=((k+1)j)!/((kj)!j!)$.
This term would be added to the free energy given in 
Eq.~\eqref{eq:energyterm_tri2} for a triangular two-dimensional system.
In addition, to account for the misfit strain, the operator 
${\cal G}_m$ that enters Eq.~\eqref{eq:F_APFC} becomes
\begin{equation}
{\cal G}_m \equiv \Lap + 2 \I {\bf G}_m \cdot {\bf} \Grad 
+1 - \alpha^2,
\end{equation}
where $\alpha=1-\vepm$.

Insight into the model can be obtained in the small deformation 
(${\bf u}$) limit,  $\eta_m = \phi \expF{-\I {\bf G}_m \cdot {\bf u}}$. The total free energy function reduces to a two dimensional 
Sine-Gordon model, i.e.,
\begin{equation}\begin{split}
F_{2d}^{\rm sg} = \int d {\bf r}\bigg[&
\frac{C_{11}}{2}\big(
(U_{xx}-\vepm)^2 
+ (U_{yy}-\vepm)^2 \big)
+2C_{44}U_{yy}^2
+C_{12}(U_{yy}-\vepm)(U_{xx}-\vepm) 
 \\ & 
+2V_0D_{kj}\phi^{(k+1)j}\sum_m^M \cos({\bf G}_m\cdot {\bf u})
\bigg],
\end{split}
\end{equation}
where $C_{11}=9A \phi^2$ and  $C_{44}=C_{12}=3A\phi^2$.  
Unfortunately this is difficult to solve for the boundary 
condition of a two dimensional triangular pattern.  
In one dimension this reduces to a Sine-Gordon model that can be solved exactly \cite{CL95}.  In 
this model the stripe to commensurate state transition occurs  when 
\begin{equation}
\frac{P}{Ka^2} = \frac{\pi^2}{16} \vepm^2,
\end{equation}
where $P$ is a measure of the potential between the film and 
substrate and $Ka^2$ is a measure of the elastic energy in 
the film.  These parameters are given by 
\begin{equation}
\frac{P}{D_{kj}\phi^{(k+1)j}V_0} = \left\{
\begin{array}{cc}
1/2 & {\rm TT} \\
4 & {\rm TH}
\end{array}
\right.,
\end{equation}
and 
\begin{equation}
\frac{K}{(C_{11}+C_{12})^2} = \left\{
\begin{array}{cc}
(C_{11}+C_{44}/3)^{-1} & {\rm TT} \\
C_{11}^{-1} & {\rm TH}
\end{array}
\right. .
\label{eq:SGpred}
\end{equation}
Details of 
these calculations can be found in Elder \etal \cite{elder2017striped}.

The full phase diagram as a function of $\vepm$ and the 
ratio of potential/elastic energy, $P/Ka^2$, can be obtained 
through numerical simulation. Sample phase diagrams 
are given for the  $\sqrt{3}\times\sqrt{3}$ R30$^{\circ}$ system 
for the TT and HT cases in 
Figs.~\ref{fig:figure12}(d) and (g)  respectively.  As can 
been seen in these figures for small $\vepm$, the 
analytic predictions (this is true for all $(k,j)$ systems) for the stripe/commensurate transition 
are quite accurate and very good for the HT case for 
all $\vepm$.  

\begin{figure}
\includegraphics[width=\textwidth]{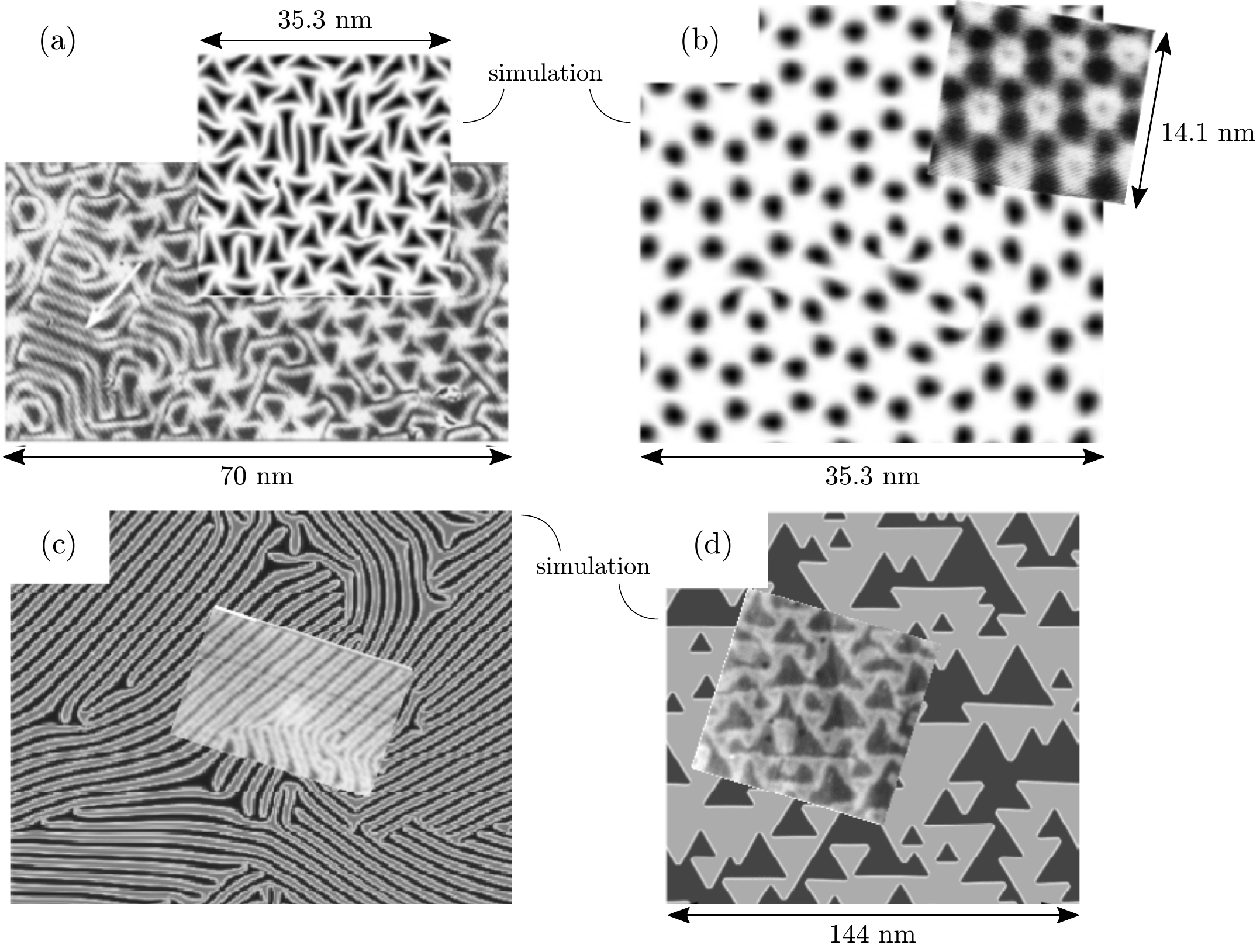}
\caption{Comparison of simulated and experimental 
patterns in Cu/Ru(0001) system. The figures correspond 
to twisted or zig-zag, honeycomb and stripe in (a), (b) and
(c) respectively.  The experimental results are 
from Gunther \etal \cite{Gunther95}.
Figure (d) compares the patterns in an 
experimentally partially filled layer with a simulation 
showing the ordering of a commensurate layer.  The 
experimental image is taken from Schmid \etal \cite{Schmid97}. Panels (d) is reprinted with permission from \cite{elder2017striped} $\copyright$ (2017) by the American Physical Society.} 
\label{fig:figure13}
\end{figure}

An interesting comparison with experiments is the Cu layers on 
a Ru(0001) substrate which is a $ 1 \times 1 $ TT system.  In 
this case, varying the number of Cu layers increases 
the film's elastic energy and the potential between 
the substrate and film. Essentially, adding more layers 
corresponds to reducing the ratio $P/Ka^2$.  One 
layer forms a completely commensurate state, two layers form 
a striped state, three layers form a twisted honeycomb 
(or zig-zag state), and four layers form a honeycomb state.  
To compare with the non-equilibrium patterns observed in 
experiments, simulations starting from random fluctuations were 
conducted.  The comparison of the experiments and 
simulations depicted in Fig.~\ref{fig:figure13}(a)-(c) shows 
a very good agreement for various patterns. In another 
experiment by Schmid \etal \cite{Schmid97} patterns of 
partially filled layers are reported. These patterns are remarkably similar to simulations of non-equilibrium patterns observed with the APFC model in the commensurate state as shown in Fig.~\ref{fig:figure13}(d).

Studies of the HT $1 \times 1$ lead to a phase diagram similar 
to that shown for the $\sqrt{3}\times \sqrt{3}$  in 
Fig.~\ref{fig:figure12}.  To compare with experiments, density functional theory (DFT) calculations were conducted by  Smirman \etal \cite{Smirman2017} to calculate the 
value of the dimensionless quantity $P/Ka^2$ for 
various $1 \times 1$ film/substrate systems.  The 
phase diagram accurately 
predicted commensurate state for twenty-five system 
mostly corresponding to films consisting of monolayers of InN or GaN on various substrates.
In addition, the phase diagram accurately predicted 
a commensurate state for graphene (G) on N, and triangular 
patterns for G on Cu, Pd, Pt, Al, Ag, and Au.  
Work was also conducted to predict the wavelength of the 
patterns as a function of misorientation with respect to 
the substrate in G/Cu(111) and G/Pt(111) 
systems. In the absence of coupling two dimensional 
patterns arise with wavelength 
$\lambda=a^{\rm f}/\sqrt{\vepm^2+2(1-\vepm)(1-\cos(\theta))}$, 
where $\theta$ is the misorientation angle. The study showed that 
as the coupling increases, the wavelength increases and 
interestingly the lowest energy states were not at 
zero degree misorientation ($0.88^{\circ}$ and $3.22^{\circ}$ for G/Cu(111) and G/Pt(111) respectively), which is unfortunately 
difficult to measure experimentally.
However, the predicted wavelengths were consistent 
with the experiments of Marino \etal \cite{Marino2011} 
for G/Cu(111).  

Other predictions of the APFC model involve the influence of defects and edges on pattern formation in the $\sqrt{3}\times\sqrt{3}$  R$30^{\circ}$ which corresponds to systems such as Xe/Pt(111) or Xe and Kr on graphite.

\subsubsection{Epitaxial growth: island formation and defect nucleation} 

\begin{figure}
\includegraphics[width=\textwidth]{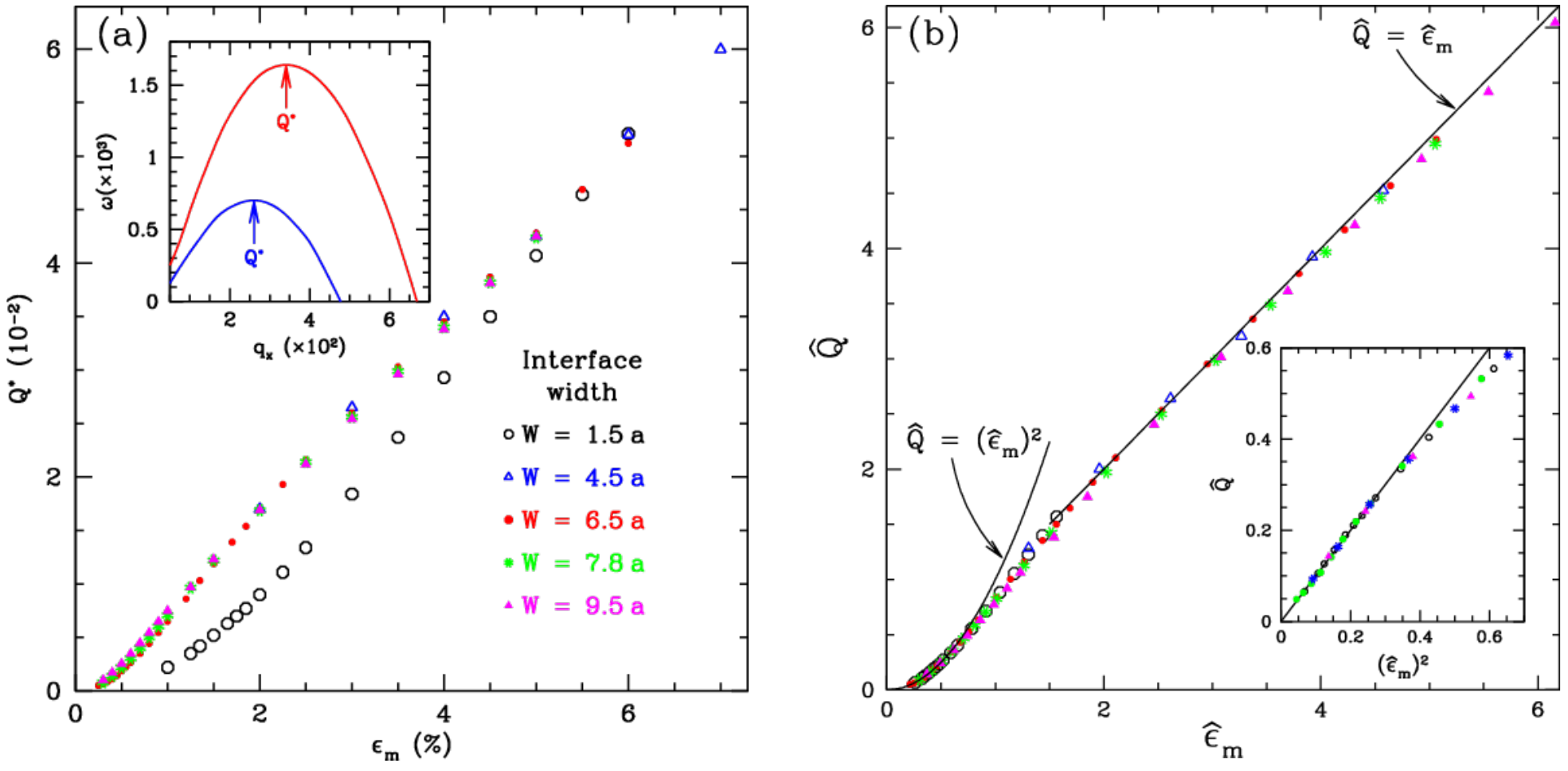}
\caption{(a) most unstable wavevector ($Q^*$) is shown as 
a function of misfit strain ($\vepm$) for various interface widths. 
In the inset dispersion relations are shown for $\vepm=4$ \% (red) 
and $3$ \% (blue).
(b) the $Q^*$ and $\vepm$ are rescaled to give rise 
to a universal curve as described in the text. In the inset $\hat{Q}$ is 
shown as a function of $\hat{\varepsilon}_{\rm m}^2$.  Details of the 
calculations can be found in reference \cite{HuangPRB2010}. 
Reconstructed from \cite{Huang2008,HuangPRB2010}.}
\label{fig:figure14}
\end{figure}

    When a material is grown epitaxially on a substrate with a 
mismatch strain, $\vepm$, the film will tend to buckle 
and form islands or mounds as it grows due to the so-called 
linear Asaro-Tiller-Grinfeld (ATG) instability \cite{Asaro72,Grinfeld86,Srolovitz89}. 
Recall that the APFC model is ideal for examining these phenomena, featuring relatively uniform amplitudes suited for adaptive meshing.
In addition, it is possible to reduce the  
study of an ATG instability in a 2D film to a 1D problem \cite{Huang2008,HuangPRB2010}. Consider expanding about the strained film such 
that $\eta'_m = \eta_m \expF{-\I\delta\qv_m\cdot \rv} $ where  
$\delta\qv_m$ is responsible for the mismatch strain imposed by the 
substrate.  For a triangular lattice with a strain imposed in the 
$x$ direction ($y$ being the growth direction) 
$\delta\qv_1 \cdot \rv=-\delta_x x - \delta_y y/2$, 
$\delta\qv_2 \cdot \rv=\delta_y y$, 
$\delta\qv_3 \cdot \rv=\delta_x x - \delta_y y/2$, 
$\delta_x = \sqrt{3}/2\vepm$ and $\delta_y$ is determined 
by lattice relaxation.  The strained amplitudes can now be expanded 
about a one dimensional profile, $\eta_j^0(y)$ as follows
\begin{equation}
\eta_j'(x,y,t) = \eta_j^0(y)+\sum_{q_x} \hat{\eta}_j(q_x,y,t)\expF{iq_x x},
\end{equation}
and similarly for the average density about $\no^0(y)$
\begin{equation}
\no(x,y,t) = \no^0(y)+\sum_{q_x} \hat{n}_{\rm o}(q_x,y,t)\expF{iq_x x}.
\end{equation}
The profiles $\eta_j^0(y)$ and $\no^0(y)$ must be determined 
numerically. The linearized equation of motion for the perturbed quantities 
$\hat{\eta}_j$ and $\hat{n}_{\rm o}$ are quite complex but are easily solved 
numerically to obtain a dispersion relation $(\omega(q_x))$ for the position of the liquid/solid front, i.e., 
the results can be fit to the form $|\hat{\eta}_j|$, 
$\hat{n}_{\rm o} \sim \expF{\omega t}$.  Dispersion relations 
are shown in the inset of Fig.~\ref{fig:figure14}(a).  Various 
analytic studies have lead to different forms of the dispersion relation depending on what physical mechanisms are included.  
Surface diffusion leads to $\omega\approx \alpha_3 q_x^3-\alpha_4 q_x^4$ \cite{Srolovitz89,Spencer91,Spencer93}, 
wetting to $\omega=-\alpha_2q_x^2+\alpha_3 q_x^3-\alpha_4q_x^4$ 
\cite{Levine07,Eisenberg00}, 
evaporation-condensation to $\omega=\alpha_1q_x-\alpha_2q_x^2$ 
\cite{Muller99,Kassner01}
and bulk diffusion to $\omega=\alpha_2q_x^2-\alpha_3q_x^3$ \cite{Wu2009}.  
In the APFC simulations, $\omega$ can be fit to a fourth order polynomial 
in $q_x$ however none of the fits are consistent with any of the 
prior results. This is due to the fact that the APFC model 
cannot separate each of the mechanisms individually.

From these studies the most unstable $q_x$, $Q^*$, can be extracted 
as a function of  misfit strain and interface width ($W$) 
as shown in Fig.~\ref{fig:figure14}(a).  
The width, in the notation of Eq.~\eqref{eq:tanh}, was altered through the variable $\Bxpar$
since $W \sim \sqrt{\Bxpar/|\DBpar|}$ \cite{Galenko2015}.  
For small values of $\vepm$ it was found that 
$Q^* \sim \vepm^2$ and for larger values 
$Q^* \sim \vepm$ for all interface widths. ATG theory gives 
$Q^*\approx (E/\gamma)\vepm^2$, where $E=\Bxpar \phi^2/2$ is Young's modulus, $\phi$ is the magnitude of the amplitudes in equilibrium, 
$\gamma$ is the surface energy which can be calculated numerically. 
The numerical results fit the small $\vepm$ to $Q^*=4E\vepm^2/3\gamma$.
The linear  behavior at large $\vepm$ can be understood by considering 
the wavelength at which the insertion of a dislocation would lead to 
perfect relaxation (i.e., the addition or 
subtraction of a lattice point every $\lambda$ returns 
the lattice constant of the film to its equilibrium value).  This occurs when $Q^*=2\pi/\lambda=q_x|\vepm|$.  It is interesting to note that 
this linear relationship was observed in experiments on 
SiGe/Si(001) growth \cite{Sutter2000,Tromp2000} although 
other explanations may exist as this is a binary system \cite{Bergamaschini2016}.

The continuum (ATG)  calculation fails  when the most unstable 
wavelength ($2\pi/Q^*$)  becomes  comparable with the interfacial 
thickness.  If one supposes that the crossover occurs at 
$\vepm^{\rm c}$ when $4E\vepm^{\rm c}/3\gamma=q_x\vepm^{\rm c}$ then 
$\vepm^{\rm c}=3\gamma q_x/4E$ and  $Q^{\rm c}=3\gamma q_x^2/4E$.  Defining 
the scaled  quantities $\hat{\varepsilon}_{\rm m}=\vepm/\vepm^{\rm c}$ $\hat{Q}=Q^*/Q^{\rm c}$ 
gives rise to the universal behavior shown in Fig.~14 (b).  
That is, the relationship between $\hat{\varepsilon}_{\rm m}$ and $\hat{Q}$ is 
independent of the interfacial thickness.  It was found numerically that $1/Q^{\rm c} \sim \Bxpar \sim W^2$.

\begin{figure}
\includegraphics[width=\textwidth]{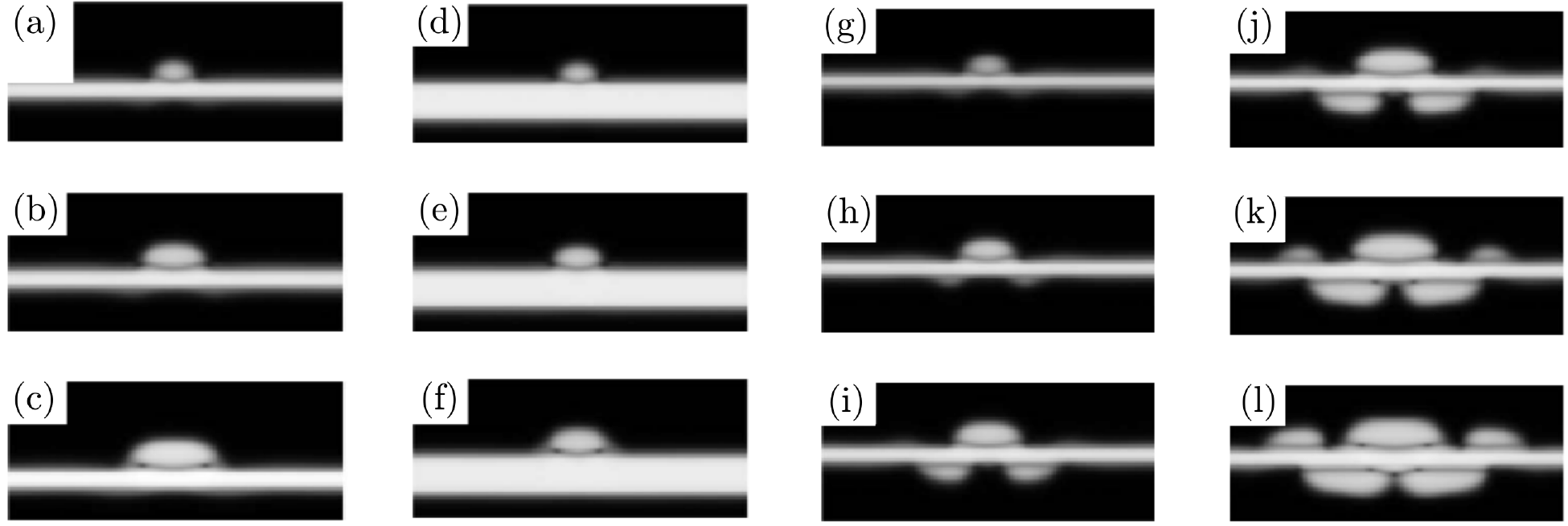}
\caption{In this figure the magnitude of the sum of the amplitudes 
is shown for an island of one material grown on another. 
In (a)-(c) the time evolution of one island is shown.  Similarly 
in (d)-(f) an island growth is illustrated for a thicker ribbon. 
In (g)-(l) the time evolution of island growth and nucleation 
is shown.  In (a)-(f) a flux of material only came from the 
top, while in (g)-(l) it came from both sides of the ribbon.
Reconstructed from \cite{ElderPRE2010}.}
\label{fig:figure15}
\end{figure}

An APFC study of the growth of islands of one material on a 
ribbon of another was conducted by Elder 
\etal \cite{ElderPRE2010,Elder_2010PFCmembranes}.  Several 
experiments \cite{Huang2005,KimLee2009,Huang2009} had to be undertaken to examine whether the 
growth of islands (or quantum dots) on thin ribbons may be 
exploited for better control of island sizes and correlations.
When an island 
of one material grows on an island of another material, the misfit 
strain will eventually lead to the nucleation of dislocation 
at the island/film/vapour junction. On very thin ribbons, the strain 
in the island can be somewhat reduced by bending the ribbons, 
leading the possibility of growing larger defect-free islands. 
An example is shown in Fig.~\ref{fig:figure15}. Figures (a)-(c) and 
(d)-(f) show the growth of an island for two different ribbon 
thicknesses. In (c) and (f), the final island size ($L_f$) at which dislocations 
appear indicates that $L_f$ is larger for 
the thinner ribbons.  Depending on conditions it was shown 
in reference \cite{Elder_2010PFCmembranes} that decreasing the 
ribbon size could almost double $L_f$.   Another interesting 
feature emerges when the island starts to 
grow. It bends the ribbon such that preferential regions 
for island nucleation appear on the other side near the 
triple junctions, leading to correlated growth as shown in 
Fig.~\ref{fig:figure15}(g)-(l). This 
correlation could potentially be exploited to create 
uniform arrays of islands.  

In summary, the binary and pure APFC models provide an excellent 
platform for studying heteroepitaxial growth. Coupled with adaptive mesh schemes as illustrated in Sec.~\ref{sec:numerical-methods}, very large simulations should be possible in both two and three dimensions.

\section{Conclusions and outlook}
\label{sec:conclusions}

In recent years, bridging-scale modeling has become crucial to comprehensively investigate crystalline systems, explore macroscopic effects of microscopic details, and unveil general properties and behaviors for further scale-specific characterizations. 
Here, an overview is provided of the model(s) obtained through the amplitude expansion of the phase-field crystal (APFC), which combines the description of crystals on relatively large (diffusive) time scales, conveyed by the PFC model \cite{Elder2002,Elder2004,Emmerich2012}, with a spatial coarse-graining. The concepts underlying its derivation have been illustrated, focusing on practical aspects such as explicit formulas, generalizations, and examples, along with presenting different formulations.

Computational aspects have also been outlined. The fields (amplitudes) to solve for within the APFC model are suited for inhomogeneous spatial discretizations, a feature that motivated its development in the first place \cite{Athreya2006}. Recently, a few optimized methods have been developed to allow for large-scale calculations and, in particular, paving the way for extensive three-dimensional calculations. 

The APFC model emerges as one of a kind among mesoscale approaches: it handles the description of crystalline systems through slowly varying continuous fields, so without resolving atoms, but retains details of the crystal structure such as anisotropies and lattice defects. Namely, it merges different aspects addressed by micro- and macroscopic approaches within a single model rather than coupling models working at different time- and length scales (like other remarkable approaches as, e.g., the quasi-continuum approach \cite{SHENOY1999611,Curtin_2003}). Among its key aspects, special attention has been given to the mesoscale description of elasticity and plasticity, being the primary goal of many coarse-grained descriptions (as the phase-field crystal itself \cite{Elder2002,Elder2004}). 
As a pivotal example, the elastic field generated by dislocations within the APFC model matches classical continuous descriptions and encodes a core regularization related to the lattice parameter. Moreover, it is expected to be affected by lattice symmetry and encodes nonlinearities. Amplitudes also allow for characterizing plasticity and defect dynamics. This description can be exploited within the broader context of PFC models as amplitudes fully characterize deformations therein \cite{HeinonenPRE2014}.

Like every other model, APFC has its range of applicability, strengths, and weaknesses.  One weakness is the ability to accurately predict the 
precise structure of atomic-scale structures such as dislocations 
and interfaces, similar to the drawbacks of traditional phase field models. 
However, it may be employed to investigate long-range effects for such systems, and extensions have been provided to improve the mesoscale descriptions with respect to the standard formulation (see, e.g., the control of energies for defects and interfaces and the modeling of Peierls barriers). Like PFC, the variational, overdamped formulation of the APFC model conveys a lack of separation among different timescales, affecting the competition among diffusion mechanisms and elastic relaxation. This issue, however, has been solved by a few different extensions, which are expected to become the standard approaches for phenomena when the separation of timescales is relevant. The most critical aspect for applications of the APFC model remains the limitation to small rotations with respect to a reference crystal orientation (see \textit{the problem of beats}  \cite{AthreyaPRE2007,Spatschek2010,Huter2017}). It prevents the thorough investigation of high-angle grain boundaries and polycrystalline systems. Therefore, providing a solution for this issue is a crucial challenge for achieving a general mesoscale description of crystals. To date, this aspect has been only partially addressed through a covariant formulation with respect to
rotation of the crystals, which still needs to be assessed for the description of elasticity and plasticity and its compatibility with other extensions. 

It is worth mentioning that in light of the limitation(s) mentioned above, the currently available APFC models should be considered valid for relatively small deformation and rotation only, \textit{de-facto} for every crystalline system where defects as dislocations can be described as separated objects. However, systems featuring such conditions are common, widely studied, and exploited in several technology-relevant applications, such as single crystals, alloys, and homo-/ heteroepitaxial systems, besides small angle-grain boundaries. The overview and discussion of the main applications addressed so far in the literature illustrate this aspect.

In conclusion, this review has attempted to collect the basics and the recent 
developments of the APFC model. While it has been used to study several physical phenomena, its potential still has not been fully exploited. 
Potential applications include the investigation of three-dimensional 
mesoscale tracking of defects and interfaces (e.g., for heteroepitaxial 
systems). 
Moreover, besides the challenges already mentioned above, a few aspects 
can be identified 
which will improve the approach further: i) direct connections with advanced 
continuum theory for elasticity and plasticity, closing the gap with methods such 
as  dislocation dynamics;
ii) description of complex crystal symmetries beyond simple ones to broaden the application to technology-relevant systems; iii) extending the parametrization to include physical parameters extracted from experiments and/or other methods; iv) connections and coupling to both microscopic, fully atomistic (e.g., PFC or Molecular Dynamics) and macroscopic (e.g., phase-field, continuum elasticity) models; v) extended boundary conditions to enable investigations beyond bulk-like systems and simple geometries; vi) further development of numerical methods, keeping up with state-of-the-art numerical techniques.

\section*{Acknowledgements}
M.S. acknowledges support from the Emmy Noether Programme of the German Research Foundation (DFG) under Grant No. SA4032/2-1. K.R.E. acknowledges support from the National Science Foundation (NSF) under Grant No. DMR-MPS-2006456. Computing resources have been provided by the Center for Information Services and High-Performance Computing (ZIH) at TU Dresden. The authors also acknowledge useful discussions with Zhi-Feng Huang, Axel Voigt, Rainer Backofen, Simon Praetorius, Lucas Benoit-Marechal, Luiza Angheluta, Vidar Skogvoll, and Jorge Vi\~nals. 

\section*{References}

\providecommand{\newblock}{}


\begin{thebibliography}{183}
\expandafter\ifx\csname url\endcsname\relax
  \def\url#1{{\tt #1}}\fi
\expandafter\ifx\csname urlprefix\endcsname\relax\def\urlprefix{URL }\fi
\providecommand{\eprint}[2][]{\url{#2}}

\bibitem{Elder2002}
Elder K~R, Katakowski M, Haataja M and Grant M 2002 {\em Phys. Rev. Lett.\/}
  \href{http://dx.doi.org/10.1103/PhysRevLett.88.245701}{{\bf 88} 245701}

\bibitem{Langer80}
Langer J~S 1980 {\em Rev. Mod. Phys.\/}
  \href{http://dx.doi.org/10.1103/RevModPhys.52.1}{{\bf 52} 1}

\bibitem{Karma96}
Karma A and Rappel W~J 1996 {\em Phys. Rev. E\/}
  \href{http://dx.doi.org/10.1103/PhysRevE.53.R3017}{{\bf 53} 3017}

\bibitem{Karma98}
Karma A and Rappel W~J 1998 {\em Phys. Rev. E\/}
  \href{http://dx.doi.org/10.1103/PhysRevE.57.4323}{{\bf 57} 4323}

\bibitem{Elder2001}
Elder K~R, Grant M, Provatas N and Kosterlitz J~M 2001 {\em Phys. Rev. E\/}
  \href{http://dx.doi.org/10.1103/PhysRevE.64.021604}{{\bf 64} 021604}

\bibitem{ElderPRE2010}
Elder K~R, Huang Z~F and Provatas N 2010 {\em Phys. Rev. E\/}
  \href{http://dx.doi.org/10.1103/PhysRevE.81.011602}{{\bf 81} 011602}

\bibitem{Elder2004}
Elder K~R and Grant M 2004 {\em Phys. Rev. E\/}
  \href{http://dx.doi.org/10.1103/PhysRevE.70.051605}{{\bf 70} 051605}

\bibitem{mot11}
Wu K~A and Voorhees P~W 2012 {\em Acta Mater.\/}
  \href{http://dx.doi.org/10.1016/j.actamat.2011.09.035}{{\bf 60} 407--409}

\bibitem{Hirvonen2016}
Hirvonen P, Ervasti M~M, Fan Z, Jalalvand M, Seymour M, {Vaez Allaei} S~M,
  Provatas N, Harju A, Elder K~R and Ala-Nissila T 2016 {\em Phys. Rev. B\/}
  \href{http://dx.doi.org/10.1103/PhysRevB.94.035414}{{\bf 94} 035414}

\bibitem{mot3}
Yamanaka A, McReynolds K and Voorhees P~W 2017 {\em Acta Mater.\/}
  \href{http://dx.doi.org/10.1016/j.actamat.2017.05.022}{{\bf 133} 160--171}

\bibitem{Berry2012}
Berry J, Provatas N, Rottler J and Sinclair C~W 2012 {\em Phys. Rev. B\/}
  \href{http://dx.doi.org/10.1103/PhysRevB.86.224112}{{\bf 86} 224112}

\bibitem{SkaugenPRB2018}
Skaugen A, Angheluta L and Vi\~nals J 2018 {\em Phys. Rev. B\/}
  \href{http://dx.doi.org/10.1103/PhysRevB.97.054113}{{\bf 97} 054113}

\bibitem{SalvalaglioPRL2021}
Salvalaglio M, Voigt A, Huang Z~F and Elder K~R 2021 {\em Phys. Rev. Lett.\/}
  \href{http://dx.doi.org/10.1103/PhysRevLett.126.185502}{{\bf 126} 185502}

\bibitem{SalvalaglioNPJ2019}
Salvalaglio M, Voigt A and Elder K~R 2019 {\em npj Comput. Mater.\/}
  \href{http://dx.doi.org/10.1038/s41524-019-0185-0}{{\bf 5} 48}

\bibitem{SkaugenPRL2018}
Skaugen A, Angheluta L and Vi{\~{n}}als J 2018 {\em Phys. Rev. Lett.\/}
  \href{http://dx.doi.org/10.1103/PhysRevLett.121.255501}{{\bf 121} 255501}

\bibitem{SalvalaglioJMPS2020}
Salvalaglio M, Angheluta L, Huang Z~f, Voigt A, Elder K~R and Vi{\~{n}}als J
  2020 {\em J. Mech. Phys. Solids\/}
  \href{http://dx.doi.org/10.1016/j.jmps.2019.103856}{{\bf 137} 103856}

\bibitem{DD0}
Amodeo R~J and Ghoniem N~M 1990 {\em Phys. Rev. B\/}
  \href{http://dx.doi.org/10.1103/PhysRevB.41.6958}{{\bf 41} 6958}

\bibitem{DD1}
Ghoniem N~M, Tong S~H and Sun L~Z 2000 {\em Phys. Rev. B\/}
  \href{http://dx.doi.org/10.1103/PhysRevB.61.913}{{\bf 61} 913}

\bibitem{DD2}
Arsenlis A, Cai W, Tang M, Rhee M, Oppelstrup T, Hommes G, Pierce T~G and
  Bulatov V~V 2007 {\em Model. Simul. Mater. Sci. Eng.\/}
  \href{http://dx.doi.org/10.1088/0965-0393/15/6/001}{{\bf 15} 553}

\bibitem{Greenwood2010}
Greenwood M, Provatas N and Rottler J 2010 {\em Phys. Rev. Lett.\/}
  \href{http://dx.doi.org/10.1103/PhysRevLett.105.045702}{{\bf 105} 045702}

\bibitem{xtal2}
Greenwood M, Rottler J and Provatas N 2011 {\em Phys. Rev. E\/}
  \href{http://dx.doi.org/10.1103/PhysRevE.83.031601}{{\bf 83} 031601}

\bibitem{Goldenfeld2005}
Goldenfeld N, Athreya B~P and Dantzig J~A 2005 {\em Phys. Rev. E\/}
  \href{http://dx.doi.org/10.1103/PhysRevE.72.020601}{{\bf 72} 020601}

\bibitem{Athreya2006}
Athreya B~P, Nigel G and Dantzig J~A 2006 {\em Phys. Rev. E\/}
  \href{http://dx.doi.org/10.1103/PhysRevE.74.011601}{{\bf 74} 011601}

\bibitem{elder2007}
Elder K~R, Provatas N, Berry J, Stefanovic P and Grant M 2007 {\em Phys. Rev.
  B\/} \href{http://dx.doi.org/10.1103/PhysRevB.75.064107}{{\bf 75} 064107}

\bibitem{Khachaturyan1983}
Khachaturyan A~G 1983 {\em The Theory of Structural Tranformation in Solids\/}
  (\href{http://www.worldcat.org/oclc/797439874}{Wiley, New York})

\bibitem{Khachaturyan1996}
Khachaturyan A~G 1996 {\em Philos. Mag. A\/}
  \href{http://dx.doi.org/10.1080/01418619608239686}{{\bf 74} 3--14}

\bibitem{cross1993pattern}
Cross M~C and Hohenberg P~C 1993 {\em Rev. Mod. Phys.\/}
  \href{http://dx.doi.org/10.1103/RevModPhys.65.851}{{\bf 65} 851}

\bibitem{Smirman2017}
Smirman M, M~Taha D, Singh A~K, Huang Z~F and Elder K~R 2017 {\em Phys. Rev.
  B\/} \href{http://dx.doi.org/10.1103/PhysRevB.95.085407}{{\bf 95} 085407}

\bibitem{Huang2010}
Huang Z~F, Elder K~R and Provatas N 2010 {\em Phys. Rev. E\/}
  \href{http://dx.doi.org/10.1103/PhysRevE.82.021605}{{\bf 82} 021605}

\bibitem{vanTeeffelen2009}
van Teeffelen S, Backofen R, Voigt A and L\"owen H 2009 {\em Phys. Rev. E\/}
  \href{http://dx.doi.org/10.1103/PhysRevE.79.051404}{{\bf 79} 051404}

\bibitem{RY79}
Ramakrishnan T~V and Yussouff M 1979 {\em Phys. Rev. B\/}
  \href{http://dx.doi.org/10.1103/PhysRevB.19.2775}{{\bf 19} 2775}

\bibitem{Tupper_2008}
Tupper P~F and Grant M 2008 {\em {EPL} (Europhysics Letters)\/}
  \href{http://dx.doi.org/10.1209/0295-5075/81/40007}{{\bf 81} 40007}

\bibitem{Emmerich2012}
Emmerich H, L{\"{o}}wen H, Wittkowski R, Gruhn T, T{\'{o}}th G~I, Tegze G and
  Gr{\'{a}}n{\'{a}}sy L 2012 {\em Adv. Phys.\/}
  \href{http://dx.doi.org/10.1080/00018732.2012.737555}{{\bf 61} 665--743}

\bibitem{Berry2014}
Berry J, Provatas N, Rottler J and Sinclair C~W 2014 {\em Phys. Rev. B\/}
  \href{http://dx.doi.org/10.1103/PhysRevB.89.214117}{{\bf 89} 214117}

\bibitem{Backofen14}
Backofen R, Barmak K, Elder K~R and Voigt A 2014 {\em Acta Mater.\/}
  \href{http://dx.doi.org/10.1016/j.actamat.2013.11.034}{{\bf 64} 72--77}

\bibitem{GRANASY2019}
Gránásy L, Tóth G~I, Warren J~A, Podmaniczky F, Tegze G, Rátkai L and
  Pusztai T 2019 {\em Prog. Mater. Sci.\/}
  \href{http://dx.doi.org/10.1016/j.pmatsci.2019.05.002}{{\bf 106} 100569}

\bibitem{Alaimo2016}
Alaimo F, Praetorius S and Voigt A 2016 {\em New J. Phys.\/}
  \href{http://dx.doi.org/10.1088/1367-2630/18/8/083008}{{\bf 18} 083008}

\bibitem{Alaimo2018}
Alaimo F and Voigt A 2018 {\em Phys. Rev. E\/}
  \href{http://dx.doi.org/10.1103/PhysRevE.98.032605}{{\bf 98} 032605}

\bibitem{Huang2020}
Huang Z~F, Menzel A~M and L\"{o}wen H 2020 {\em Phys. Rev. Lett.\/}
  \href{http://dx.doi.org/10.1103/PhysRevLett.125.218002}{{\bf 125} 218002}

\bibitem{Menzel2013}
Menzel A~M and L\"{o}wen H 2013 {\em Phys. Rev. Lett.\/}
  \href{http://dx.doi.org/10.1103/PhysRevLett.110.055702}{{\bf 110} 055702}

\bibitem{Menzel2014}
Menzel A~M, Ohta T and L\"{o}wen H 2014 {\em Phys. Rev. E\/}
  \href{http://dx.doi.org/10.1103/PhysRevE.89.022301}{{\bf 89} 022301}

\bibitem{Praetorius2015}
Praetorius S and Voigt A 2015 {\em J. Chem. Phys.\/}
  \href{http://dx.doi.org/10.1063/1.4918559}{{\bf 142} 154904}

\bibitem{Aland2012}
Aland S, R\"atz A, R\"oger M and Voigt A 2012 {\em Multiscale Model. Simul.\/}
  \href{http://dx.doi.org/10.1137/110834718}{{\bf 10} 82--110}

\bibitem{ashcroft1976solid}
Ashcroft N~W and Mermin N~D 1976 {\em {S}olid {S}tate {P}hysics\/}
  (\href{http://www.worldcat.org/oclc/779499471}{New York: Holt, Rinehart and
  Winston})

\bibitem{Galenko2015}
Galenko P, Sanches F~I and Elder K 2015 {\em Phys. D: Nonlinear Phenom.\/}
  \href{http://dx.doi.org/10.1016/j.physd.2015.06.002}{{\bf 308} 1--10}

\bibitem{Shiwa2011}
Shiwa Y 2011 {\em Prog. Theor. Phys.\/}
  \href{http://dx.doi.org/10.1143/PTP.125.871}{{\bf 125} 871--878}

\bibitem{Oono2012}
Oono Y and Shiwa Y 2012 {\em Phys. Rev. E\/}
  \href{http://dx.doi.org/10.1103/PhysRevE.86.061138}{{\bf 86} 061138}

\bibitem{Provatas2010}
Provatas N and Elder K 2010 {\em Phase-Field Methods in Materials Science and
  Engineering\/}
  (\href{https://onlinelibrary.wiley.com/doi/book/10.1002/9783527631520}{Wiley-VCH
  Verlag GmbH})

\bibitem{Yeon2010}
Yeon D~H, Huang Z~F, Elder K and Thornton K 2010 {\em Philos. Mag.\/}
  \href{http://dx.doi.org/10.1080/14786430903164572}{{\bf 90} 237--263}

\bibitem{HohHal}
Hohenberg P~C and Halperin B~I 1977 {\em Rev. Mod. Phys.\/}
  \href{http://dx.doi.org/10.1103/RevModPhys.49.435}{{\bf 49} 435}

\bibitem{Grossmann93}
Grossmann B, Elder K~R, Grant M and Kosterlitz J 1993 {\em Phys. Rev. Lett.\/}
  \href{http://dx.doi.org/10.1103/PhysRevLett.71.3323}{{\bf 71} 3323}

\bibitem{Drolet00}
Drolet F, Elder K~R, Grant M and Kosterlitz J 2000 {\em Phys. Rev. E\/}
  \href{http://dx.doi.org/10.1103/PhysRevE.61.6705}{{\bf 61} 6705}

\bibitem{Elder94}
Elder K~R, Drolet F, Grant M and Kosterlitz J 1994 {\em Phys. Rev. Lett.\/}
  \href{http://dx.doi.org/10.1103/PhysRevLett.72.677}{{\bf 72} 677}

\bibitem{HeinonenPRL2016}
Heinonen V, Achim C~V, Kosterlitz J~M, Ying S~C, Lowengrub J and Ala-Nissila T
  2016 {\em Phys. Rev. Lett.\/}
  \href{http://dx.doi.org/10.1103/PhysRevLett.116.024303}{{\bf 116} 024303}

\bibitem{MajaniemiPRB2007}
Majaniemi S and Grant M 2007 {\em Phys. Rev. B\/}
  \href{http://dx.doi.org/10.1103/PhysRevB.75.054301}{{\bf 75} 054301}

\bibitem{Majaniemi2009}
Majaniemi S and Provatas N 2009 {\em Phys. Rev. E\/}
  \href{http://dx.doi.org/10.1103/PhysRevE.79.011607}{{\bf 79} 011607}

\bibitem{Chan2009}
Chan P~Y and Goldenfeld N 2009 {\em Phys. Rev. E\/}
  \href{http://dx.doi.org/10.1103/PhysRevE.80.065105}{{\bf 80} 065105}

\bibitem{Ofori-Opoku2013}
Ofori-Opoku N, Stolle J, Huang Z~F and Provatas N 2013 {\em Phys. Rev. B\/}
  \href{http://dx.doi.org/10.1103/PhysRevB.88.104106}{{\bf 88}(10) 104106}

\bibitem{XuPRB2016}
Xu Y~C, Geslin P~A and Karma A 2016 {\em Phys. Rev. B\/}
  \href{http://dx.doi.org/10.1103/PhysRevB.94.144106}{{\bf 94} 144106}

\bibitem{Salvalaglio2017}
Salvalaglio M, Backofen R, Voigt A and Elder K~R 2017 {\em Phys. Rev E\/}
  \href{http://dx.doi.org/10.1103/PhysRevE.96.023301}{{\bf 96} 023301}

\bibitem{AnkudinovPRE2020}
Ankudinov V, Elder K~R and Galenko P~K 2020 {\em Phys. Rev. E\/}
  \href{http://dx.doi.org/10.1103/PhysRevE.102.062802}{{\bf 102} 062802}

\bibitem{Mkhonta2013}
Mkhonta S~K, Elder K~R and Huang Z~F 2013 {\em Phys. Rev. Lett.\/}
  \href{http://dx.doi.org/10.1103/PhysRevLett.111.035501}{{\bf 111} 035501}

\bibitem{Greenwood2011binary}
Greenwood M, Ofori-Opoku N, Rottler J and Provatas N 2011 {\em Phys. Rev. B\/}
  \href{http://dx.doi.org/10.1103/PhysRevB.84.064104}{{\bf 84} 064104}

\bibitem{Alster17}
Alster E, Elder K~R, Hoyt J~J and Voorhees P~W 2017 {\em Phys. Rev. E\/}
  \href{http://dx.doi.org/10.1103/PhysRevE.95.022105}{{\bf 95} 022105}

\bibitem{Oono87}
Oono Y and Puri S 1987 {\em Phys. Rev. Lett.\/}
  \href{http://dx.doi.org/10.1103/PhysRevLett.58.836}{{\bf 58} 836}

\bibitem{AthreyaPRE2007}
Athreya B~P, Goldenfeld N, Dantzig J~A, Greenwood M and Provatas N 2007 {\em
  Phys. Rev. E\/} \href{http://dx.doi.org/10.1103/PhysRevE.76.056706}{{\bf 76}
  056706}

\bibitem{Bercic2018}
Ber{\v{c}}i{\v{c}} M and Kugler G 2018 {\em Phys. Rev. E\/}
  \href{http://dx.doi.org/10.1103/PhysRevE.98.033303}{{\bf 98} 033303}

\bibitem{Bercic2020}
Ber\ifmmode \check{c}\else \v{c}\fi{}i\ifmmode~\check{c}\else \v{c}\fi{} M and
  Kugler G 2020 {\em Phys. Rev. E\/}
  \href{http://dx.doi.org/10.1103/PhysRevE.101.043309}{{\bf 101} 043309}

\bibitem{Geslin2015}
Geslin P~a, Xu Y and Karma A 2015 {\em Phys. Rev. Lett.\/}
  \href{http://dx.doi.org/10.1103/PhysRevLett.114.105501}{{\bf 114} 105501}

\bibitem{Guan16}
Guan Z, Heinonen V, Lowengrub J, Wang C and Wise S~M 2016 {\em J. Comp.
  Phys.\/} \href{http://dx.doi.org/10.1016/j.jcp.2016.06.007}{{\bf 321} 1026}

\bibitem{cooley1965algorithm}
Cooley J~W and Tukey J~W 1965 {\em Math. Comp.\/}
  \href{http://dx.doi.org/10.1090/S0025-5718-1965-0178586-1}{{\bf 19} 297--301}

\bibitem{Chen98}
Chen L~Q and Shen J 1998 {\em Comp. Phys. Comm.\/}
  \href{http://dx.doi.org/10.1016/S0010-4655(97)00115-X}{{\bf 108} 147}

\bibitem{Huter2017}
H{\"{u}}ter C, Neugebauer J, Boussinot G, Svendsen B, Prahl U and Spatschek R
  2017 {\em Contin. Mech. Thermodyn.\/}
  \href{http://dx.doi.org/10.1007/s00161-015-0424-7}{{\bf 29} 895--911}

\bibitem{Spatschek2010}
Spatschek R and Karma A 2010 {\em Phys. Rev. B\/}
  \href{http://dx.doi.org/10.1103/PhysRevB.81.214201}{{\bf 81} 214201}

\bibitem{HeinonenPRE2014}
Heinonen V, Achim C~V, Elder K~R, Buyukdagli S and Ala-Nissila T 2014 {\em
  Phys. Rev. E\/} \href{http://dx.doi.org/10.1103/PhysRevE.89.032411}{{\bf 89}
  032411}

\bibitem{Jreidini2021}
Jreidini P, Pinomaa T, Wiezorek J~M~K, McKeown J~T, Laukkanen A and Provatas N
  2021 {\em Phys. Rev. Lett.\/}
  \href{http://dx.doi.org/10.1103/PhysRevLett.127.205701}{{\bf 127} 205701}

\bibitem{Praetorius2019}
Praetorius S, Salvalaglio M and Voigt A 2019 {\em Model. Simul. Mater. Sci.
  Eng.\/} \href{http://dx.doi.org/10.1088/1361-651X/ab1508}{{\bf 27} 044004}

\bibitem{Xiaoting2022}
Luo X, Huang Z, Wang S, Xiao M, Meng Y, Yan H, Li Q and Wang G 2022 {\em
  Electronics\/} \href{http://dx.doi.org/10.3390/electronics11020221}{{\bf 11}
  221}

\bibitem{backofen07}
Backofen R, Rätz A and Voigt A 2007 {\em Philos. Mag. Lett.\/}
  \href{http://dx.doi.org/10.1080/09500830701481737}{{\bf 87} 813--820}

\bibitem{Gomez2012}
Gomez H and Nogueira X 2012 {\em Comput. Methods Appl. Mech. Eng.\/}
  \href{http://dx.doi.org/10.1016/j.cma.2012.03.002}{{\bf 249-252} 52--61}

\bibitem{Vignal2015}
Vignal P, Dalcin L, Brown D, Collier N and Calo V 2015 {\em Comput. Struct.\/}
  \href{http://dx.doi.org/10.1016/j.compstruc.2015.05.029}{{\bf 158} 355--368}

\bibitem{Ruihan2016}
Guo R and Xu Y 2016 {\em SIAM J. Sci. Comput.\/}
  \href{http://dx.doi.org/10.1137/15M1038803}{{\bf 38} A105--A127}

\bibitem{Liupeng2020}
Wang L, Huang Y and Jiang K 2020 {\em Numer. Math. Theor. Meth. Appl.\/}
  \href{http://dx.doi.org/10.4208/nmtma.OA-2019-0110}{{\bf 13} 372--399}

\bibitem{Vey2007}
Vey S and Voigt A 2007 {\em Comput. Visual. Sci.\/}
  \href{http://dx.doi.org/10.1007/s00791-006-0048-3}{{\bf 10} 57--67}

\bibitem{WitkowskiACM2015}
Witkowski T, Ling S, Praetorius S and Voigt A 2015 {\em Adv. Comput. Math.\/}
  \href{http://dx.doi.org/10.1007/s10444-015-9405-4}{{\bf 41} 1145--1177}

\bibitem{Praetorius2014}
Praetorius S and Voigt A 2015 {\em SIAM J. Sci. Comput.\/}
  \href{http://dx.doi.org/10.1137/140980375}{{\bf 37} B425--B451}

\bibitem{PraetoriusThesis2015}
Praetorius S 2015 {\em Efficient Solvers for the Phase-Field Crystal Equation -
  Development and Analysis of a Block-Preconditioner\/}
  \href{http://nbn-resolving.de/urn:nbn:de:bsz:14-qucosa-195532}{PhD Thesis}
  \href{http://nbn-resolving.de/urn:nbn:de:bsz:14-qucosa-195532}{Technische
  Universit{\"a}t Dresden}
  \href{http://nbn-resolving.de/urn:nbn:de:bsz:14-qucosa-195532}{Germany}

\bibitem{Backofen2022}
Backofen R, Salvalaglio M and Voigt A 2022 {\em
  \href{https://arxiv.org/abs/2202.06654}{arXiv:2202.06654}\/}

\bibitem{seth1961generalized}
Seth B 1961 Generalized strain measure with applications to physical problems.
  Tech. rep. \href{https://apps.dtic.mil/sti/citations/AD0266913}{Wisconsin
  Univ-Madison Mathematics Research Center}

\bibitem{hill1968constitutive}
Hill R 1968 {\em J. Mech. Phys. Solids\/}
  \href{http://dx.doi.org/10.1016/0022-5096(68)90031-8}{{\bf 16} 229--242}

\bibitem{hill1970constitutive}
Hill R 1970 {\em Proc. R. Soc. A: Math\/}
  \href{http://dx.doi.org/10.1098/rspa.1970.0018}{{\bf 314} 457--472}

\bibitem{bruhns2015multiplicative}
Bruhns O~T 2015 The multiplicative decomposition of the deformation gradient in
  plasticity—origin and limitations {\em From Creep Damage Mechanics to
  Homogenization Methods\/}
  (\href{https://www.doi.org/10.1007/978-3-319-19440-0_3}{Springer}) pp 37--66

\bibitem{Neff2016}
Neff P, Eidel B and Martin R~J 2016 {\em Arch. Ration. Mech. Anal.\/}
  \href{http://dx.doi.org/10.1007/s00205-016-1007-x}{{\bf 222} 507--572}

\bibitem{LLEL}
Landau L~D, Lifshitz E~M, Kosevich A~M and Pitaevskii L~P 1986 {\em Theory of
  elasticity\/}
  (\href{https://www.sciencedirect.com/book/9780080570693/theory-of-elasticity}{Elsevier})

\bibitem{Skogvoll_stress2021}
Skogvoll V, Skaugen A and Angheluta L 2021 {\em Phys. Rev. B\/}
  \href{http://dx.doi.org/10.1103/PhysRevB.103.224107}{{\bf 103} 224107}

\bibitem{Skogvoll2021}
Skogvoll V, Skaugen A, Angheluta L, Salvalaglio M and Vi\~nals J 2021 {\em
  \href{https://arxiv.org/abs/2110.03476}{arXiv:2110.03476}\/}

\bibitem{anderson2017}
Anderson P, Hirth J and Lothe J 2017 {\em Theory of Dislocations\/}
  (\href{http://www.worldcat.org/oclc/1132912878}{Cambridge University Press})

\bibitem{Mazenko1997}
Mazenko G~F 1997 {\em Phys. Rev. Lett.\/}
  \href{http://dx.doi.org/10.1103/PhysRevLett.78.401}{{\bf 78} 401}

\bibitem{Mazenko2001}
Mazenko G~F 2001 {\em Phys. Rev. E\/}
  \href{http://dx.doi.org/10.1103/PhysRevE.64.016110}{{\bf 64} 016110}

\bibitem{Angheluta2012}
Angheluta L, Jeraldo P and Goldenfeld N 2012 {\em Phys. Rev. E\/}
  \href{http://dx.doi.org/10.1103/PhysRevE.85.011153}{{\bf 85} 011153}

\bibitem{Huter2016}
H{\"{u}}ter C, Fri{\'{a}}k M, Weikamp M, Neugebauer J, Goldenfeld N, Svendsen B
  and Spatschek R 2016 {\em Phys. Rev. B\/}
  \href{http://dx.doi.org/10.1103/PhysRevB.93.214105}{{\bf 93} 214105}

\bibitem{Cai2006}
Cai W, Arsenlis A, Weinberger C and Bulatov V 2006 {\em J. Mech. Phys.
  Solids\/} \href{http://dx.doi.org/10.1016/j.jmps.2005.09.005}{{\bf 54}
  561--587}

\bibitem{Head_1953}
Head A~K 1953 {\em Proc. Roy. Soc. Lond. B\/}
  \href{http://dx.doi.org/10.1088/0370-1301/66/9/309}{{\bf 66} 793--801}

\bibitem{Marzegalli2013}
Marzegalli A, Brunetto M, Salvalaglio M, Montalenti F, Nicotra G, Scuderi M,
  Spinella C, De~Seta M and Capellini G 2013 {\em Phys. Rev. B\/}
  \href{http://dx.doi.org/10.1103/PhysRevB.88.165418}{{\bf 88} 165418}

\bibitem{Lazar2005}
Lazar M and Maugin G~A 2005 {\em Int. J. Eng. Sci.\/}
  \href{http://dx.doi.org/https://doi.org/10.1016/j.ijengsci.2005.01.006}{{\bf
  43} 1157--1184}

\bibitem{Lazar2017}
Lazar M 2017 {\em Philos. Mag.\/}
  \href{http://dx.doi.org/10.1080/14786435.2017.1375608}{{\bf 97} 3246--3275}

\bibitem{Mindlin1964}
Mindlin R~D 1964 {\em Arch. Ration. Mech. Anal.\/}
  \href{http://dx.doi.org/10.1007/BF00248490}{{\bf 16} 51--78}

\bibitem{Mindlin1968}
Mindlin R and Eshel N 1968 {\em Int. J. Solids Struct.\/}
  \href{http://dx.doi.org/10.1016/0020-7683(68)90036-X}{{\bf 4} 109 -- 124}

\bibitem{Lazar2018}
Lazar M and Po G 2018 {\em J. Micromech. Mol. Phys.\/}
  \href{http://dx.doi.org/10.1142/s2424913018400088}{{\bf 03} 1840008}

\bibitem{Chockalingam2021}
Salvalaglio M, Chockalingam K, Voigt A and D\"orfler W 2022 {\em Examples and
  Counterexamples\/} \href{http://dx.doi.org/10.1016/j.exco.2022.100067}{{\bf
  2} 100067}

\bibitem{Kinoshita1971}
Kinoshita N and Mura T 1971 {\em Phys. Status Solidi A\/}
  \href{http://dx.doi.org/10.1002/pssa.2210050332}{{\bf 5} 759--768}

\bibitem{eshelby1957determination}
Eshelby J~D 1957 {\em Proc. Roy. Soc. Lond. A\/}
  \href{http://dx.doi.org/10.1098/rspa.1957.0133}{{\bf 241} 376--396}

\bibitem{eshelby1959elastic}
Eshelby J~D 1959 {\em Proc. Roy. Soc. Lond. A\/}
  \href{http://dx.doi.org/10.1098/rspa.1959.0173}{{\bf 252} 561--569}

\bibitem{mura2013micromechanics}
Mura T 1987 {\em Micromechanics of defects in solids\/}
  (\href{https://link.springer.com/book/10.1007/978-94-009-3489-4}{Springer})

\bibitem{HeinonenThesis}
Heinonen V 2016 {\em Phase field crystal models and fast dynamics\/}
  \href{https://aaltodoc.aalto.fi/handle/123456789/20772?show=full}{PhD Thesis}
  \href{https://aaltodoc.aalto.fi/handle/123456789/20772?show=full}{Aalto
  University}
  \href{https://aaltodoc.aalto.fi/handle/123456789/20772?show=full}{Finland}

\bibitem{StefanovicPRL2006}
Stefanovic P, Haataja M and Provatas N 2006 {\em Phys. Rev. Lett.\/}
  \href{http://dx.doi.org/10.1103/PhysRevLett.96.225504}{{\bf 96} 225504}

\bibitem{GalenkoPRE2009}
Galenko P, Danilov D and Lebedev V 2009 {\em Phys. Rev. E\/}
  \href{http://dx.doi.org/10.1103/PhysRevE.79.051110}{{\bf 79} 051110}

\bibitem{Huang2013}
Huang Z~F 2013 {\em Phys. Rev. E\/}
  \href{http://dx.doi.org/10.1103/PhysRevE.87.012401}{{\bf 87} 012401}

\bibitem{Cahn1958}
Cahn J~E and Hilliard J~W 1958 {\em J. Chem. Phys.\/}
  \href{http://dx.doi.org/10.1063/1.1744102}{{\bf 28} 258--267}

\bibitem{Kocher2015}
Kocher G and Provatas N 2015 {\em Phys. Rev. Lett.\/}
  \href{http://dx.doi.org/10.1103/PhysRevLett.114.155501}{{\bf 114} 155501}

\bibitem{Guo2016}
Guo C, Wang J, Wang Z, Li J, Guo Y and Huang Y 2016 {\em Soft Matter\/}
  \href{http://dx.doi.org/10.1039/C6SM00774K}{{\bf 12} 4666--4673}

\bibitem{Hwa1991}
Hwa T, Kardar M and Paczuski M 1991 {\em Phys. Rev. Lett.\/}
  \href{http://dx.doi.org/10.1103/PhysRevLett.66.441}{{\bf 66} 441}

\bibitem{Huang2016}
Huang Z~F 2016 {\em Phys. Rev. E\/}
  \href{http://dx.doi.org/10.1103/PhysRevE.93.022803}{{\bf 93} 022803}

\bibitem{kundin2014bridging}
Kundin J, Choudhary M and Emmerich H 2014 {\em Eur. Phys. J.: Spec. Top.\/}
  \href{http://dx.doi.org/10.1140/epjst/e2014-02096-y}{{\bf 223} 363--372}

\bibitem{chen2002phase}
Chen L~Q 2002 {\em Annu. Rev. Mater. Res.\/}
  \href{http://dx.doi.org/10.1146/annurev.matsci.32.112001.132041}{{\bf 32}
  113--140}

\bibitem{boettinger2002phase}
Boettinger W~J, Warren J~A, Beckermann C and Karma A 2002 {\em Annu. Rev.
  Mater. Res.\/}
  \href{http://dx.doi.org/10.1146/annurev.matsci.32.101901.155803}{{\bf 32}
  163--194}

\bibitem{Steinbach2009}
Steinbach I 2009 {\em Model. Simul. Mater. Sci. Eng.\/}
  \href{http://dx.doi.org/10.1088/0965-0393/17/7/073001}{{\bf 17} 073001}

\bibitem{Chen1991}
Chen L~Q, Wang Y and Khachaturyan A~G 1991 {\em Phil. Mag. Lett.\/}
  \href{http://dx.doi.org/10.1080/09500839108214618}{{\bf 64} 241--251}

\bibitem{Bugaev2002}
Bugaev V~N, Reichert H, Shchyglo O, Udyansky A, Sikula Y and Dosch H 2002 {\em
  Phys. Rev. B\/} \href{http://dx.doi.org/10.1103/PhysRevB.65.180203}{{\bf 65}
  180203}

\bibitem{Tewary2004}
Tewary V~K 2004 {\em Phys. Rev. B\/}
  \href{http://dx.doi.org/10.1103/PhysRevB.69.094109}{{\bf 69} 094109}

\bibitem{Varvenne2012}
Varvenne C, Finel A, {Le Bouar} Y and F{\`{e}}vre M 2012 {\em Phys. Rev. B\/}
  \href{http://dx.doi.org/10.1103/PhysRevB.86.184203}{{\bf 86} 184203}

\bibitem{Varvenne2017}
Varvenne C and Clouet E 2017 {\em Phys. Rev. B\/}
  \href{http://dx.doi.org/10.1103/PhysRevB.96.224103}{{\bf 96} 224103}

\bibitem{Nizovtseva2017}
Nizovtseva I~G and Galenko P~K 2018 {\em Philos. Trans. R. Soc. A\/}
  \href{http://dx.doi.org/10.1098/rsta.2017.0202}{{\bf 376} 20170202}

\bibitem{Ofori-Opoku2018}
Ofori-Opoku N, Warren J~A and Voorhees P~W 2018 {\em Phys. Rev. Materials\/}
  \href{http://dx.doi.org/10.1103/PhysRevMaterials.2.083404}{{\bf 2} 083404}

\bibitem{Wheeler2006}
Wheeler A 2006 {\em Proc. Roy. Soc. Lond. A\/}
  \href{http://dx.doi.org/10.1098/rspa.2006.1721}{{\bf 462} 3363--3384}

\bibitem{Tor2009}
Torabi S, Lowengrub J, Voigt A and Wise S 2009 {\em Proc. Roy. Soc. Lond. A\/}
  \href{http://dx.doi.org/10.1098/rspa.2008.0385}{{\bf 465} 1337--1359}

\bibitem{Sal2015b}
Salvalaglio M, Backofen R, Bergamaschini R, Montalenti F and Voigt A 2015 {\em
  Cryst Growth Des.\/} \href{http://dx.doi.org/10.1021/acs.cgd.5b00165}{{\bf
  15} 2787--2794}

\bibitem{Salvalaglio2018}
Salvalaglio M, Backofen R, Elder K~R and Voigt A 2018 {\em Phys. Rev.
  Materials\/} \href{http://dx.doi.org/10.1103/PhysRevMaterials.2.053804}{{\bf
  2} 053804}

\bibitem{Adland2013}
Adland A, Karma A, Spatschek R, Buta D and Asta M 2013 {\em Phys. Rev. B\/}
  \href{http://dx.doi.org/10.1103/PhysRevB.87.024110}{{\bf 87} 024110}

\bibitem{Pinomaa2022}
Pinomaa T, Lindroos M, Jreidini P, Haapalehto M, Ammar K, Wang L, Forest S,
  Provatas N and Laukkanen A 2022 {\em Phil. Trans. R. Soc. A\/}
  \href{http://dx.doi.org/10.1098/rsta.2020.0319}{{\bf 380} 20200319}

\bibitem{Cottrell1949}
Cottrell A~H, Jaswon M~A and Mott N~F 1949 {\em Proc. Roy. Soc. Lond. A\/}
  \href{http://dx.doi.org/10.1098/rspa.1949.0128}{{\bf 199} 104--114}

\bibitem{Cottrell1949_2}
Cottrell A~H and Bilby B~A 1949 {\em Proc. Roy. Soc. Lond. A\/}
  \href{http://dx.doi.org/10.1088/0370-1298/62/1/308}{{\bf 62} 49--62}

\bibitem{CottrellBOOK}
Cottrell A~H 1953 {\em {Dislocations and plastic flow in crystals}\/} Internat.
  Ser. Mono. Phys. ({\href{http://www.worldcat.org/oclc/299614439}{Oxford}}:
  {\href{http://www.worldcat.org/oclc/299614439}{Clarendon Press}})

\bibitem{Zhang_2008}
Zhang F and Curtin W~A 2008 {\em Model. Simul. Mater. Sci. Eng.\/}
  \href{http://dx.doi.org/10.1088/0965-0393/16/5/055006}{{\bf 16} 055006}

\bibitem{Sills2016}
Sills R~B and Cai W 2016 {\em Philos. Mag.\/}
  \href{http://dx.doi.org/10.1080/14786435.2016.1142677}{{\bf 96} 895--921}

\bibitem{Gu2020}
Gu Y and El-Awady J~A 2020 {\em Mater. Theory\/}
  \href{http://dx.doi.org/10.1186/s41313-020-00020-2}{{\bf 4} 1}

\bibitem{MISHIN2019}
Mishin Y 2019 {\em Acta Mater.\/}
  \href{http://dx.doi.org/https://doi.org/10.1016/j.actamat.2019.08.046}{{\bf
  179} 383 -- 395}

\bibitem{Koju2020DirectAM}
Koju R~K and Mishin Y 2020 {\em Acta Mater.\/}
  \href{http://dx.doi.org/10.1016/j.actamat.2020.07.052}{{\bf 198} 111--120}

\bibitem{DarvishiKamachali2020}
{Darvishi Kamachali} R, {Kwiatkowski da Silva} A, McEniry E, Ponge D, Gault B,
  Neugebauer J and Raabe D 2020 {\em npj Comput. Mater.\/}
  \href{http://dx.doi.org/10.1038/s41524-020-00456-7}{{\bf 6} 191}

\bibitem{Elder_2010PFCmembranes}
Elder K~R and Huang Z~F 2010 {\em J. Condens. Matter Phys.\/}
  \href{http://dx.doi.org/10.1088/0953-8984/22/36/364103}{{\bf 22} 364103}

\bibitem{Kubstrup1996}
Kubstrup C, Herrero H and P\'erez-Garc\'{\i}a C 1996 {\em Phys. Rev. E\/}
  \href{http://dx.doi.org/10.1103/PhysRevE.54.1560}{{\bf 54} 1560}

\bibitem{Kundin_2017}
Kundin J and Choudhary M~A 2017 {\em Model. Simul. Mater. Sci. Eng.\/}
  \href{http://dx.doi.org/10.1088/1361-651x/aa6e48}{{\bf 25} 055004}

\bibitem{Marino2011}
Merino P, Svec M, Pinardi A~L, Otero G and Martin-Gago J~A 2011 {\em ACS
  Nano\/} \href{http://dx.doi.org/10.1021/nn201200j}{{\bf 5} 5627--5634}

\bibitem{Roos2011}
Roos M, Uhl B, K\"unzel D, Hoster H~E, Gro\ss A and Behm R~J 2011 {\em
  Beilstein J. Nanotechnol.\/}
  \href{http://dx.doi.org/https://doi.org/10.3762/bjnano.2.42}{{\bf 2}
  365--373}

\bibitem{Balog2010}
Balog R, J{\o}rgensen B, Nilsson L, Andersen M, Rienks E, Bianchi M, Fanette M,
  L{\ae}gsgaard E, Baraldi A, Lizzit S, Sljivancanin Z, Besenbacher F, Hammer
  B, Pedersen T~G, Hofmann P and Hornek{\ae}r 2010 {\em Nat. Mat.\/}
  \href{http://dx.doi.org/10.1038/nmat2710}{{\bf 9} 315--319}

\bibitem{Asaro72}
Asaro R~J and Tiller W~A 1972 {\em Metall. Trans.\/}
  \href{http://dx.doi.org/10.1007/BF02642562}{{\bf 3} 1789--1796}

\bibitem{Grinfeld86}
Grinfeld M 1993 {\em J. Nonlinear Sci.\/}
  \href{http://dx.doi.org/10.1007/BF02429859}{{\bf 31} 35--83}

\bibitem{Srolovitz89}
Srolovitz D~J 1989 {\em Acta Metall.\/}
  \href{http://dx.doi.org/10.1016/0001-6160(89)90246-0}{{\bf 37} 621--625}

\bibitem{Gunther95}
Gunther C, Vrijmoeth J, Hwang R~Q and Behm R~J 1995 {\em Phys. Rev. Lett.\/}
  \href{http://dx.doi.org/10.1103/PhysRevLett.74.754}{{\bf 74} 754}

\bibitem{Schmid97}
Schmid A~K, Bartelt N~C, Hamilton J~C, Carter C~B and Hwang R~Q 1997 {\em Phys.
  Rev. Lett.\/} \href{http://dx.doi.org/10.1103/PhysRevLett.78.3507}{{\bf 78}
  3507--3510}

\bibitem{elder2017striped}
Elder K~R, Achim C~V, Granato E, Ying S~C and Ala-Nissila T 2017 {\em Phys.
  Rev. B\/} \href{http://dx.doi.org/10.1103/PhysRevB.96.195439}{{\bf 96}
  195439}

\bibitem{Elder16}
Elder K~R, Achim C~V, Granato E, Ying S~C and Ala-Nissila T 2016 {\em EPL\/}
  \href{http://dx.doi.org/10.1209/0295-5075/116/56002}{{\bf 116} 56002}

\bibitem{Elder16B}
Elder K~R, Chen Z, Elder K~L~M, Hirvonen P, Mkhonta S~K, Ying S~C, Granato E,
  Huang Z~F and Ala~Nissila T 2016 {\em J. Chem. Phys.\/}
  \href{http://dx.doi.org/10.1063/1.4948370}{{\bf 144} 174703}

\bibitem{Elder13}
Elder K~R, Rossi G, Kanerva P, Sanches F, Ying S~C, Granato E, Achim C~V and
  Ala-Nissila T 2013 {\em Phys. Rev. B.\/}
  \href{http://dx.doi.org/10.1103/PhysRevB.88.075423}{{\bf 88} 075423}

\bibitem{Elder12}
Elder K~R, Rossi G, Kanerva P, Sanches F, Ying S~C, Granato E, Achim C~V and
  Ala-Nissila T 2012 {\em Phys. Rev. Lett.\/}
  \href{http://dx.doi.org/10.1103/PhysRevLett.108.226102}{{\bf 108} 226102}

\bibitem{CL95}
Chaikin P~M and Lubensky T~C 1995 {\em {Principles of Condensed Matter
  Physics}\/} (\href{https://doi.org/10.1017/CBO9780511813467}{Cambridge
  University Press})

\bibitem{Huang2008}
Huang Z~F and Elder K~R 2008 {\em Phys. Rev. Lett.\/}
  \href{http://dx.doi.org/10.1103/PhysRevLett.101.158701}{{\bf 101} 158701}

\bibitem{HuangPRB2010}
Huang Z~F and Elder K~R 2010 {\em Phys. Rev. B\/}
  \href{http://dx.doi.org/10.1103/PhysRevB.81.165421}{{\bf 81} 165421}

\bibitem{Spencer91}
Spencer B~J, Voorhees P~W and Davis S~H 1991 {\em Phys. Rev. Lett.\/}
  \href{http://dx.doi.org/10.1103/PhysRevLett.67.3696}{{\bf 67} 3696}

\bibitem{Spencer93}
Spencer B~J, Voorhees P~W and Davis S~H 1993 {\em J. Appl. Phys.\/}
  \href{http://dx.doi.org/10.1063/1.353815}{{\bf 73} 4955}

\bibitem{Levine07}
Levine M~S, Golovin A~A, Davis S~H and Voorhees P~W 2007 {\em Phys. Rev. B\/}
  \href{http://dx.doi.org/10.1103/PhysRevB.75.205312}{{\bf 75} 205312}

\bibitem{Eisenberg00}
Eisenberg H~R and Kandel D 2000 {\em Phys. Rev. Lett.\/}
  \href{http://dx.doi.org/10.1103/PhysRevLett.85.1286}{{\bf 85} 1286}

\bibitem{Muller99}
M\"uller J~M and Grant M 1999 {\em Phys. Rev. Lett.\/}
  \href{http://dx.doi.org/10.1103/PhysRevLett.82.1736}{{\bf 82} 1736}

\bibitem{Kassner01}
Kassner K, Misbah C, M\"uller J, Kappey J and Kohlert P 2001 {\em Phys. Rev.
  E\/} \href{http://dx.doi.org/10.1103/PhysRevE.63.036117}{{\bf 63} 036117}

\bibitem{Wu2009}
Wu K~A~W and Voorhees P~W 2009 {\em Phys. Rev. B\/}
  \href{http://dx.doi.org/10.1103/PhysRevB.80.125408}{{\bf 80} 125408}

\bibitem{Sutter2000}
Sutter P and Lagally M~G 2000 {\em Phys. Rev. Lett.\/}
  \href{http://dx.doi.org/10.1103/PhysRevLett.84.4637}{{\bf 84} 4637}

\bibitem{Tromp2000}
Tromp R~M, Ross F~M and Reuter M~C 2000 {\em Phys. Rev. Lett.\/}
  \href{http://dx.doi.org/10.1103/PhysRevLett.84.4641}{{\bf 84} 4641}

\bibitem{Bergamaschini2016}
Bergamaschini R, Salvalaglio M, Backofen R, Voigt A and Montalenti F 2016 {\em
  Adv. Phys. X\/} \href{http://dx.doi.org/10.1080/23746149.2016.1181986}{{\bf
  1} 331--367}

\bibitem{Huang2005}
Huang M, Rugheimer P, Lagally M~G and Liu F 2005 {\em Phys. Rev. B\/}
  \href{http://dx.doi.org/10.1103/PhysRevB.72.085450}{{\bf 72} 085450}

\bibitem{KimLee2009}
Kim-Lee H~J, Savage D~E, Ritz C~S, Lagally M~G and Turner K~T 2009 {\em Phys.
  Rev. Lett.\/} \href{http://dx.doi.org/10.1103/PhysRevLett.102.226103}{{\bf
  102} 226103}

\bibitem{Huang2009}
Huang M~H, Ritz C~S, Novakovic B, Yu D, Zhang Y, Flack F, Savage D~E, Evans
  P~G, Knezevic I, Liu F and Lagally M~G 2009 {\em ACS Nano\/}
  \href{http://dx.doi.org/10.1021/nn8008883}{{\bf 3} 721--727}

\bibitem{SHENOY1999611}
Shenoy V, Miller R, Tadmor E, Rodney D, Phillips R and Ortiz M 1999 {\em J.
  Mech. Phys. Solids\/}
  \href{http://dx.doi.org/https://doi.org/10.1016/S0022-5096(98)00051-9}{{\bf
  47} 611--642}

\bibitem{Curtin_2003}
Curtin W~A and Miller R~E 2003 {\em Model. Simul. Mater. Sci. Eng.\/}
  \href{http://dx.doi.org/10.1088/0965-0393/11/3/201}{{\bf 11} R33--R68}

\end{thebibliography}
\end{document}